\input harvmac 
\input epsf.tex
\def\IN{\relax{\rm I\kern-.18em N}}
\def\IR{\relax{\rm I\kern-.18em R}}
\font\cmss=cmss10 \font\cmsss=cmss10 at 7pt
\def\IZ{\relax\ifmmode\mathchoice
{\hbox{\cmss Z\kern-.4em Z}}{\hbox{\cmss Z\kern-.4em Z}}
{\lower.9pt\hbox{\cmsss Z\kern-.4em Z}}
{\lower1.2pt\hbox{\cmsss Z\kern-.4em Z}}\else{\cmss Z\kern-.4em Z}\fi}

\overfullrule=0mm
\def\file#1{#1}
\newcount\figno \figno=0
\newcount\figtotno      \figtotno=0
\newdimen\captionindent \captionindent=1cm
\def\figbox#1#2{\epsfxsize=#1\vcenter{\epsfbox{\file{#2}}}}
\newcount\figno
\figno=0
\def\fig#1#2#3{
\par\begingroup\parindent=0pt\leftskip=1cm\rightskip=1cm\parindent=0pt
\baselineskip=11pt
\global\advance\figno by 1
\midinsert
\epsfxsize=#3
\centerline{\epsfbox{#2}}
\vskip 12pt
{\bf Fig. \the\figno:} #1\par
\endinsert\endgroup\par
}
\def\figlabel#1{\xdef#1{\the\figno}}
\def\encadremath#1{\vbox{\hrule\hbox{\vrule\kern8pt\vbox{\kern8pt
\hbox{$\displaystyle #1$}\kern8pt}
\kern8pt\vrule}\hrule}}
\def\enca#1{\vbox{\hrule\hbox{\vrule\kern8pt\vbox{\kern8pt
\hbox{$\displaystyle #1$}\kern8pt}
\kern8pt\vrule}\hrule}}


\def\IR{\relax{\rm I\kern-.18em R}}
\font\cmss=cmss10 \font\cmsss=cmss10 at 7pt
\def\IZ{\relax\ifmmode\mathchoice
{\hbox{\cmss Z\kern-.4em Z}}{\hbox{\cmss Z\kern-.4em Z}}
{\lower.9pt\hbox{\cmsss Z\kern-.4em Z}}
{\lower1.2pt\hbox{\cmsss Z\kern-.4em Z}}\else{\cmss Z\kern-.4em Z}\fi}
\def\buildrel#1\under#2{\mathrel{\mathop{\kern0pt #2}\limits_{#1}}}


\Title{SPhT/96-008}
{{\vbox {
\bigskip
\centerline{Meanders and the Temperley-Lieb algebra}
}}}
\bigskip
\centerline{P. Di Francesco,}
\medskip
\centerline{O. Golinelli}
\medskip
\centerline{and} 
\medskip
\centerline{E. 
Guitter\footnote*{e-mails: philippe,golinel,guitter@amoco.saclay.cea.fr},}

\bigskip

\centerline{ \it Service de Physique Th\'eorique,}

\centerline{C.E.A. Saclay,}

\centerline{ \it F-91191 Gif sur Yvette Cedex, France}

\vskip .5in


The statistics of meanders is studied in connection with the Temperley-Lieb algebra.
Each (multi-component) meander corresponds to a pair of reduced elements of the
algebra.  The assignment of a weight $q$ per connected component of meander 
translates into a bilinear form on the algebra, with a Gram matrix encoding the
fine structure of meander numbers.  
Here, we calculate the associated Gram determinant as a function of $q$, and
make use of the orthogonalization process to derive alternative expressions
for meander numbers as sums over correlated random walks.

\noindent
\Date{}

\nref\HMRT{K. Hoffman, K. Mehlhorn, P. Rosenstiehl and
R. Tarjan, {\it Sorting Jordan sequences in linear time using level-linked
search trees}, Information and Control {\bf 68} (1986) 170-184.}
\nref\PHI{A. Phillips, {\it La topologia dei labirinti}, in M. Emmer, ed.
{\it L' occhio di Horus: itinerario nell'immaginario matematico}, Istituto
della Enciclopedia Italia, Roma (1989) 57-67.}
\nref\ARNO{V. Arnold, {\it The branched covering of $CP_2 \to S_4$,
hyperbolicity and projective topology}, 
Siberian Math. Jour. {\bf 29} (1988) 717-726.}
\nref\KOSMO{K.H. Ko, L. Smolinsky, {\it A combinatorial matrix in
$3$-manifold theory}, Pacific. J. Math {\bf 149} (1991) 319-336.}
\nref\LZ{S. Lando and A. Zvonkin, {\it Plane and Projective Meanders}, Theor. Comp.
Science {\bf 117} (1993) 227-241, 
and {\it Meanders}, Selecta Math. Sov. {\bf 11} (1992) 117-144.}
\nref\DGG{P. Di Francesco, O. Golinelli and E. Guitter, {\it Meander,
folding and arch statistics}, to appear in Journal of Mathematical and Computer
Modelling (1996).}
\nref\MAK{Y. Makeenko, {\it Strings, Matrix Models and Meanders}, proceedings
of the 29th Inter. Ahrenshoop Symp., Germany (1995).}
\nref\TOU{J. Touchard, {\it Contributions \`a l'\'etude du probl\`eme des
timbres poste}, Canad. J. Math. {\bf 2} (1950) 385-398.}
\nref\LUN{W. Lunnon, {\it A map--folding problem}, 
Math. of Computation {\bf 22} 
(1968) 193-199.}
\nref\TLA{H. Temperley and E. Lieb, {\it Relations between the percolation and 
coloring problem and other graph-theoretical problems associated with regular 
planar lattices: some exact results for the percolation problem}, Proc. Roy. Soc. 
{\bf A322} (1971) 251-280.}
\nref\MARTIN{P. Martin, {\it Potts models and related problems in statistical
mechanics}, World Scientific (1991).}


\newsec{Introduction}

The meander problem is one of these fundamental combinatorial problems 
with a simple formulation, which resist the repeated attempts to solve
them.  The problem is to count the number $M_n$ of meanders of order $n$, 
i.e. of inequivalent configurations
of a closed non-self-intersecting
loop crossing an infinite line through $2n$ points.  
The infinite line may be
viewed as a river flowing from east to west, and the loop as a closed circuit
crossing this river through $2n$ bridges.  
Two configurations are considered as equivalent if they are smooth deformations
of one another. 

Apparently, the meander problem dates back the work of Poincar\'e about
differential geometry. Since then, it arose in various domains such as
mathematics, physics, computer science \HMRT\
and fine arts \PHI. In the late 80's, Arnold reactualized
this problem in relation with Hilbert's 16th problem, concerning the enumeration of
ovals of planar algebraic curves \ARNO. 
Meanders also emerged in the classification
of 3-manifolds \KOSMO.  More recently, random matrix model techniques, 
borrowed from quantum field theory, were applied to this problem \LZ\ \DGG.
As such, the meander problem seems to belong to the same class as large N QCD \MAK.

In a previous paper \DGG, we made our first incursion into the meander problem,
in trying to solve the {\it compact folding} problem of a polymer chain.
Considering indeed a long closed polymer chain of say $2n$ identical monomers,
we ask the question of counting the inequivalent ways of folding the whole
chain onto itself, forbidding interpenetration of monomers. 
By compact folding, we mean that all the monomers are packed on top of each other.
Accordingly, folding is a simple realization of objects with self-avoiding
constraints. 
The reader may bear in mind the simple image of the folding of a closed strip
of $2n$ stamps, with all stamps piled up on top of each other \TOU\ \LUN.

\fig{A compactly folded polymer (a) with $2n=6$ monomers, and
the associated meander (b), obtained by drawing a line (river) horizontally
through the monomers. Each monomer becomes a bridge, and each hinge
a segment of road between two bridges.}{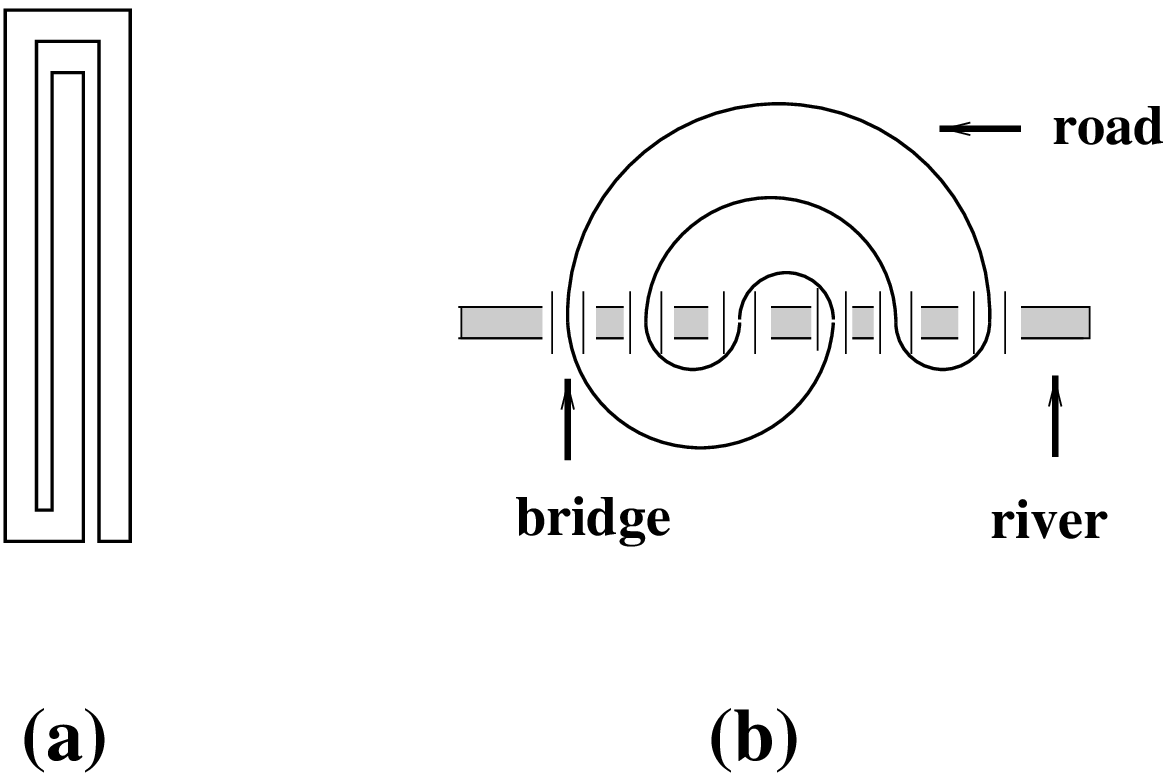}{7.cm}
\figlabel\comp

\noindent{}The equivalence between this folding problem and the meander 
problem may be
seen as follows.
As illustrated in Fig.\comp, drawing a line (river) across the 
$2n$ constituents (bridges)
of the folded polymer, and pulling them apart, 
produces a meander of order $n$. 
The folding of a {\it closed} polymer chain and the meander problem
are therefore completely identical.
By analogy, we were led to define 
the meander counterpart of the folding problem of an {\it open} polymer chain:
the semi-meanders. The latter are defined in the same way as meanders, 
except that the river is now semi-infinite, i.e. it has a source, around which
the semi-meander is allowed to wind freely. We denote by ${\bar M}_n$ the
number of semi-meanders of order $n$, namely with $n$ bridges.

In this paper, we reconsider the meander and semi-meander problems in the
framework of the Temperley-Lieb algebra \TLA. 
This is based on a one-to-one
correspondence between (multicomponent) semi-meanders and reduced elements 
of the Temperley-Lieb algebra. 
Similarly, (multicomponent) meanders are associated
to {\it pairs} of such elements.
More precisely, the Temperley-Lieb algebra is endowed with a bilinear structure 
out of which a Gram matrix can be constructed.  In our language, the bilinear
form associates to each pair of elements of the algebra
a weight $q^c$, where $c$ denotes the number of connected 
components of the associated meander.  
In particular, the Gram matrix, as a polynomial of $q$,
encodes all the relevant information about meander and semi-meander numbers. 

Here we obtain as a main result an exact compact expression 
for the determinant of the Gram matrix, referred to as the 
{\it meander determinant}. 
Far from solving the question of enumerating meanders, this gives however some 
partial information on the problem, and produces an exact solution to a
meander-flavored issue.
This result is summarized in Eq.(5.6), and proved by explicit
orthogonalization of the Gram matrix.
In a second step, we make use of the precise form of the change of basis 
in the orthogonalization process to derive various expressions
for the semi-meander (Eq.(6.62))
and meander (Eq.(6.63)) numbers as statistical sums over paths, with
an interpretation as Solid On Solid (SOS) model partition functions.

The paper is organized as follows. We start in Sect.2 by giving basic 
definitions of (multi-component) meander (Eq.(2.1)) and semi-meander 
(Eq.(2.3)) numbers and associated
polynomials in which a weight $q$ is assigned to each connected component.
The relation between (semi-)meanders and both arch configurations and walk 
diagrams is then discussed, and known results for $q=\pm 1$ are reviewed
(Eqs.(2.6)-(2.8)).
Various conjectured and/or numerical asymptotic behaviors for large $n$
are given (Eqs.(2.11)-(2.18)).  
In Sect.3, we introduce the Temperley-Lieb algebra $TL_n(q)$,
and discuss its relation with walk diagrams and arch configurations, 
in one-to-one correspondence with reduced elements of the algebra. 
These reduced elements form a natural basis (basis 1) of $TL_n(q)$.
The contact with meanders is
made through the introduction of a trace and a bilinear form on $TL_n(q)$
(Eqs.(3.11) and (3.14)).
When evaluated on pairs of reduced elements (of the basis 1), this form
generates the Gram matrix (Eq.(3.15)), which encodes the fine structure of meander
numbers.  
In Sect.4, we make a change from basis 1 to a new basis 2,
in which the Gram matrix is diagonal. This allows for the calculation of the 
Gram determinant as a function of $q$ (Eq.(5.6)), and the identification of 
its zeros (Eq.(5.10))
and their multiplicities (Eq.(5.23)). 
These results, together with a complete 
combinatorial proof are detailed in Sect.5.
The matrix for the change of basis $1 \to 2$ is studied in 
great detail in Sect.6,
where it is shown to obey a simple recursion relation (Eq.(6.29)).
This equation is explicitly solved, in the form of matrix elements between
two walk diagrams,
factorized into a selection rule (with value $0$ or $1$, see Eq.(6.38)) 
multiplied
by some weight, with a local dependence on the heights of the walk diagrams
(Eq.(6.43)).
This leads to expressions for the meander and semi-meander polynomials as 
sums over selected walk diagrams (Eqs.(6.62) and (6.63)). 
Analogous formulas are derived within
the framework of SOS models (Eq.(6.90)), leading to various conjectures 
as to the asymptotic form of the meander and semi-meander polynomials 
for $q\geq 2$.
Sect.7 is devoted to a refinement of the meander determinant for 
semi-meanders with fixed number of windings around the source of the river 
(Eq.(7.5)). 
A few concluding remarks are gathered in Sect.8.  
Some technical ingredients are detailed in Appendices A,B and C.

\newsec{Definitions}

\subsec{Meanders}

\fig{The four meanders of order $n=2$, i.e. with $2n=4$ bridges. 
The two first ones have $k=1$
connected component, the two other have $k=2$ 
connected components.}{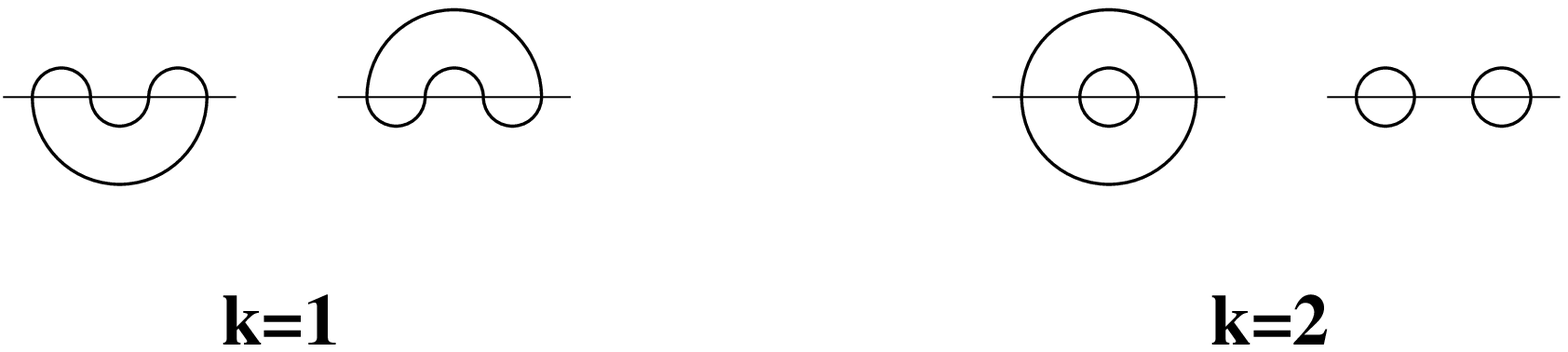}{10.cm}
\figlabel\xmean

A {\it meander} of order $n$ is a planar configuration of a closed 
non-self-intersecting loop (road) crossing an infinite oriented
line (river flowing from east to west) through $2n$ points (bridges). 
We denote by $M_n$ the number of 
topologically inequivalent meanders of order $n$.
We extend the definition to a set of $k$ roads (i.e., a meander
with $k$ possibly interlocking 
connected components). 
The number of meanders with $k$ connected components 
is denoted by $M_n^{(k)}$. Note that necessarily $1\leq k \leq n$.
These numbers are summarized in the 
meander polynomial
\eqn\mepo{\enca{ m_n(q)~=~ \sum_{k=1}^n M_n^{(k)}~q^k } }
The various meanders corresponding to $n=2$ are depicted in Fig.\xmean.
They correspond to the polynomial
\eqn\expome{ m_2(q)~=~ 2q+2 q^2 }
The numbers $M_n^{(k)}$ are listed in \DGG\ for $1 \leq k \leq n \leq 12$.

\subsec{Semi-meanders}

\fig{The five semi-meanders of order $n=3$, arranged according to 
their numbers $k=1,2,3$ of connected components.}{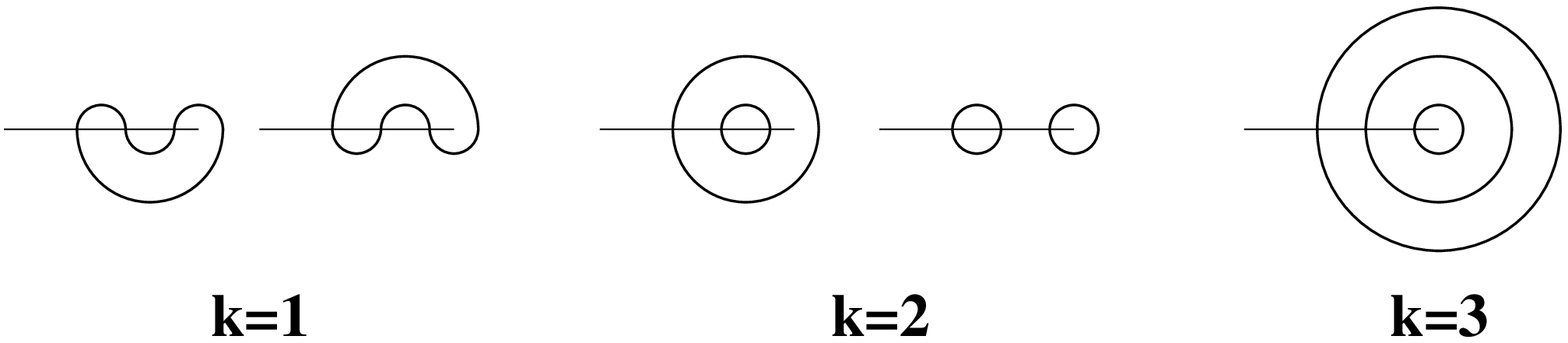}{10.cm}
\figlabel\xsmean

A {\it semi-meander} of order $n$ is a planar configuration of a closed
non-self-intersecting loop (road) crossing a semi-infinite line 
(river with a source) through $n$ points (bridges). 
Note that, in a semi-meander, the road may wind around the 
source of the river.
We denote by ${\bar M}_n$ the number of 
topologically inequivalent semi-meanders of order $n$, and by
${\bar M}_n^{(k)}$ the number of semi-meanders with $k$ connected 
components, $1 \leq k \leq n$. We also have the semi-meander polynomial
\eqn\sempo{\enca{{\bar m}_n(q)~=~ \sum_{k=1}^n {\bar M}_n^{(k)}~q^k }}
The various semi-meanders corresponding to $n=3$ are depicted in Fig.\xsmean.
They correspond to the polynomial
\eqn\exposme{ {\bar m}_3(q)~=~ 2q+2 q^2+q^3 }
The numbers ${\bar M}_n^{(k)}$ are listed in \DGG\ for $1 \leq k \leq n \leq 14$.

\subsec{Arch configurations and (semi) meanders}

\fig{Any meander is obtained as the superimposition of a top
($a$) and bottom ($b$) arch configurations of same order ($n=5$ here).
An arch configuration is a planar pairing
of the $(2n)$ bridges through $n$ non-intersecting arches 
lying above the river (by convention, we represent
the lower configuration $b$ reflected with respect to the 
river).}{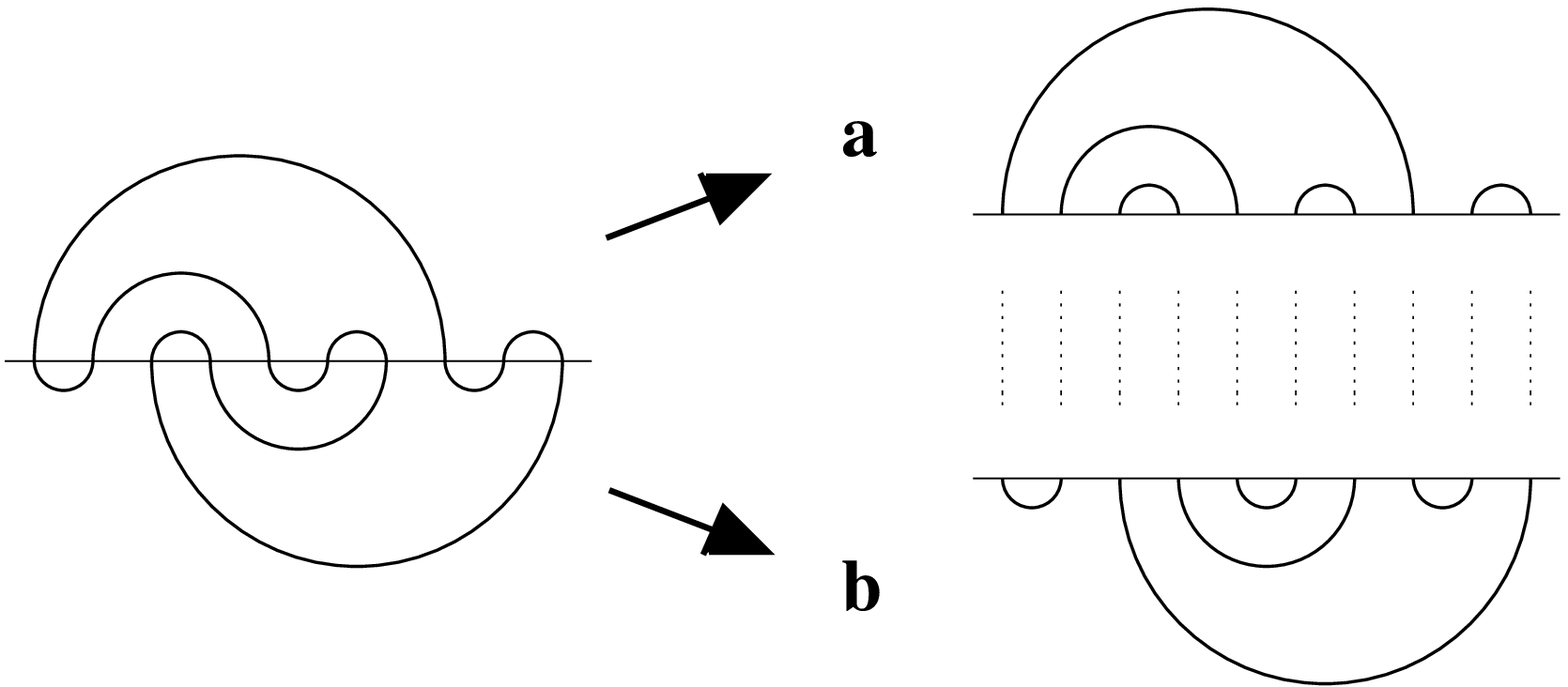}{10.cm}
\figlabel\archm

A multicomponent meander may be viewed as the superimposition of two
(top and bottom)
{\it arch configurations} of order $n$, corresponding respectively 
to the configurations of the road on both sides of the river, 
as shown in Fig.\archm.
An arch configuration is simply a configuration
of $n$ planar non-intersecting arches (lying, say, above the river)
linking the $2n$ bridges by pairs.
The number of arch configurations of order $n$ is given by the Catalan number
\eqn\cata{ c_n~=~ {(2n)! \over (n+1)! n!} }
The set of arch configurations of order $n$ is denoted by $A_n$.

As an immediate consequence, as arbitrary multicomponent meanders
are obtained by superimpositions of arbitrary top and bottom arch
configurations, we have
\eqn\sume{ m_n(1)~=~ (c_n)^2 }

\fig{Any semi-meander may be viewed as a particular meander
by opening the semi-infinite river as indicated by the arrows.
This doubles the number of bridges in the resulting meander,
hence the order is conserved ($n=5$ here). By construction,
the lower arch configuration of the meander is always a rainbow 
arch configuration of same order. The number of connected components
($k=3$ here) is conserved in the transformation.}{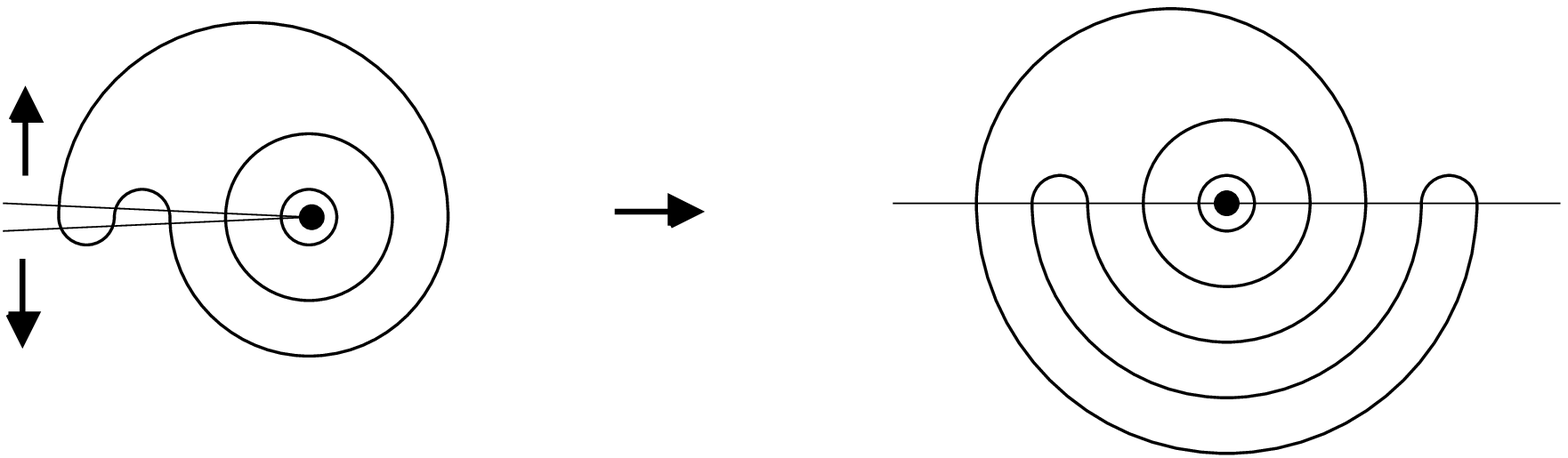}{10.cm}
\figlabel\semimean

As for semi-meanders, upon opening the semi-infinite river and dedoubling
the bridges (cf. Fig.\semimean), 
they can also be viewed as the superimposition of a top arch
configuration of order $n$, and of a particular bottom ``rainbow" 
arch configuration (namely that linking the $i$-th bridge to the
$(2n+1-i)$-th one, $i=1,2,...,n$). 
Therefore arbitrary multicomponent semi-meanders may be obtained by
superimposing an arbitrary arch configuration with a rainbow of order $n$,
leading to
\eqn\sumem{ {\bar m}_n(1)~=~ c_n}

In ref.\DGG, we have also proved the following results
\eqn\smin{\eqalign{m_n(-1)~
&=~\left\{\matrix{ 0 &\  \hbox{if $n=2p$} \cr
-(c_p)^2 & \ \hbox{if $n=2p+1$} \cr}\right. ~\cr
{\bar m}_n(-1)~&=~\left\{\matrix{ 0 &\  \hbox{if $n=2p$} \cr
-(c_p) & \ \hbox{if $n=2p+1$} \cr}\right. ~\cr } }

Note that the one-component meander and semi-meander numbers are 
recovered in the $q \to 0$ limit of respectively $m_n(q)/q$ and 
${\bar m}_n(q)/q$.

\subsec{Walk diagrams}

An arch configuration of order $n$ may be viewed as
a closed random walk of $2n$ steps
on a semi-infinite line, or equivalently its two-dimensional extent,
which we call a {\it walk diagram}, defined as follows.
Let us first label the segments of river between consecutive bridges,
namely the segment $i$ lies between the $i$-th and the $(i+1)$-th bridge,
for $i=1,2,...,2n-1$.
Let us also label by $0$ and $2n$, the semi-infinite portions of river
respectively to the left of the first bridge and to the right of the last one.
To each portion of river $i$, we attach a height
$\ell_i$ equal to the number of arches passing at the vertical
of $i$.  
The nonnegative integers $\ell_i$ satisfy the following
conditions
\eqn\heigprop{\eqalign{ 
\ell_0~&=~\ell_{2n}~=~0 \cr
\ell_{i+1} -\ell_i~&\in ~\{ \pm 1 \} \quad i=0,1,...,2n-1 \cr} }
The diagram formed by the broken line joining the successive
points $(i,\ell_i)$, $i=0,1,...,2n$, is the walk diagram corresponding 
to the initial arch configuration. This diagram represents the two-dimensional
extent of a walk of $2n$ steps on the semi-infinite line $\ell\geq 0$
starting and ending at its origin.

\fig{A walk diagram of $18$ steps, and the corresponding arch configuration. Each dot corresponds to a segment of river. The height on
the walk diagram is given by the number of arches intersected by the 
vertical dotted line.}{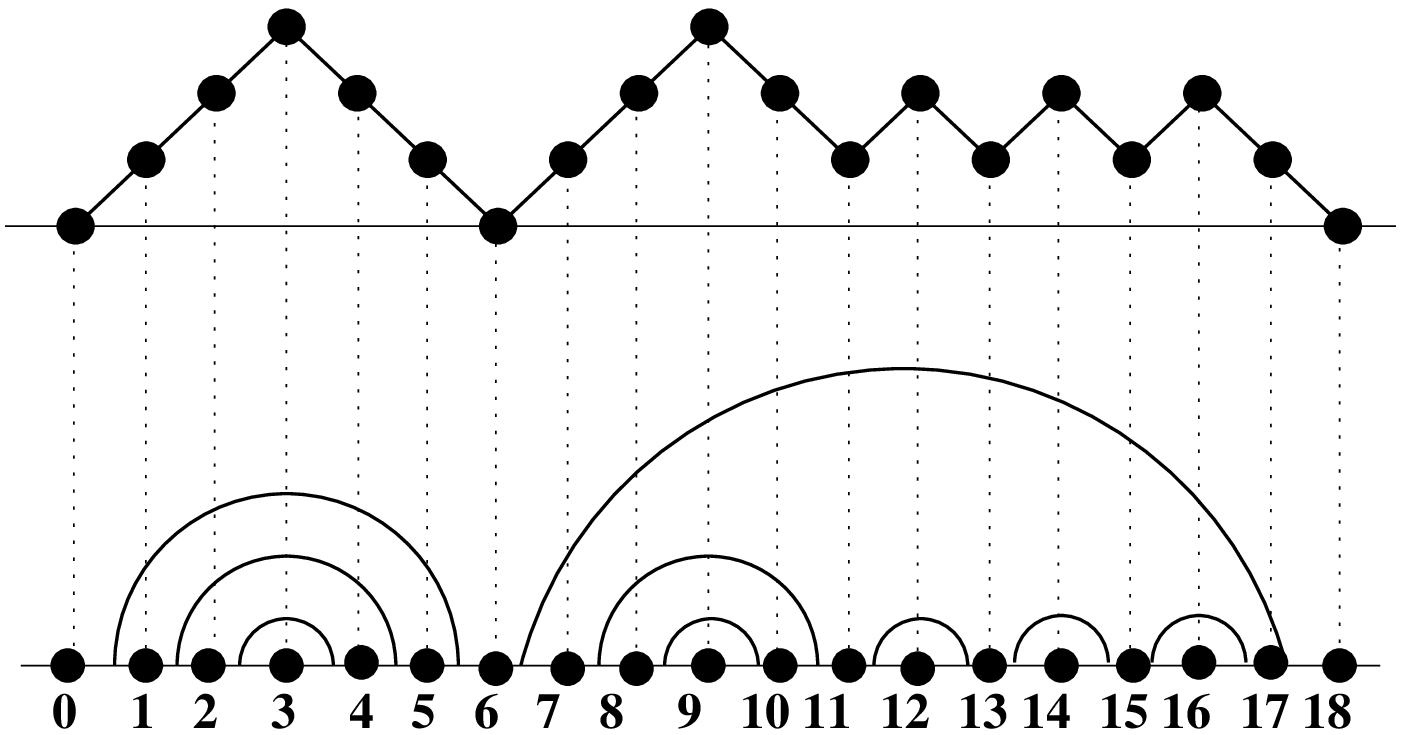}{9.cm}
\figlabel\walkyrie

Conversely, any walk diagram of $2n$ steps, characterized by integer heights 
$\ell_i \geq 0$, $i=0,...,2n$, satisfying \heigprop, corresponds to a unique
arch configuration of order $n$. To construct the 
arch configuration corresponding to a walk
diagram, notice that, going from left to right along the river,
whenever $\ell_{i}-\ell_{i-1}=1$, a
new arch originates from the bridge $i$, whereas when 
$\ell_{i}-\ell_{i-1}=-1$, an arch terminates at the bridge $i$.
We denote by $W_n$ the set of walk diagrams of $2n$ steps.
We have the identification
\eqn\eqset{ W_n~ \equiv ~ A_n}

In this paper, we will alternatively use the arch configuration
and walk diagram pictures.

\subsec{Asymptotics}

Earlier numerical work \LUN\ \LZ\ \DGG\ suggests that the (one-component)
meander and semi-meander numbers behave in the large $n$ limit as respectively
\eqn\asymp{\eqalign{M_n~&\sim ~ {R^n \over n^\alpha} \cr
{\bar M}_n~&\sim ~ {{\bar R}^n \over n^\gamma} \cr}}
with
\eqn\asexp{\eqalign{ {\bar R}\simeq 3.5... &\qquad R= {\bar R}^2 \cr
\alpha~=~7/2 &\qquad \gamma~=~ 2 \cr}}
The values of the exponents $\alpha$ and $\gamma$ are conjectured to be
exact. 
The relation $R={\bar R}^2$ is a consequence of the polymer folding 
interpretation \DGG:
the entropy per monomer is the same for the open and closed polymer folding
problems. Note however that
the configuration exponents $\alpha$ and $\gamma$ depend 
on the boundary conditions (open or closed).
A natural quantity of interest for the study of semi-meanders is the 
{\it winding}, namely the number of times the road winds around 
the source of the river in the river/road picture of the semi-meander. 
In the arch configuration picture,
the winding of a semi-meander is the number of arches of the upper configuration
passing at the vertical of 
the middle point; representing the upper arch configuration as a walk diagram 
$a$, the winding of the semi-meander is simply $\ell_n^a$.
Denoting by $c(a)$ the number of connected components of the superimposition
of the arch configuration $a$ and of a rainbow configuration of order $n$, the
average winding in semi-meanders of order $n$ reads
\eqn\avwind{   w_n(q)~=~ {\sum_{a \in W_n} \ell_n^a \, q^{c(a)} \over
\sum_{a \in W_n} q^{c(a)} }~
{\ \atop {\displaystyle\sim \atop n \to \infty} }~n^{\nu(q)} }
where we have identified a winding exponent $\nu(q)\in[0,1]$.
In this paper, we give strong analytical evidence that $\nu(q)=1$ for 
all $q \geq 2$. For $0<q<2$, numerical work seems to indicate that 
$1/2 \leq \nu(q) <1$.

More generally, we expect the meander and semi-meander polynomials 
to behave for large $n$ as
\eqn\poas{\eqalign{ m_n(q)~&\sim~ {R(q)^n \over n^{\alpha(q)}} \cr
{\bar m}_n(q)~&\sim ~ {{\bar R}(q)^n \over n^{\gamma(q)} }\cr}}
where
$R(q)={\bar R}(q)^2$ like in the $q=0$ 
case, but only for $q<2$\foot{This relation is only expected insofar as $\nu(q)<1$.
Indeed, in this case, comparing the numbers $m_n(q)$ and ${\bar m}_{2n}(q)$
of respectively meanders and semi-meanders with $(2n)$ bridges, we see that 
the semi-meanders with significative winding (i.e., $w_n \sim n$) 
are negligible, hence we expect the two numbers to be of the same order,
namely
$$ m_n(q)~\sim~ R(q)^n ~\sim {\bar m}_{2n}(q)~\sim ~ {\bar R}(q)^{2n} $$
hence $R(q)={\bar R}(q)^2$.  According to the previous discussion,
this fails for $q \geq 2$, where $\nu(q)=1$. Indeed, it is easy to see that, for
large $q$, $R(q)/{\bar R}(q)^2 \sim 4/q\to 0$, as $m_n(q)\sim c_n q^n\sim (4q)^n$ and
${\bar m}_{2n}(q)\sim q^{2n}$.}

As an element of comparison, by using Stirling's formula for factorials, 
we have
\eqn\asycat{\eqalign{ {\bar m}_n(1)~&=~c_n ~\sim ~{4^n \over n^{3/2} }\cr
m_n(1)~&=~(c_n)^2~\sim~ {4^{2n} \over n^3}\cr}}
hence ${\bar R}(1)=4$ and
\eqn\expocat{ \alpha(1)~=~3 \qquad \gamma(1)~=~3/2 }
We also have the obvious large $q$ asymptotics
\eqn\larqas{\eqalign{ {\bar m}_{n}(q)~&\sim~ q^{n} \cr
m_n(q)~&\sim~ c_n \, q^n~\sim~ {(4q)^n \over n^{3/2}} \cr} }
hence $R(q)\sim 4q$ and ${\bar R}(q)\sim q$, whereas
\eqn\expoinf{ \alpha(\infty)~=~ {3 \over 2} \qquad \gamma(\infty)~=~0}

\newsec{Temperley-Lieb algebra and meanders}

\subsec{The Temperley-Lieb algebra and arch configurations}

The Temperley-Lieb algebra of order $n$ and parameter $q$, denoted by
$TL_n(q)$, is defined through its $n$ generators
$1,e_1,e_2,...,e_{n-1}$ subject to the relations
\eqn\tla{\eqalign{(i)\ \ \ \ \ \ \ \ \  e_i^2 ~&=~ 
q \, e_i \quad i=1,2,...,n-1\cr
(ii)\ \ \ \ [e_i,e_j]~&=~0 \quad {\rm if}\ |i-j|>1 \cr
(iii)\ e_i\, e_{i \pm 1}\, e_i~&=~ e_i  \quad i=1,2,...,n-1\cr}}
This definition becomes clear in the ``braid" pictorial representation,
where the generators act on $n$ parallel strings as follows:
\eqn\braid{ 1~=~\figbox{2.cm}{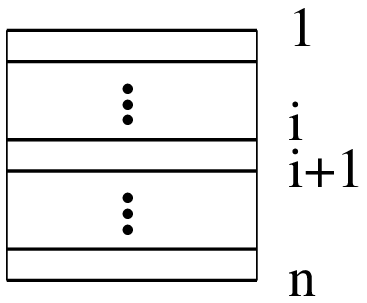} \qquad e_i~=~\figbox{2.cm}{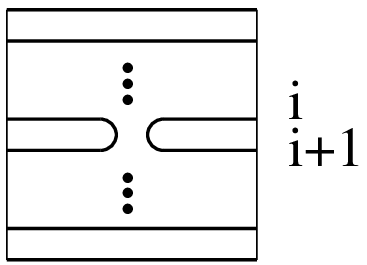} }
and a product of elements is represented by the juxtaposition of the 
corresponding braid diagrams.
The relation (ii) expresses the locality of the $e$'s, 
namely that the $e$'s commute whenever they involve distant strings.
The relations (i) and (iii) read respectively
\eqn\unbraid{\eqalign{(i)\ \ \ \ \ \ \ \ \ e_i^2~&=~ 
\figbox{2cm}{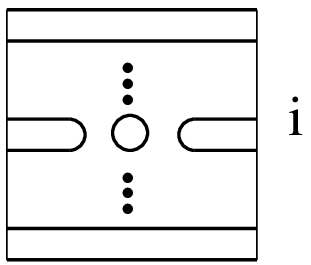}~=~q~\figbox{2.4cm}{ei.eps}~=~q \, e_i\cr
(iii)\ e_i\, e_{i+1}\, e_i~&=~ 
\figbox{2.4cm}{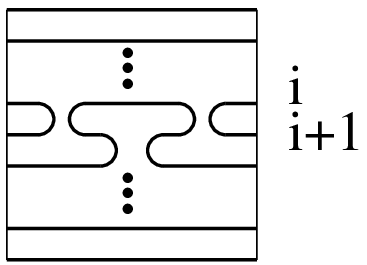}~=~\figbox{2.4cm}{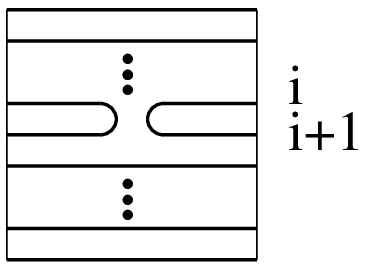}~=~ e_i\cr}}
In the relation (i), the loop has been erased, but affected the weight $q$.
The relation (iii) is simply obtained by stretching the $(i+2)$-th string.

\subsec{The basis 1}

The algebra $TL_n(q)$ is built out of arbitrary products of generators $e_i$.
Up to numerical factors depending on $q$, any such product
can be reduced by using the relations (i)-(iii). The algebra
$TL_n(q)$, as a real vector space, is therefore naturally endowed with
the basis formed by all the distinct reduced elements
of the algebra. This basis will be referred to as {\it basis 1} 
in the following.
For illustration, the reduced elements of 
$TL_3(q)$ read
\eqn\redthree{\eqalign{ 1~=~ \figbox{1.5cm}{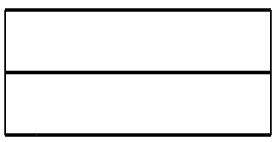} \quad 
e_1~&=~ \figbox{1.5cm}{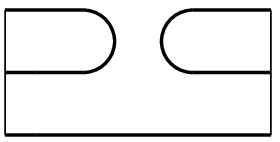}  \quad 
e_2~=~ \figbox{1.5cm}{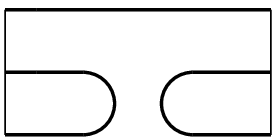}  \cr
e_1 e_2~=~ \figbox{2.cm}{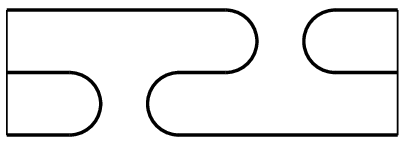}  &\quad 
e_2 e_1~=~ \figbox{2.cm}{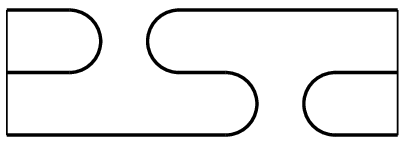}\cr} }

\fig{The transformation of a reduced element of $TL_9(q)$
into an arch configuration of order $9$. The reduced
element reads $e_3 e_4 e_2 e_5 e_3 e_1 e_6 e_4 e_2$.}{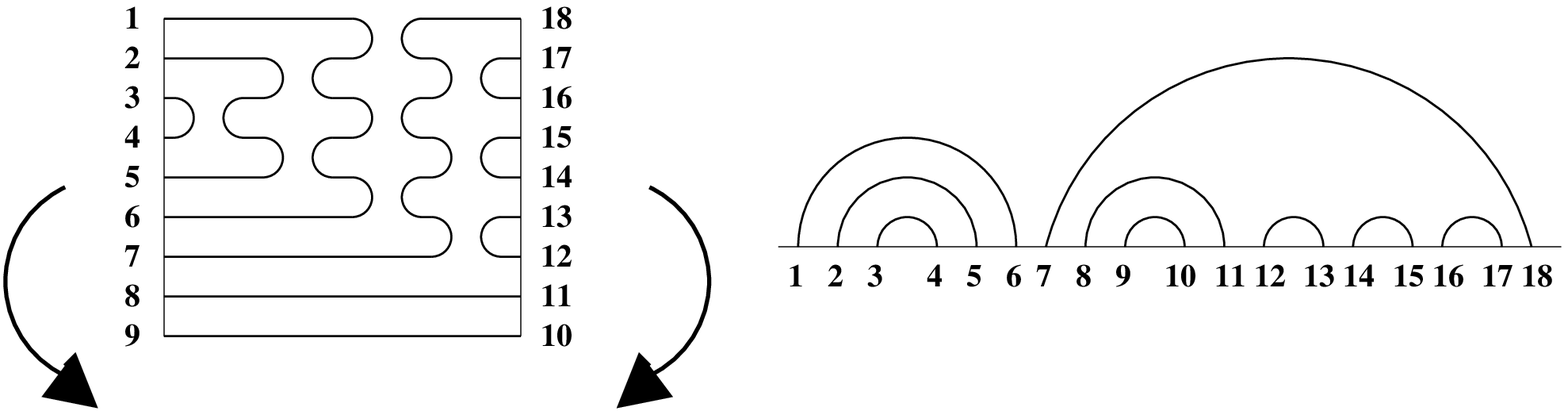}{10.cm}
\figlabel\tlarch

Let us now show that the reduced elements of $TL_n(q)$ are in one to one 
correspondence with arch configurations of order $n$.
This is most clearly seen by considering the braid pictorial
representation of a reduced element. Such a diagram has no internal
loop (by virtue of (i)), and all its strings are stretched (using (iii)).
As shown in Fig.\tlarch, one can construct a unique arch configuration
of order $n$ by deforming
the diagram so as to bring the ($2n$) ends of the strings on a line. This
deformation is invertible, and we conclude that, as a vector space,
$TL_n(q)$ has dimension 
\eqn\dimtla{ {\rm dim} ( TL_n(q) )~=~ c_n }

The basis 1 is best expressed in the language of walk diagrams. The
walk diagrams of $2n$ steps are arranged according to their
middle height $\ell_n=h$, where $h=n-2p$, $0 \leq p \leq n/2$. 
For each value of $h$,
the basic reduced element 
\eqn\redbas{f_h^{(n)}~=~e_1 e_3 e_5 ... e_{2p-1} \qquad f_n^{(n)}~=~1}
corresponds to the lowest walk diagram ${\cal W}_h^{(n)}$
with middle height $h$, namely 
\eqn\deffh{ {\cal W}_h^{(n)}~=~ \figbox{6.cm}{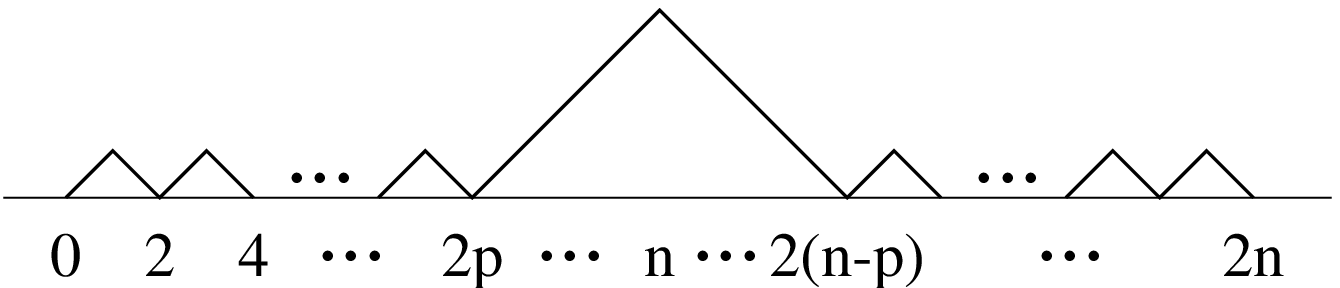}}
with
\eqn\lowmid{\eqalign{\ell_0~&=~\ell_2~=~...~=~\ell_{2p}=0 \cr
\ell_1~&=~\ell_3~=~...~=~\ell_{2p-1}~=~1\cr
\ell_{2p+j}~&=~j \quad j=1,2,...,h \cr
\ell_{2n-j}~&=~\ell_j \quad j=0,1,2,...,n \cr}}

\fig{An example of allowed left multiplication by $e_i$. 
The initial walk diagram must 
have a minimum at the vertical of the point $i$. 
This operation adds a box to the walk diagram at the vertical of 
the point $i<n$.}{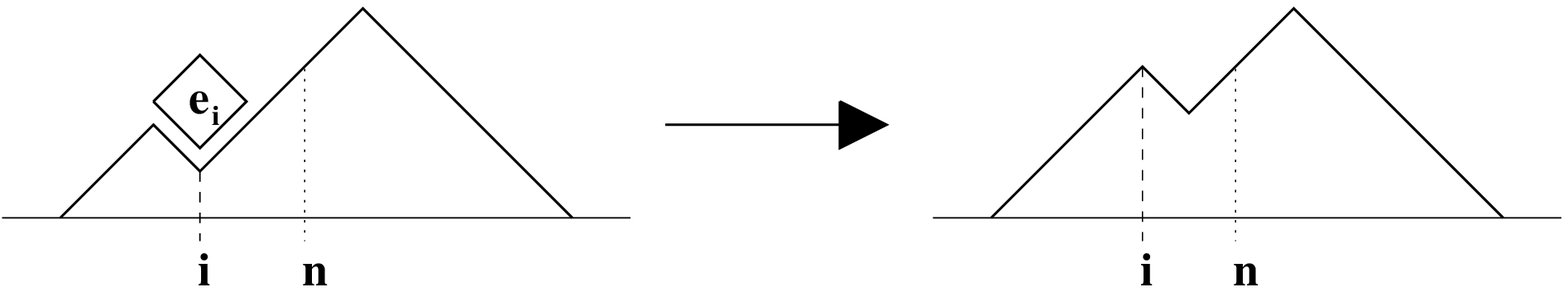}{10.cm}
\figlabel\addbox

It is then easy to see that any reduced element corresponding to a walk
diagram with middle height $\ell_n=h$ is obtained by repeated appropriate
multiplications to the left or to the right of $f_h^{(n)}$ with $e$'s. 
The walk diagrams of middle height $h$ are constructed univocally
by adding ``boxes" to the diagram ${\cal W}_h^{(n)}$. 
As illustrated on Fig.\addbox,
adding a box to a diagram $\cal W$ at the vertical of the point $i$
is allowed only if $i$ is a minimum of $\cal W$, 
namely $\ell_{i+1}=\ell_{i-1}=\ell_i +1$, in which case the 
new diagram, with the box added, 
has $\ell_i \to \ell_i+2$.
For the associated basis 1 elements, this addition
of a box corresponds to the left (resp. right) multiplication by $e_i$ (resp.
$e_{2n-i}$) when $i<n$
(resp. $i>n$). 
This does not affect the middle height $\ell_n=h$.
For illustration, we list the elements of the basis 1 for $TL_3(q)$
together with the corresponding walk diagram (the middle height $\ell_3$
takes only the values $1$ (in $4$ diagrams) and $3$ (in $1$ diagram))
\eqn\walkex{\eqalign{ 
e_1~=~f_1^{(3)}~&=~\figbox{3.cm}{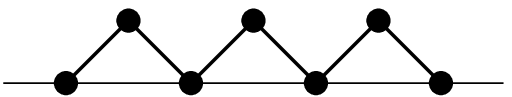} \cr
e_2 e_1~=~e_2 f_1^{(3)}~&=~\figbox{3.cm}{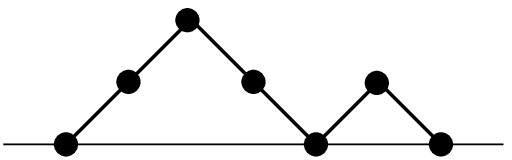} \cr
e_1 e_2 ~=~ f_1^{(3)} e_2~&=~\figbox{3.cm}{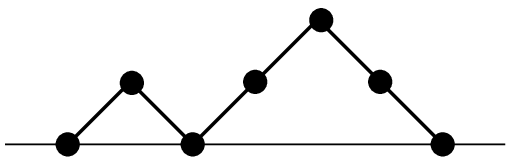} \cr
e_2~=~e_2 f_1^{(3)} e_2~&=~\figbox{3.cm}{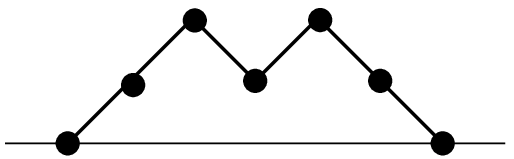} \cr
1~=~f_3^{(3)}~&=~\figbox{3.cm}{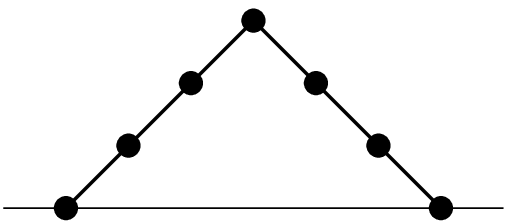} \cr} }

To avoid confusion, we will denote by $(a)_1$ the basis 1 element
corresponding to the walk diagram (or arch configuration) $a\in W_n$
($\equiv A_n$).

\subsec{Scalar product and meanders}

\fig{The trace of an element $e\in TL_6(q)$ is obtained by identifying
the left and right ends of its strings (dashed lines). 
In the arch configuration picture, this amounts to closing the 
upper configuration by a rainbow of order $6$. The corresponding semi-meander
has $3$ connected components, hence ${\rm Tr}(e)=q^3$.}{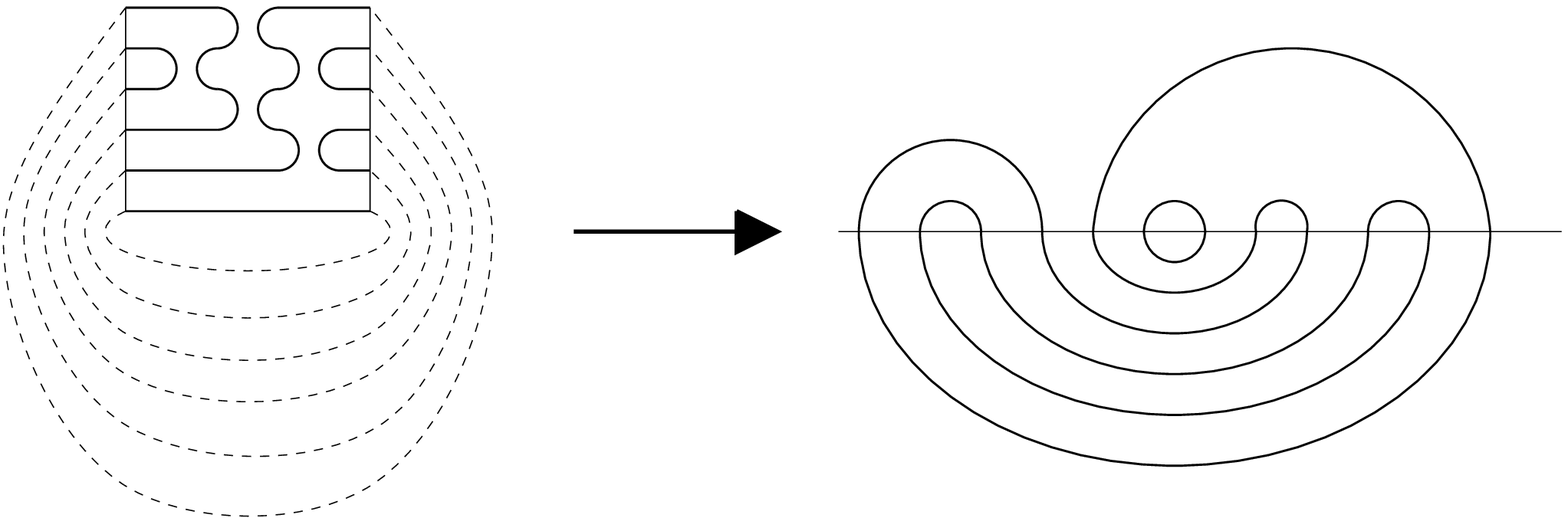}{10.cm}
\figlabel\trac

The standard scalar product on $TL_n(q)$ is defined as follows. First one
introduces a trace over $TL_n(q)$. From the relation (i) of \tla, 
we see that in any element $e$ of $TL_n(q)$
each closed loop may be erased and replaced by a prefactor $q$.  
Taking the trace of a basis 1 element
$e$ corresponds to identifying the left and right ends of each string as in Fig.\trac, 
and assigning an analogous factor to each closed loop, which results in
a factor
\eqn\tra{ {\rm Tr}( e )~=~ q^{c(e)} }
where $c(e)$ is the number of connected components of the closure of $e$.
The definition of the trace is extended to any linear combination of 
basis elements by linearity. Note that, with this definition, the trace is cyclic,
namely ${\rm Tr}(ef)={\rm Tr}(fe)$.
In the arch configuration picture, $c(e)$ is easily identified
as the number of connected components of the semi-meander obtained
by superimposing the arch configuration $a$ corresponding to $e$ 
and the rainbow of order $n$:
indeed, the rainbow connects the $i$-th bridge to the $(2n+1-i)$-th, which
exactly corresponds to the above identification  of string ends.
In particular, this permits to identify the semi-meander polynomial
\sempo\ as
\eqn\idensempo{\enca{ {\bar m}_n(q)~=~\sum_{e \in \ {\rm basis}\ 1} q^{c(e)}~=~
\sum_{a \in W_n} {\rm Tr}((a)_1) } }

We also define the transposition on $TL_n(q)$, by its action
on the generators $e_i^t=e_i$, and the relation $(ef)^t=f^t e^t$ for any
$e,f\in TL_n(q)$. The definition extends to real linear combinations
by $(\lambda e+ \mu f)^t = \lambda e^t + \mu f^t$. 
In the arch configuration picture, this corresponds to 
the reflection $i \to (2n+1-i)$ of the bridges.  In the walk diagram picture,
this is the reflection $i \to (2n-i)$.

\fig{The scalar product $(e,f)$ is obtained by first multiplying $e$
with $f^t$, and then identifying the left and right ends of the strings
(by the dashed lines). Here we have $(e,f)=q^3$. The corresponding
meander is obtained by superimposition of the upper arch configuration
$a$ corresponding to $e$
and lower arch configuration $b$ corresponding to $f$ (the transposition
of $f$ is crucial to recover $b$ as lower arch configuration). 
Here the meander has $c(a,b)=c(e,f)=3$ connected 
components.}{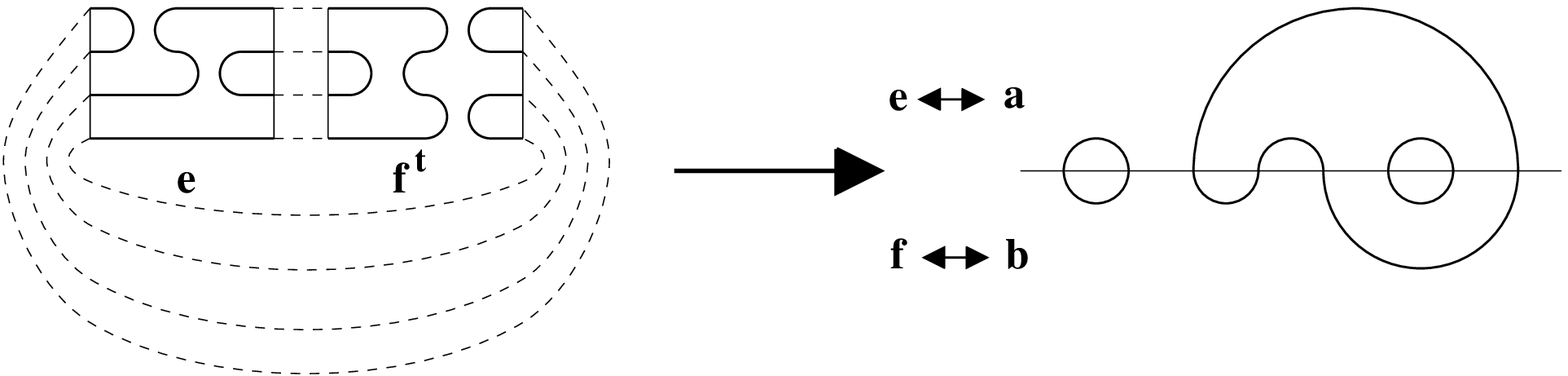}{10.cm}
\figlabel\trame

For any two elements $e$ and $f\in TL_n(q)$, the scalar product is defined as
\eqn\defsca{ (e,f)~=~ {\rm Tr}(e \, f^t) }
This has a simple interpretation
in terms of meanders. We have indeed 
\eqn\tramea{ (e,f)~=~ q^{c(e,f)}~=~q^{c(a,b)} }
where $c(e,f)=c(a,b)$ is the number of connected components of 
the meander obtained
by superimposing the $a$ and $b$ arch configurations 
corresponding respectively to
$e$ and $f$ (see Fig.\trame\ for an
example).  
This permits to identify the meander polynomial as
\eqn\idenmepo{\enca{ m_n(q)~=~ \sum_{a,b \in A_n} q^{c(a,b)}~=~
\sum_{a,b \in W_n} \big((a)_1,(b)_1\big) } }
Note that $(e,1)={\rm Tr}(e)$, hence the semi-meander expression \idensempo\
corresponds to taking $(b)_1=1$ in the above and summing over $a\in W_n$ only. 
This agrees with the
abovementioned fact that the semi-meanders are particular meanders, namely with 
lower arch configuration fixed to be a rainbow. Indeed, the unit  $1\in TL_n(q)$
corresponds in the arch configuration picture to the rainbow of order $n$,
$(r_n)_1=1$.

\subsec{Gram matrix}

The Gram matrix ${\cal G}_n(q)$
of the basis 1 of $TL_n(q)$ is the $c_n \times c_n$ symmetric
matrix with entries
equal to the scalar products of the basis elements, namely
\eqn\gram{\enca{ \big[ {\cal G}_n(q)\big]_{a,b}~=~ \big((a)_1,(b)_1\big) 
~=~ q^{c(a,b)} \qquad 
\forall \ a,b \in \ A_n\equiv W_n} }
For instance, ${\cal G}_3(q)$ reads, in the basis 1 \walkex:
\eqn\exdn{ {\cal G}_3(q)~=~\pmatrix{ q^3 & q^2 & q^2 & q & q^2\cr
q^2 & q^3 & q & q^2 & q\cr
q^2 & q & q^3 & q^2 & q\cr
q & q^2 & q^2 & q^3 & q^2\cr
q^2 & q & q & q^2 & q^3\cr}}

The meander and semi-meander polynomials are easily expressed in terms
of the Gram matrix. Arranging the elements of basis 1 by growing middle
height of the walk diagrams
(in particular, the unit $1$ is the last element), and
defining the $c_n$-dimensional vectors 
\eqn\devec{\vec{u}~=~(1,1,1,\cdots,1)\qquad \vec{v}~=~(0,0,\cdots,0,1)}
we have
\eqn\relagrame{ \eqalign{ 
m_n(q)~&=~  \vec{u} \cdot {\cal G}_n(q) \vec{u} \cr
{\bar m}_n(q) ~&=~ \vec{v} \cdot {\cal G}_n(q) \vec{u} \cr }}
where $\vec{x} \cdot \vec{y}$ denotes the ordinary Euclidian scalar product
of $\IR^{c_n}$.
Moreover, we also have
\eqn\also{ m_n(q^2)~=~ {\rm tr}\big({\cal G}_n(q)^2\big) }

The Gram matrix  ${\cal G}_n(q)$ contains therefore all the information we
need about meanders. The remainder of the paper is devoted to a thorough
study of this matrix and of the consequences on meanders.

\newsec{The basis 2}

The multiplication of elements of the basis 1 involves many reductions,
and therefore is quite complicated.
In this section, we describe another basis for $TL_n(q)$, which we refer to
as {\it basis 2}, in which the products of basis
elements are trivialized, namely the product of any two basis 2 elements
is either $0$ or equal to another basis element.  
This second basis, described in detail in \MARTIN,
will be instrumental in writing alternative expressions
of the meander and semi-meander polynomials.

\subsec{Definition of the basis 2}

We need a few preliminary definitions. The Chebishev polynomials 
of the second kind are defined by the initial data $U_0(x)=1$
and $U_1(x)=x$ and the recursion relation 
\eqn\tche{ U_{n+1}(x)~=~ x\, U_n(x) - U_{n-1}(x) }
or equivalently by 
\eqn\eqtc{ U_n(z+{1\over z})~=~ { z^{n+1} -z^{-n-1} \over z-z^{-1}}}
We also introduce the fractions
\eqn\defmu{ \mu_n~=~ { U_{n-1}(q) \over U_n(q)} }
subject to the recursion relation
\eqn\recumu{ {1 \over \mu_{n+1}}~=~{1 \over \mu_1} - \mu_n }

To describe the basis 2, we use a walk diagram picture analogous to that
for basis 1. Each basis element will be attached to a walk diagram
of $2n$ steps.  
As in the case of basis 1, we start from the definition of the fundamental element 
$\varphi_h^{(n)}$, corresponding
to ${\cal W}_h^{(n)}$, the lowest walk diagram with middle height 
$\ell_n=h=n-2p$ \deffh, namely
\eqn\deffi{ \varphi_h^{(n)}~=~ (\mu_1)^p e_1 e_3 \cdots e_{2p-1} E_h(e_{2p+1},e_{2p+2},...,e_{n-1}) }
where the elements $E_h$ are defined recursively by
\eqn\recuE{\eqalign{
E_0~&=~E_1~=~1 \cr
E_{h+1}&(e_i,e_{i+1},...,e_{i+h-1})\cr 
&=~ E_h(e_i,e_{i+1},...,e_{i+h-2})
(1-\mu_{h} e_{i+h-1}) E_h(e_i,e_{i+1},...,e_{i+h-2}) \cr}}
For instance, we have
\eqn\instE{\eqalign{ E_2(e_i)~&=~ 1 -\mu_1 e_i \cr
E_3(e_i,e_{i+1})~&=~ (1-\mu_1 e_i) (1- \mu_2 e_{i+1}) (1-\mu_1 e_i) \cr
&=~ 1 - \mu_2 (e_i+e_{i+1}) +\mu_1 \mu_2 (e_i e_{i+1} + e_{i+1} e_i) \cr}}
Note that $E_h$ is a projector\foot{This is easily proved by recursion on $h$, by
simultaneously proving that $E_h^2=E_h$ and
$(E_h(e_i,...,e_{i+h-2}) e_{i+h-1})^2= \mu_h^{-1} E_h(e_i,...,e_{i+h-2}) e_{i+h-1}$.}
($E_h^2=E_h$), and that
the normalization factor in \deffi\ ensures that $\varphi_h^{(n)}$ is 
a projector too.

In a second step, we construct the other basis elements corresponding
to walk diagrams with middle height $h$. The latter are obtained by repeated
left and right additions of boxes on the basic diagram ${\cal W}_h^{(n)}$.
To define the corresponding basis 2 elements, it is sufficient to give
the multiplication rule corresponding to a box addition
(see Fig.\addbox).  The rule reads as follows.
If a box is added on a minimum ($\ell_{i+1}=\ell_{i-1}=\ell_i+1$)
of the walk diagram at the vertical of the point $i<n$
(resp. $2n-i>n$), the corresponding basis element is multiplied to the left (resp. right)
by the quantity
\eqn\quanti{ \sqrt{\mu_{\ell_i+2} \over \mu_{\ell_i+1}} 
(e_i -\mu_{\ell_i+1}) }

Applying these rules in the case of $TL_3(q)$, we find the following basis 2 elements
\eqn\batwo{\eqalign{
\figbox{3.cm}{e1w.eps}~&=~\varphi_1^{(3)}~=~\mu_1 e_1  \cr
\figbox{3.cm}{e21w.eps}~&=~\sqrt{\mu_2\over \mu_1}(e_2 -\mu_1) \mu_1 e_1 \cr
&=~\sqrt{\mu_1 \mu_2} (e_2 e_1 -\mu_1 e_1)\cr
\figbox{3.cm}{e12w.eps}~&=~\mu_1 e_1
\sqrt{\mu_2\over \mu_1}(e_2 -\mu_1)  \cr
&=~\sqrt{\mu_1 \mu_2} ( e_1 e_2 -\mu_1 e_1) \cr
\figbox{3.cm}{e2w.eps}~&=~\sqrt{\mu_2\over \mu_1}(e_2 -\mu_1) \mu_1 e_1
\sqrt{\mu_2\over \mu_1}(e_2 -\mu_1)  \cr
&=~ \mu_2 ( e_2 -\mu_1 (e_1 e_2 + e_2 e_1) +\mu_1^2 e_1) \cr
\figbox{3.cm}{onew.eps}~&=~ \varphi_3^{(3)}~=~ E_3(e_1,e_2) \cr
&=~ 1 - \mu_2( e_1+ e_2) + \mu_1 \mu_2 (e_1 e_2+ e_2 e_1) \cr} }

\subsec{Properties of the basis 2}

The construction of the basis 2 basic elements $\varphi_h^{(n)}$ is entirely
dictated by the requirement that 
\eqn\defproE{ e_j \, E_h(e_i,e_{i+1},...,e_{i+h-1})~=~0 \quad 
{\rm for}\ j=i,i+1,...,i+h-1}
These relations were indeed used in \MARTIN\ as a defining
property for the $E_h$'s.

The multiplication rule \quanti\ ensures that 
whenever the multiplication by $e_i$ acts on a slope of the corresponding 
walk diagram (i.e., when $\ell_{i+1}+\ell_{i-1}-2\ell_i=0$), the result
vanishes.  In other words,
\eqn\vanslop{ e_i \, (a)_2~=~0 \qquad 
{\rm whenever} \ \ell_{i+1}^a+\ell_{i-1}^a-2\ell_i^a~=~0}

These rules are also responsible for the following main property 
of the basis 2 elements.
To write it explicitly, we need a more detailed notation for the walk diagrams
of middle height $\ell_n=h$, and the associated basis 2 elements. Such a diagram 
will be denoted $a=lr$, where
$l$ (resp. $r$) denotes the left (resp. right) half of the walk diagram, 
with $i=0,1,...,n$ (resp. $i=2n,2n-1,...,n$), namely
\eqn\nota{ l~=~ \{ (i,\ell_i) \} \qquad r= \{ (i, \ell_{2n-i}) \}  }
for $i=0,1,2...,n$.  Note that $l$ is read from left to right on $a$ and 
that $r$ is read from right to left. Moreover, 
\eqn\symlr{(lr)^t~=~(rl)} 
Both half-walks start at height
$\ell_0=\ell_{2n}=0$ and end at height $\ell_n=h$. To avoid confusion, we
will denote the corresponding basis 1,2 elements by $(lr)_1,(lr)_2$
respectively.

The main property satisfied by the basis 2 elements reads, for any elements
$(a)_2,(a')_2$ of the basis 2, $a=lr$ and $a'=l'r'$:
\eqn\mainpro{ (lr)_2 \, (l'r')_2 ~=~ \delta_{r,l'}\, (lr')_2 }

On this relation, we learn that all the self-transposed elements 
(i.e., with $(a)_2=(a)_2^t$), namely those attached to symmetric 
walk diagrams (i.e., with $l=r$), are projectors. In particular,
we recover the fact that 
$\varphi_h^{(n)}=({\cal W}_h^{(n)})_2$ is a projector.
As we shall see in the next section, the relation \mainpro\ 
implies also that the basis 2 is orthogonal with respect to the 
scalar product \defsca.

\newsec{The meander determinant}

\subsec{The Gram matrix for basis 2}

Thanks to the main property \mainpro, the Gram matrix $\Gamma_n(q)$
of the basis 2 elements
takes a particularly 
simple diagonal form.
Its $c_n \times c_n$ entries read 
\eqn\gratwo{ \big[ \Gamma_n(q)\big]_{a,a'}~=~ \big((a)_2,(a')_2
\big) }

Let us compute the scalar product
\eqn\scapin{\eqalign{
\big((a)_2,(a')_2\big)~&=~ {\rm Tr}( (lr)_2 (l'r')_2^t)
~=~{\rm Tr}( (lr)_2 (r'l')_2)
~=~ \delta_{r,r'}\, {\rm Tr}(  (ll')_2 )\cr
&=~ {\rm Tr} \big( (l'r')_2^t (lr)_2 \big)~=~{\rm Tr}((r'l')_2 (lr)_2 )
~=~ \delta_{l,l'}\, {\rm Tr}(  (rr')_2 ) \cr
&=~ \delta_{a,a'}\,  {\rm Tr}((a)_2(a)_2^t)\cr}}
by direct application of \mainpro\ and use of the cyclicity of the trace
and of \symlr.
Hence the matrix $\Gamma_n(q)$ is diagonal. 
Moreover
\eqn\track{ {\rm Tr}((a)_2(a)_2^t)~=~
{\rm Tr}(  (rr)_2 )~=~{\rm Tr}(  (ll)_2 )}
for any $r$, $l$, 
does not depend on the half-path $r$ of final height $\ell_n=h$. It
may be evaluated on the left half-path $\rho_h$ corresponding  to the
walk diagram ${\cal W}_h^{(n)}$ of \deffh. A simple calculation shows that
\eqn\evalr{ {\rm Tr}(  (\rho_h \rho_h)_2 )~=~ 
{\rm Tr}( \varphi_h^{(n)})~=~U_h(q)}
where $U$ denotes the Chebishev polynomial \tche.
Hence $\Gamma_n(q)$ is simply the diagonal matrix with the 
$c_n$ entries
\eqn\resdel{ \big[ \Gamma_n(q) \big]_{a,a}~=~ U_{\ell_n^a}(q)}
where $\ell_n^a$ denotes the middle height of the walk diagram $a$.

We conclude that the basis 2 is orthogonal with respect to the scalar
product $(\ ,\ )$.

\subsec{Main result}

This remarkable property of the basis 2 will enable us to compute 
the determinant $D_n(q)$ of the Gram matrix ${\cal G}_n(q)$ for the basis 1,
also referred to as {\it meander determinant}.
The result reads\foot{Ref. \KOSMO\ presents a 
recursive algorithm for computing this determinant, 
which relies on direct manipulations of lines and columns of ${\cal G}_n$.
The main result of \KOSMO\ is the identification of the zeros of
$D_n(q)$. Here we also give their multiplicities.}
\eqn\mainres{\enca{\eqalign{
D_n(q)~&=~ \det \big({\cal G}_n(q) \big)~=~\prod_{i=1}^n
U_i(q)^{a_{n,i}}  \cr
a_{n,i}~&=~ {2n \choose n-i}-2 {2n\choose n-i-1}+
{2n \choose n-i-2} \cr }} }
where $U_i(q)$ are the Chebishev polynomials \tche, and we use
the convention that ${j \choose k}=0$ if $j<0$.
For instance, the determinant of the matrix ${\cal G}_3(q)$ \exdn\
reads
\eqn\exthr{  D_3(q)~=~U_1(q)^4 \, U_2(q)^4 \, U_3(q)~=~
q^5\, (q^2-1)^4\, (q^2-2) }

As a nontrivial check, let us first compute the degree of $D_n(q)$
as a polynomial in $q$
\eqn\degdn{ {\rm deg}(D_n(q))~=~ \sum_{i=1}^n i a_{n,i}~=~
{2n \choose n-1}~=~ n\, c_n}
which is in agreement with the definition of the Gram matrix
${\cal G}_n$: the term with highest degree in the expansion 
of the determinant comes from the product of the diagonal elements
of ${\cal G}_n$, namely
\eqn\checdet{ \prod_{a \in W_n} q^{c(a,a)}~=~
\prod_{a \in W_n} q^n ~=~ q^{n c_n}}
as all the meanders with identical top and bottom arch configurations
have the maximal number $n$ of connected components.

\subsec{The zeros of the meander determinant and their multiplicities}

Before going into the proof of the formula \mainres,
let us describe a few consequences of this result.
The zeros $z_{k,l}$ of the polynomial $D_n(q)$ are those of the 
$U_k(q)$, for $k=1,...,n$, namely, using \eqtc\
\eqn\zero{\enca{ z_{k,l}~=~2 \cos \pi {l \over k+1} \qquad 1 
\leq l \leq k \leq n} }
hence we may rewrite
\eqn\rewdet{ D_n(q)~=~ \prod_{ 1 \leq l \leq k \leq n }
\left(q -2 \cos \pi {l \over k+1}\right)^{a_{n,k}} }
This yields the multiplicity $d_n(z_{k,l})$
of each zero $z_{k,l}$, when 
$(k+1)$ and $l$ are coprime integers,  and $l \leq (k+1)/2$ 
($z_{k,k+1-l}=-z_{k,l}$ has the same multiplicity as $z_{k,l}$)
\eqn\multizero{ d_n(z_{k,l})~=~ \sum_{m=1}^{[(n+1)/(k+1)]} a_{n,m(k+1)-1} }
For $k=1,l=1$ this yields the multiplicity of the zero $q=0$
\eqn\mulzer{ d_n(0)~=~ \sum_{m=1}^{[(n+1)/2]} a_{n,2m-1}~=~
{2n \choose n} -{2n \choose n-1}~=~ c_n}
The fact that the zero $q=0$ of $d_n(q)$ has multiplicity $c_n$ 
enables us to
write, in the limit $q \to 0$
\eqn\limdet{ D_n(q) ~\sim q^{c_n} D_n'(0) }
where $D_n'(0)\neq 0$ is the determinant of the matrix ${\cal G}_n'(0)$
with entries 
\eqn\entrig{ \big[ {\cal G}_n'(0) \big]_{a,b}~=~ \left\{ \matrix{1&
{\rm if}\ c(a,b)=1 \cr
0& {\rm otherwise} \cr} \right. }
hence ${\cal G}_n'(0)$ is the one-connected 
component piece of the Gram matrix ${\cal G}_n(q)$.
For instance, 
\eqn\extrpr{{\cal G}_3'(0)~=~\pmatrix{ 0&0&0&1&0\cr
0&0&1&0&1\cr
0&1&0&0&1\cr
1&0&0&0&0\cr
0&1&1&0&0\cr}\qquad D_3'(0)~=~-2 }
Noting that in the limit $q \to 0$
\eqn\protch{ U_{2i}(q)\to (-1)^i \qquad U_{2i-1}(q)/q \to i (-1)^{i-1}}  
the limit $q \to 0$ of \mainres\ yields
\eqn\calcudep{\eqalign{
D_n'(0) ~&=~ \prod_{i=1}^{[n/2]}(-1)^{i a_{n,2i}}
\prod_{i=1}^{[( n+1)/2]}\big[ i (-1)^{i-1} \big]^{a_{n,2i-1}} \cr
&=~(-1)^{(n-1)c_n/2} \prod_{i=1}^{[(n+1)/2]} i^{a_{n,2i-1}}    \cr} }
Therefore
\eqn\lodet{\eqalign{ \log |D_n'(0)|~&=~\sum_{i=1}^{[(n+1)/2]}
\big[{2n \choose n-2i}-2{2n \choose n-2i-1}+{2n \choose n-2i-2}   
\big] \log i \cr
&\sim~ {4^n \over 2 n}~\sim~ {\sqrt{\pi n} \over 2} \, c_n  \cr} }
where the asymptotic estimate results from a saddle-point approximation
to the sum.
If most of the eigenvalues $\lambda$ of the matrix ${\cal G}_n'(0)$ were 
of the same order 
\eqn\ordet{ \lambda ~\sim~(D_n'(0))^{1/c_n}~\sim~ e^{\sqrt{\pi n}/2} }
we would have a meander polynomial, expressed through \also, of the order
\eqn\estifalse{ m_n(q^2)~\sim~ q^2\, \sum_\lambda (\lambda)^2 ~\sim~ 
q^2 e^{\sqrt{\pi n} } c_n}
which clearly is incompatible with the numerical estimate
\eqn\truas{ m_n(q^2)/q^2 ~{\ \atop {\displaystyle\sim \atop q \to 0}}~
M_n \sim~ {{\bar R}^{2n} \over n^{7/2}} }
We conclude that the eigenvalues $\lambda$ of ${\cal G}_n'(0)$ do not have
a localized distribution when $n$ becomes large.
This is also the case when $q=1$.
Indeed, the matrix ${\cal G}_n(q=1)$ is simply the $c_n \times c_n$ matrix
with all entries equal to $1$. It has the eigenvalue $0$, with degeneracy 
$(c_n-1)$, and the nondegenerate eigenvalue $c_n$. This permits 
to recover the sum rules \sume\-\sumem\ easily. In this case, the 
distribution of eigenvalues
of the Gram matrix is certainly not localized
when $n \to \infty$, as the only eigenvalue which matters
diverges while all the other eigenvalues remain $0$.

More generally, the expression \multizero\ can be resummed to yield
\eqn\gendn{\enca{ c_n-d_n(z_{k,l})~=~{1 \over 2(k+1)} \sum_{m=1}^{k} 
\big( 2 \sin {\pi m \over k+1} \big)^2
\, \big( 2 \cos {\pi m  \over k+1} \big)^{2n} } }
(see Appendix A for a detailed proof).
The result is independent of $l$, under the requirement that $l$ and $k$ be coprime.
For instance, for $k=2,3,4,5$ and $n$,
we find
\eqn\mulsq{\eqalign{ 
d_n(\pm 1)~&=~ c_n -1 \cr
d_n(\pm \sqrt{2})~&=~ c_n -2^{n-1} \cr
d_n(\pm{\sqrt{5}\pm 1 \over 2})~&=~ c_n -
{1 \over \sqrt{5}}
\bigg[ 
\left( {\sqrt{5}+1 \over 2}\right)^{2n-1} +
\left( {\sqrt{5}-1 \over 2}\right)^{2n-1} \bigg] \cr
d_n(\pm \sqrt{3})~&=~ c_n- {3^{n-1}+1 \over 2} \cr 
d_n(2 \cos {\pi l\over n+1} )~&=~ 1 \quad {\rm for}\ l \ {\rm and} \ 
(n+1)\ {\rm coprime} \cr} }

The r.h.s. of \gendn\ appears to be an integer in the following interpretation.
Let ${\cal A}_k$ be the $k \times k$ symmetric matrix, with entries
\eqn\entria{ \big[{\cal A}_k\big]_{r,s}~=~ \delta_{s,r+1}+\delta_{s,r-1} }
for $r$, $s=1,2,...,k$. 
This matrix diagonalizes in the orthonormal basis
$\{ {\bf v}_r \ , \ r=1,2,...,k\}$, where
\eqn\eigenv{ \big[ {\bf v}_r \big]_s ~=~ \sqrt{2 \over k+1} \sin {\pi rs \over k+1} } 
are the entries of the eigenvector ${\bf v}_r$ of ${\cal A}_k$, for the eigenvalue
$\beta_r^k=2 \cos \pi r/(k+1)$.
Hence, the r.h.s. of \gendn\ is nothing but 
\eqn\reexdn{ c_n- d_n(z_{k,l})~=~ 
\sum_{m=1}^{k} \big[{\bf v}_m\big]_1 (\beta_m^k)^{2n}  \big[{\bf v}_m\big]_1~=~
\big[ \big({\cal A}_k\big)^{2n} \big]_{1,1}}
This expression is clearly an integer,
as a matrix element of the $(2n)$-th power of an integral matrix.
Moreover, this permits to interpret the number $c_n-d_n(z_{k,l})$ as
counting the number of distinct closed walks of $(2n)$ steps on a
{\it segment} of size $k$, which start and end up at a fixed end of the segment.
Indeed, ${\cal A}_k$ is the adjacency matrix of a chain of $k$ vertices, labeled
$1,2,...,k$.  The quantity $\big[ \big({\cal A}_k\big)^{2n} \big]_{1,1}$
counts the number of distinct paths of length $(2n)$ on the chain which 
start and end up at the vertex $1$. 
In the language of walk diagrams, this is the number of
walk diagrams $w\in W_n$
whose heights do not exceed $(k-1)$.
Denoting by 
\eqn\defmaxhei{W_{n,j}~=~\{ a \in W_n\ \vert \ \ell_i^a \leq j\ 
{\rm for}\ i=0,1,...,2n \} }
eq.\reexdn\ may be rephrased into
\eqn\reph{c_n- d_n(z_{k,l})~=~{\rm card}(W_{n,k-1})}

\fig{Any walk on ${\cal A}_{k+1}$ may be viewed as the prolongation
of a walk on ${\cal A}_2$, by one or several walks on ${\cal A}_k$, 
at each visit of the vertex $2$. Here we have represented a walk on ${\cal A}_2$, 
of length $6$, corresponding to the term $x^3$ in $G_2(x)$. Each of its three
visits of the vertex $2$ may be arbitrarily prolongated by walks on 
${\cal A}_k$ (vertices $2$, $3$, ...,$k+1$ on the figure), to generate 
all the walks on ${\cal A}_{k+1}$,
resulting in the substitution $x^3 \to \big(x G_k(x)\big)^3$ in the corresponding 
generating function.}{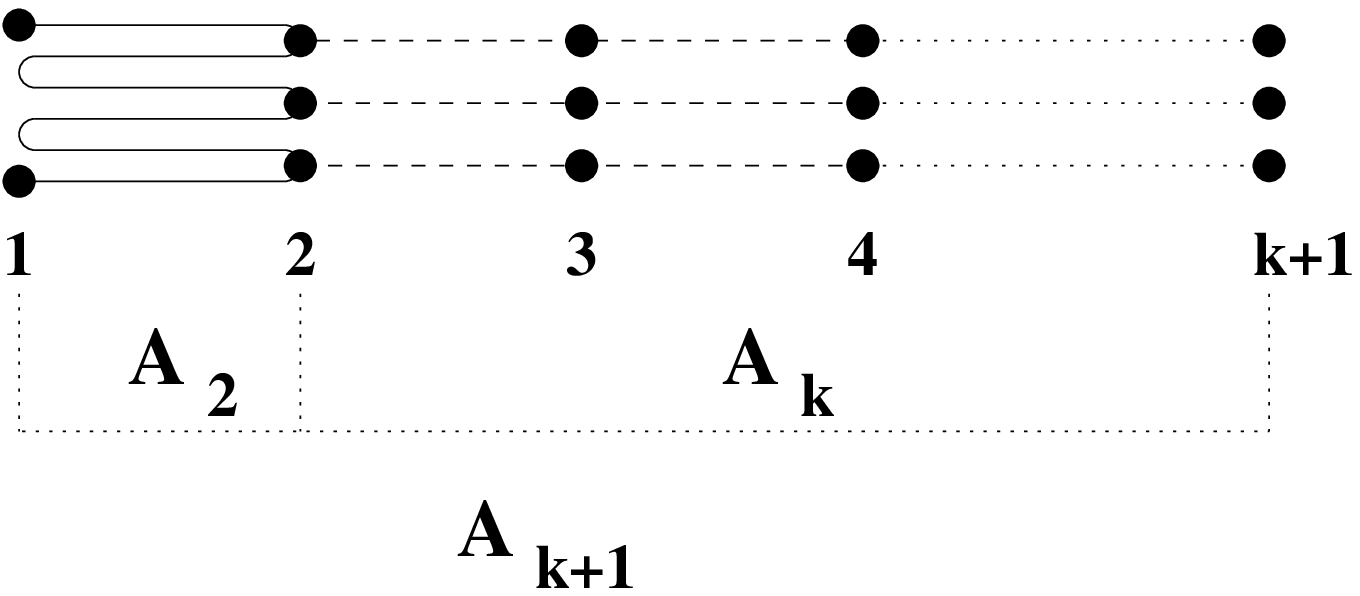}{8.cm}
\figlabel\backandforth

This interpretation permits to write a very simple generating function for
the multiplicities $d_n(z_{k,l})$. Indeed, for $k \geq 1$, let
\eqn\genednf{ G_k(x)~=~\sum_{n=0}^{\infty} x^n \, \big(c_n- d_n(z_{k,l})\big)}
When $k=1$, we set $G_1(x)=1$, deciding by convention that $d_0(0)=0$, 
whereas $c_0=1$.  When $k=2$, we are simply counting the only path of length $2n$,
going back and forth between the origin and the other vertex of the chain.
Each such come and go picks a factor of $x$, resulting in the generating function
\eqn\gtwo{ G_2(x)~=~1+x+x^2+\cdots ~=~ {1 \over 1-x} }
in agreement with the first line of \mulsq.
To compute $G_{k+1}(x)$, knowing $G_k(x)$,
we may view all the walks on ${\cal A}_{k+1}$, as an arbitrary insertion of walks 
on ${\cal A}_k$ at each visit of the vertex $2$ by arbitrary walks on ${\cal A}_2$,
as indicated in Fig.\backandforth. 
This leads to the following composition of generating functions
\eqn\compogen{ G_{k+1}(x)~=~ G_2(x G_k(x))~=~ {1 \over 1 -x G_k(x)} 
~=~{1 \over {1-{x \over {1- {\scriptstyle x \over {\ \ \ {
\ddots_{1 -x}}}}}}}} }
where the fraction is iterated $k$ times.
Together with the initial condition $G_1(x)=1$, eq.\compogen\
completely determines $G_k(x)$ for all $k$.  In fact, we find the following 
simple expression in terms of the Chebishev polynomials \eqtc\
and the function $\mu_k(q)$ \defmu\
\eqn\exsitc{\enca{ G_k(x)~=~ {U_1(1/\sqrt{x}) U_{k-1}(1/\sqrt{x}) \over
U_k(1/\sqrt{x})} ~=~ {1 \over \sqrt{x}}\, \mu_k\big( {1 \over \sqrt{x}}\big)} }
where we have identified the recursion relation \compogen\ with \recumu\
upon a mutliplicative redefinition of $\mu$ (which also gives $G_1(x)=1$) 
and the change of variable $q=1/\sqrt{x}$. Note that the expression \exsitc\
is valid as a series expansion in powers of $x$ for small enough $x$.
This in turn translates into the following expression for the generating function 
for the multiplicities $d_n(z_{k,l})$
\eqn\genmudn{\enca{ F_k(x)~=~\sum_{n=0}^\infty x^n\, d_n(z_{k,l})~=~
C(x)-{1 \over \sqrt{x}}\, \mu_k\big( {1 \over \sqrt{x}}\big)}}
where
\eqn\catge{ C(x)~=~ \sum_{n=0}^\infty x^n \, c_n ~=~{1 - \sqrt{1-4x} \over 2x}}
denotes the generating function of the Catalan numbers \cata. 
The results \mulsq\ may be easily recovered from the expression \genmudn,
for $k=2,3,4,5$.  Note that the series $F_k(x)$ has the valuation $k$,
namely $F_k(x)\sim x^{k}$ when $x \to 0$, as $d_{n}(z_{k,l})=0$, 
for $n \leq k-1$, and $d_k(z_{k,l})=1$ for $n=k$. 
Hence $\lim_{k \to \infty} F_k(x)=0$ uniformly for small enough $x$, 
which means that 
$G_k(x)$ converges to $C(x)$ uniformly when $k\to \infty$: this establishes a
link between the Chebishev polynomials and the Catalan numbers.

Remarkably, the formula \gendn, together with the
above interpretation \reexdn, suggest a relation between the multiplicities 
$d_n(z_{k,l})$
of the zeros of $D_n(q)$ and the rank $r_n(z_{k,l})={\rm dim} \, {\rm Im}\, 
{\cal G}_n(z_{k,l})$
of the matrix  ${\cal G}_n(q=z_{k,l})$, namely that
\eqn\mulrank{ d_n(z_{k,l}) + r_n(z_{k,l}) ~=~ c_n ~=~{\rm dim}(TL_n(q)) }
or in other words that
\eqn\ranker{{\rm dim} \, {\rm Ker}\, {\cal G}_n(z_{k,l})~=~ d_n(z_{k,l})}
Indeed, the matrix ${\cal G}_n(0)=0$ has rank $0$, whereas ${\cal G}_n(1)$
has rank $1$ as all its lines are identical and non-vanishing. 
We also checked that ${\cal G}_n(\sqrt{2})$ has rank $2^{n-1}$ for 
$n=1,2,3,4,5$.
Eqs. \reexdn\ and \mulrank\ would imply in general that the rank of the matrix 
${\cal G}_n(z_{k,l})$ (for $l$ and $(k+1)$ coprime)
is equal to the number of walk diagrams $w\in W_n$,
whose heights are {\it less or equal to $(k-1)$}, i.e., card$(W_{n,k-1})$. 
Assuming that \mulrank\ is true, it is tempting to conjecture that the lines
of ${\cal G}_n(z_{k,l})$ corresponding to the diagrams $a\in W_{n,k-1}$ 
form a collection of $r_n(z_{k,l})$ independent vectors, of which any other line
of ${\cal G}_n(z_{k,l})$ is a linear combination. In particular, the last line of 
${\cal G}_n(z_{k,l})$, corresponding to the diagram ${\cal W}_n^{(n)}$, should
be a linear combination (with coefficients $\lambda_a^n$)
of the lines of ${\cal G}_n(z_{k,l})$ pertaining to the diagrams $a \in W_{n,k-1}$.
This would result in a relation
\eqn\relaconj{ (z_{k,l})^{c({\cal W}_n^{(n)},b)}~=~  \sum_{a \in W_{n,k-1}}
\lambda_{a}^n \, (z_{k,l})^{c(a,b)} }
Summing this over $b \in W_n$ would give a new expression for the 
semi-meander polynomial at $q=z_{k,l}$, as a linear
combination of the polynomials corresponding to the diagrams $a \in W_{n,k-1}$,
namely
\eqn\conjex{ {\bar m}_n(z_{k,l})~=~ \sum_{a \in W_{n,k-1}} 
\lambda_a^n \, {\bar m}(a,z_{k,l}) }
where
\eqn\popoli{ {\bar m}(a,q)~=~ \sum_{b \in W_n} q^{c(a,b)} }
This conjecture is illustrated in appendix B, for $q=\sqrt{2}$ ($k=3$, $l=1$).

\subsec{Proof of the main result}

We now turn to the proof of the formula \mainres\ 
expressing the meander determinant.  
Since the Gram matrix \resdel\ is trivial in basis 2, we simply have 
to compute the determinant of the matrix of the change of basis 1 to 2.
This is done by first showing that this matrix can be put in an upper 
triangular form and computing the product of its diagonal entries. 
The result, combined with \resdel, is identified with the desired 
expression \mainres\ through a subtle mapping of walk diagrams.

\noindent{\bf Preliminaries.}
Let ${\cal P}_n(q)$ denote the matrix of the change of basis 1 to 2,
made of the column vectors of the basis 2 expressed in the basis 1.
It satisfies
\eqn\passage{ (b)_2~=~ \sum_{a\in W_n} 
\big[{\cal P}_n(q) \big]_{a,b} \, (a)_1 }
Let us show that the walk diagrams indexing the vectors of both bases
can be ordered in such a way that the matrix ${\cal P}_n(q)$ is
upper triangular. 

The basic element $\varphi_h^{(h)}$ is, according to \deffi\-\recuE, 
a linear combination of the basis 1 elements of $TL_h(q)$ of
the form
\eqn\licom{\varphi_h^{(h)}~=~\sum_{a \in W_h} \lambda_a 
\, (a)_1(e_1,...,e_{h-1})}
where the sum extends over all the diagrams of $2h$ steps, which are
all {\it included} in the walk ${\cal W}_h^{(h)}$. 
By inclusion of diagrams $a$, $b\in W_n$, we mean
\eqn\incl{ a \subset b \qquad {\rm iff} \qquad l_i^a \leq l_i^b \quad 
\forall\ i=0,1,...,2n}
Similarly, the basic element $\varphi_h^{(n)}$ is equal to
the linear combination
\eqn\licot{\varphi_h^{(n)}~=~\sum_{a \in W_h} \lambda_a 
(\mu_1)^p \, e_1 e_3 \cdots e_{2p-1}\, (a)_1(e_{2p+1},...,e_{n-1})}
which corresponds only to walk diagrams of $2n$ steps,
included in ${\cal W}_h^{(n)}$.

The other basis 2 elements with middle height $h$ are obtained by repeated 
box additions on ${\cal W}_h^{(n)}$
(see Fig.\addbox), with the corresponding multiplication
rule \quanti. It is then easy to prove recursively that any basis 2 element
with middle height $h$, of the form $(b)_2$, 
is a linear combination of basis 1 elements
whose walk diagrams are included in $b$, namely
\eqn\predia{ \big[ {\cal P}_n(q)\big]_{a,b} \neq 0 
\Rightarrow a \subset b}

Arranging the walk diagrams
by growing middle height, we see that $\varphi_h^{(n)}$ is expressed
only in terms of lower basis 1 elements: this gives only 
upper triangular entries in the matrix ${\cal P}_n(q)$. 
More generally, the walk diagrams can be ordered for each fixed middle
height $h$ is such a way that all the diagrams included in $a$ come
before $a$: it is sufficient, for instance, to order the 
diagrams by growing number of boxes added to ${\cal W}_h^{(n)}$.
With such an ordering of the bases 1 and 2, the matrix
${\cal P}_n(q)$ is upper triangular (with nonzero terms on the
diagonal). For instance, with the ordering of \walkex\ and
\batwo, we get the upper triangular matrix
\eqn\expt{ {\cal P}_3(q)~=~\pmatrix{
\mu_1 &-\mu_1\sqrt{\mu_1 \mu_2} &-\mu_1\sqrt{\mu_1 \mu_2} 
&\mu_1^2\mu_2 &-\mu_2 \cr 
0 &\sqrt{\mu_1 \mu_2} &0 &-\mu_1 \mu_2 &\mu_1 \mu_2 \cr
0 &0 &\sqrt{\mu_1 \mu_2} &-\mu_1 \mu_2 &\mu_1 \mu_2 \cr
0 &0 &0 &\mu_2 &-\mu_2 \cr
0 &0 &0 &0 &1\cr}}

Let us decompose the upper triangular matrix ${\cal P}_n(q)$ into the product 
\eqn\decpro{ {\cal P}_n(q)~=~ {\cal Q}_n(q)\, {\cal N}_n(q) }
where ${\cal N}_n(q)$ is a diagonal normalization matrix and ${\cal Q}_n(q)$ an 
upper triangular matrix, with diagonal entries equal to $1$.  
This separates the 
redefinition of basis elements (which does not affect the Gram determinant),
through the matrix ${\cal Q}_n(q)$, from the change of overall normalization of
the basis vectors (which affects the Gram determinant), through ${\cal N}_n(q)$.
For $n=3$, these matrices read
\eqn\extomat{\eqalign{
{\cal N}_3(q)~&=~\pmatrix{\mu_1 &0 &0 &0 &0\cr
0 &\sqrt{\mu_1 \mu_2} &0 &0 &0\cr
0 &0 &\sqrt{\mu_1 \mu_2} &0 &0 \cr
0 &0 &0 &\mu_2 &0\cr
0 &0 &0 &0 &1\cr} \cr
{\cal Q}_3(q)~&=~\pmatrix{1 &- \mu_1 & -\mu_1 & \mu_1^2 &-\mu_2\cr
0 &1 &0 &-\mu_1 &\mu_1 \mu_2\cr
0 &0 &1 &-\mu_1 &\mu_1 \mu_2\cr
0 &0 &0 &1 &-\mu_2\cr
0 &0 &0 &0 &1\cr} \cr} }

The change of basis 1 $\to$ 2 translates into the matrix identity
\eqn\matchan{\eqalign{ 
\Gamma_n(q)~&=~ {\cal P}_n(q)^t \, {\cal G}_n(q) \, {\cal P}_n(q) \cr
&=~ {\cal N}_n(q) \, {\cal Q}_n(q)^t \, {\cal G}_n(q) \, {\cal Q}_n(q)
{\cal N}_n(q)\cr}}
hence, as $\det {\cal Q}_n(q)=1$, we have the relation between determinants
\eqn\detrel{ \det[\Gamma_n(q)]~=~ \det[{\cal N}_n(q)]^2\, D_n(q)}
with, according to \resdel,
\eqn\detres{\eqalign{ 
\det[\Gamma_n(q)]~&=~\prod_{a \in W_n} U_{l_n^a}(q) \cr
&=~\prod_{p=0}^{[n/2]} \big[ U_{n-2p}(q) \big]^{(b_{n,n-2p})^2} \cr} }
where $b_{n,n-2p}$ is the number of half-walks of $n$ steps with final height $h=n-2p$,
and constrained by $\ell_i \geq 0$, for $i=0,1,...,n$.
The walk diagrams of middle height $h=n-2p$ are simply obtained
by taking arbitrary left and right halves of final height $h$, hence their
number is $(b_{n,n-2p})^2$.

The number $b_{n,n-2p}$ is obtained by subtracting from ${n \choose p}$, 
the total number of unconstrained 
walks with $\ell_0=0$ and $\ell_n=h$, the number of those which touch the 
line $\ell=-1$, namely ${n \choose p-1}$. Indeed, by a simple reflection 
(mirror image) with respect
to the line $\ell=-1$ of the portion of walk
between its origin and the first encounter with $\ell=-1$, 
we get a one-to-one mapping
with unconstrained walks such that $\ell'_0=-2$ and $\ell'_n=h$; 
the number of such walks
is ${n \choose p-1}$. Hence we have
\eqn\halwa{b_{n,n-2p}~=~ {n \choose p} -{n \choose p-1} }

\noindent{\bf The normalization ${\cal N}_n(q)$.}
To get $D_n(q)$ from \detrel, we are left with the calculation of $\det[{\cal N}_n(q)]$.
The diagonal entries of  ${\cal N}_n(q)$ are computed as follows. 
For the diagram ${\cal W}_h^{(n)}$, the entry reduces to the global 
normalization of the vector $\varphi_{h}^{(n)}$, namely
\eqn\debutn{ \big[ {\cal N}_n(q) \big]_{{\cal W}_h^{(n)},{\cal W}_h^{(n)}}~=~
(\mu_1)^p}

\fig{The left and right strip decomposition of a diagram of middle height $h$.
the strip lengths are given by the numbers $\ell_i^a$ corresponding to
the maxima $i$ of $a$, $i\neq n$.}{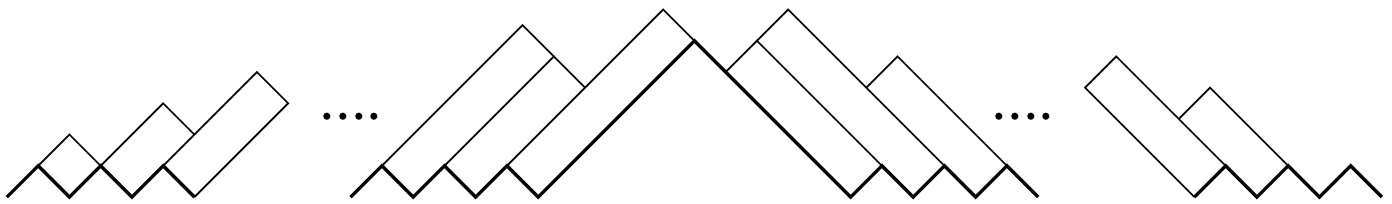}{8.cm}
\figlabel\strip

The entries corresponding to other walk diagrams of middle height $h$ are simply the 
product of this factor by
the product over all the box additions to ${\cal W}_h^{(n)}$ of the normalization factors $\sqrt{\mu_{\ell_i+2}/\mu_{\ell_i+1}}$ which enter the multiplication
rule \quanti.  In other words
\eqn\entsuite{ \big[ {\cal N}_n(q) \big]_{a,a}~=~(\mu_1)^p \, \prod_{{\rm box}\  
{\rm additions} \ i} \sqrt{\mu_{\ell_i+2} \over \mu_{\ell_i+1}} }
To make this formula more explicit, let us arrange the box additions needed to generate
$a$ from ${\cal W}_h^{(n)}$ into $p$ left and $p$ right {\it strips} of consecutive boxes, 
oriented
respectively to the right and left 
as indicated on Fig.\strip. 
This is called the {\it strip decomposition} of $a$.
Each strip ends at a local maximum of $a$, namely
at the vertical of a point $i$ with $\ell_{i+1}=\ell_{i-1}=\ell_i-1$. 
The {\it length} of the corresponding strip is defined to be 
$\ell_i$ (there are actually $\ell_i-1$ boxes
in a strip of length $\ell_i$; a strip with no box has indeed $\ell_i=1$).
The expression \entsuite\ becomes
\eqn\strien{  \big[ {\cal N}_n(q) \big]_{a,a}~=~ (\mu_1)^p\, \prod_{\rm strips} 
\sqrt{\mu_{\ell} \over \mu_1} }
where $\ell$ denotes the length of each strip. As there are $p$ left and $p$ right strips, 
the factors $\mu_1$ cancel out, and we are left with
\eqn\nor{\enca{\big[ {\cal N}_n(q) \big]_{a,a}~=~\prod_{\rm strips} \sqrt{\mu_{\ell}}} }
Hence the prefactor in \detrel\ reads
\eqn\detnum{ \det \big[{\cal N}_n(q)\big]^2~=~ \prod_{a \in W_n}
\prod_{{\rm strips}\ {\rm of}\ a}  \mu_{\ell}~=~\prod_{i=1}^n (\mu_i)^{s_{n,i}}  }
where $s_{n,i}$ denotes the {\it total} number of strips of length $i$ in the strip
decompositions of {\it all} the walk diagrams of $W_n$, or equivalently
the number of distinct diagrams of $W_n$, with a marked top of strip 
of length $i$.

Using the relation $U_i=1/(\mu_1 \mu_2 \cdots \mu_i)$, we can rewrite $ \det[\Gamma_n(q)]$
\detres\ as
\eqn\rewdetga{ \det[\Gamma_n(q)]~=~ \prod_{i=1}^n (\mu_i)^{-h_{n,i}} }
where 
\eqn\hhread{ h_{n,i}~=~ \sum_{i \leq k \leq n \atop k=n\ {\rm mod}\ 2} (b_{n,k})^2 }
is the total number of walk diagrams of $2n$ steps with middle height larger or equal to
$i$. 
We finally get
\eqn\finde{ D_n(q)~=~ \prod_{i=1}^n (\mu_i)^{-s_{n,i}-h_{n,i}} }
Let us now prove that 
\eqn\relawalk{ s_{n,i}+h_{n,i}~=~ b_{2n,2i} }
namely that the total number of walk diagrams of $2n$ steps 
with final height $2i$
is equal to the {\it total} number of strips of length $i$ plus the total 
number of walk diagrams
in $W_n$ with middle height larger or equal to $i$.
To prove this, we establish a map between the walk diagrams of length $2n$ and final
height $2i$ and (i) the walk diagrams of $W_n$ with a {\it marked} top of strip
of height $i$
or (ii) the diagrams of $W_n$ with middle height $\ell_n \geq i$.

\noindent{\bf The mapping of walk diagrams.}
Starting from a given walk diagram $w$ of $2n$ steps from $\ell_0=0$ to $\ell_{2n}=2i$, 
with $\ell_k \geq 0$, for all $k$, the construction proceeds in three steps.

\fig{Reflection-translation of the diagram $w$. The rightmost crossing point 
between $w$ and the line of constant height $\ell=i$, at an ascending slope, 
is marked by a black dot. The dot separates $w$ into a left piece $a$ and a 
right piece $b$. The reflection-translation consists in a reflection of $a\to a^t$,
followed by a translation of $a^t$ in order to glue the two walks $b$ and $a^t$.
The gluing point is indicated by a black dot on the second diagram.}{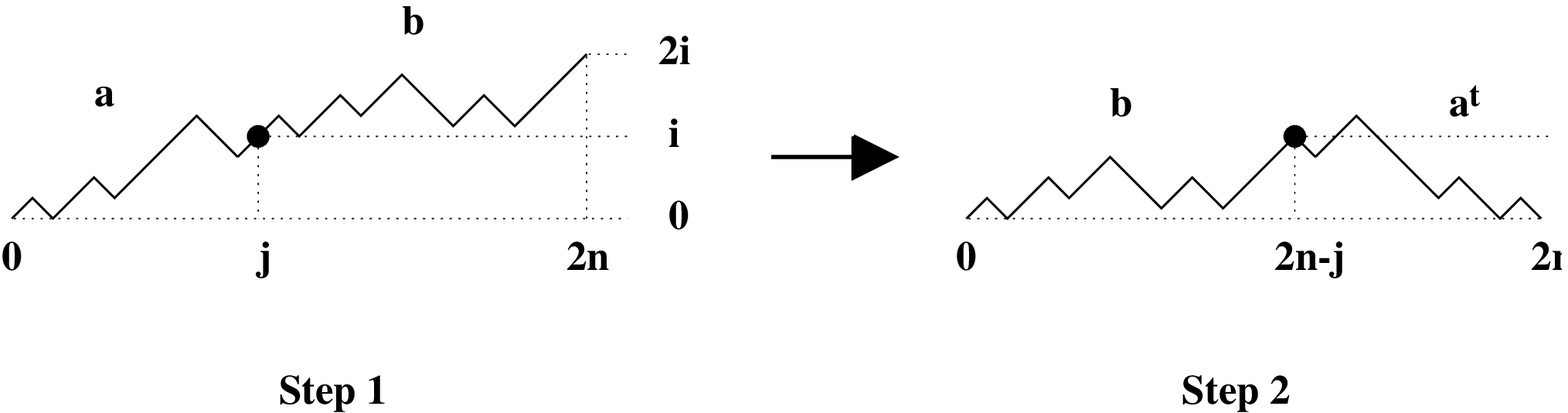}{11.cm}
\figlabel\mapone

\noindent{\bf Step 1.} Let $j$ be the largest point $0 \leq j \leq 2n$ of height $\ell_j=i$
on $w$, and such that $\ell_{j-1}=\ell_j-1=\ell_{j+1}-2$.  
As shown on Fig.\mapone,
this separates the walk diagram $w$ into a left piece $a$, 
with $j$ steps and final height $i$ (namely $\ell_0^a=0$, $\ell_j^a=i$ and
$\ell_k^a \geq 0$ for all $k$), and a right piece $b$, with $(2n-j)$ steps,
initial height $i$ and final height $2i$.  Note that the heights of $b$ remain above 
the $\ell=i$ line, by definition of $j$, hence $b$ can be considered as a walk diagram
of $(2n-j)$ steps, with extremal heights $\ell_0^b=0$, $\ell_{2n-j}^b=i$, and subject
to the constraint $\ell_k^b \geq 0$, for all $k$.

\noindent{\bf Step 2.} Let us perform the following reflection-translation on $w$, 
shown in Fig.\mapone:
reflect the $a$ diagram ($a \to a^t$) and translate it so that its height $i$ (left) end
is glued to the height $i$ (right) end of $b$. 
Note that the resulting diagram $w'$ is an element of $W_n$, as all its heights 
ly above the $\ell=0$ line.
Let us mark this gluing point on the 
resulting diagram $w' \in W_n$. 
This procedure maps the diagram $w$ onto a marked diagram $w'\in W_n$.

\noindent{\bf Step 3.} Only two possibilities may occur for the marked point,
denoted by $j$ in the following: it is
either (1) a maximum of $w'$ ($\ell_{j+1}^{w'}=\ell_{j-1}^{w'}=\ell_j^{w'}-1$) 
or (2) a descending
slope ($\ell_{j-1}^{w'}=\ell_j^{w'}+1=\ell_{j+1}^{w'}+2$) of $w'$. 
Indeed, the point $(j+1)$ on $w'$ has the height 
$\ell_{j+1}^{w'}=\ell_{j-1}^a=\ell_j^{w'}-1$.  

\noindent{\bf Case 1.} When $j$ is a maximum of $w'$, the
marked point $(j,\ell_j^{w'})$ corresponds to the top end of a
strip in the strip decomposition of $w'$ {\it unless} $j=n$. 
Therefore we have the two subcases

\item{}(1)(a): If the marked point is a maximum of $w'$, 
{\it not in the middle} of $w'$ (i.e. $j \neq n$), $w'$ is a walk 
diagram of $W_n$, with a marked (right or left)
top of strip at height $i$.

\item{}(1)(b): If $j=n$, the diagram $w'$ has a middle height $i$ (hence enters 
the category of walk diagrams of $W_n$ with middle height $\geq i$).

\noindent{\bf Case 2.} When the walk has a descending slope at $j$, the marked
point $(j,\ell_j^{w'})$ corresponds to the top end of a (left) strip in the strip decomposition
of $w'$ {\it only if} $j<n$. Therefore we have the three subcases
 
\item{}(2)(a): If the marked point has $j<n$, $w'$ is a diagram with marked top of
(left) strip.

\fig{The cases (2)(b)($\alpha$) and ($\beta$). We indicate the migration
of the marked dot in the ($\alpha$) case. In the ($\beta$) case, the diagram 
$w$ has a middle
height $\geq i$.}{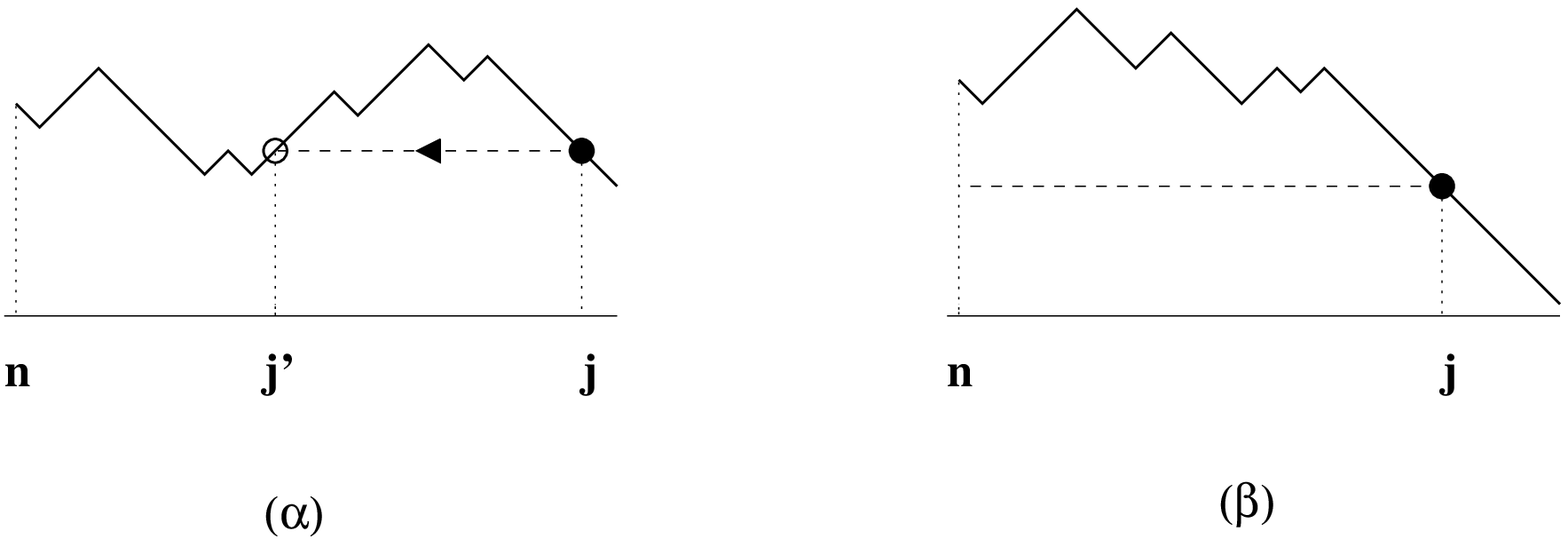}{10.cm}
\figlabel\migration

\item{}(2)(b): If $j>n$, we move the marked point to the left along a 
line of fixed height $\ell=\ell_j^{w'}$, 
until we reach a top of (right) strip (see Fig.\migration\ ($\alpha$)). 
Of course, one
may reach the middle of the diagram before crossing any top of strip 
(see Fig.\migration ($\beta$)).
This leads to two more possibilities

\item{}$\qquad$ (2)(b)($\alpha$): The line of constant height $\ell=\ell_j^{w'}$
crosses an ascending slope of $w'$ at $j'$ 
($\ell_{j'-1}=\ell_{j'}-1=\ell_{j'+1}-2$), such that $n<j'<j$. 
Taking for $j'$ the largest such integer, we
move the mark from $j$ to $j'$, and end up with a diagram $w'\in W_n$ with a
marked top of (right) strip.

\item{}$\qquad$ (2)(b)($\beta$): The line of constant height $\ell=\ell_j^{w'}$
does not cross any ascending slope of $w'$ between $n$ and $j$. The diagram
$w' \in W_n$ has therefore a middle height $\geq i$. More precisely, we have
either possibility

\item{}$\qquad\qquad$ (2)(b)($\beta$)(i): The middle height is $>i$.

\item{}$\qquad\qquad$ (2)(b)($\beta$)(ii): The middle height is $=i$.

\item{}(2)(c): The marked point is at $j=n$. The diagram $w'\in W_n$ has middle height $i$
(hence enters the category of walk diagrams of $W_n$ with middle height $\geq i$).

\noindent{}This exhausts all the diagrams with marked top of strips, 
according to whether

\item{} -the top is a maximum  (1)(a)

\item{} -the top is on a left descending slope (2)(a)

\item{} -the top is on a right ascending slope (2)(b)($\alpha$)

\noindent{}and all the diagrams with middle height $\geq i$, according to whether

\item{} -the middle height is $>i$ (2)(b)($\beta$)

\item{} -the middle height is $=i$ and is a maximum (1)(b)

\item{} -the middle height is $=i$ and is either an ascending slope 
or a minimum (2)(b)($\beta$)(ii)

\item{} -the middle height is $=i$ and is a descending slope (2)(c)

\noindent{\bf The inverse map.}
Conversely, any walk diagram $w'\in W_n$ with a marked top of
strip at height $i$ can be mapped
onto a walk $w$ of $2n$ steps with $\ell_0^w=0$, $\ell_{2n}^w=2i$ and $\ell_k^w\geq 0$ 
for all  $k$ as follows. 
The marked top of strip $(j,\ell_j^{w'}=i)$
can be either (i) a maximum, (ii) a descending slope in the
left half of $w'$ ($j<n$) or (iii) an ascending slope in the right part of 
$w'$ ($j>n$).

In the cases (i) and (ii), the 
marked point separates the walk $w'$
into a left piece $a$ (with $j$ steps and $\ell^a_0=0$, 
$\ell^a_j=i$, $\ell^a_k \geq 0$
for all $k$),  and a right piece $b$ (with $(2n-j)$ steps and $\ell^b_0=i$, 
$\ell^b_{2n-j}=0$, $\ell^b_k \geq 0$
for all $k$). The diagram $w$ is built by the inverse of the reflection-translation
of Fig.\mapone, namely by first reflecting $b \to b^t$, and then by translating 
it and gluing its right end to the left end of $a$.  
After this transformation, the gluing point, now at position $(2n-j)$
has an ascending slope on $w$ at height $i$, and is the largest such point.

In the case (iii), the marked point is first moved to the right
until the first crossing of the
line of constant height $\ell=i$ with a descending slope is reached: 
such a point always exists, because the height $\ell_{2n}^{w'}=0$
must be eventually reached.  On then applies the above inverse reflection-translation 
to this new marked diagram. This produces again a diagram $w\in W_n$ where the
gluing point is the {\it largest} point on $w$ with ascending slope and height $i$.

Finally, any walk diagram $w' \in W_n$ with middle height $\geq i$ may first be marked
as follows. Mark the first crossing $j>n$ between the line of constant height 
$\ell=i$ and the walk $w'$ at a descending slope.  Then apply the above
inverse reflection-translation. 

In all cases, we have associated a walk diagram $w$ to each diagram $w'$
with either a marked end of strip of height $i$ or a middle height $\geq i$. 
This concludes the proof of \relawalk.

\noindent{\bf Conclusion.} Eq. \finde\ implies that
\eqn\partdet{ D_n(q)~=~ \prod_{i=1}^n (\mu_i)^{-b_{2n,2i}} }
or, reexpressed in terms of $U_i$ through $\mu_i=U_{i-1}/U_i$
\eqn\detpart{  D_n(q)~=~ \prod_{i=1}^n \big[ U_i(q) \big]^{(b_{2n,2i}-b_{2n,2i+2})} }
which takes the desired form \mainres\ with
\eqn\luttefinale{\eqalign{ a_{n,i}~&=~ b_{2n,2i}-b_{2n,2i+2} \cr
&=~  {n-i \choose 2n}-2 {n-i-1\choose 2n}+
{n-i-2 \choose 2n} \cr } }

\newsec{Effective meander theory}

In this section, we study various properties of the matrix ${\cal P}_n(q)$
and its inverse, in relation with the meander and semi-meander polynomials
through \relagrame.  Indeed, rewriting \matchan\ as
\eqn\invmat{ {\cal G}_n(q)~=~ \big( {\cal P}_n(q)^t \big)^{-1}\,
\Gamma_n(q) \,  \big( {\cal P}_n(q) \big)^{-1} }
the relations \relagrame\ become
\eqn\smpop{\eqalign{ {\bar m}_n(q)~&=~ \vec{v} \, \cdot {\cal G}_n(q)
\, \vec{u}~=~\big( {\cal P}_n(q)^{-1} \vec{v} \big) \cdot 
\Gamma_n(q) \, {\cal P}_n(q)^{-1} \vec{u} \cr
m_n(q)~&=~\vec{u} \, \cdot {\cal G}_n(q)
\, \vec{u}~=~\big( {\cal P}_n(q)^{-1} \vec{u}\big) \cdot
\Gamma_n(q) \, {\cal P}_n(q)^{-1} \vec{u} \cr} }
where the vectors $\vec{u}$ and $\vec{v}$ are defined in \devec.

\subsec{The matrix ${\cal P}_n(q)^{-1}$}

By definition, the matrix ${\cal P}_n(q)^{-1}$ describes the change of basis 
2 $\to$ 1, through
\eqn\twotoone{ (a)_1~=~ \sum_{b \in W_n} \big[{\cal P}_n(q)^{-1}\big]_{b,a}\,
(b)_2 }
Multiplying both sides to the right by $(c)_2^t$, for some $c\in W_n$,
and taking the trace, we get
\eqn\calcutwo{\eqalign{
{\rm Tr}\big( (a)_1 (c)_2^t \big)~&=~ \sum_{b \in W_n} \big[
{\cal P}_n(q)^{-1}\big]_{b,a}\, {\rm Tr}\big( (b)_2 (c)_2^t\big) \cr
&=~ \big[{\cal P}_n(q)^{-1}\big]_{c,a} \, {\rm Tr}\big((c)_2(c)_2^t\big) \cr} }
where we have used the orthogonality of the basis 2 elements.
According to \track\-\evalr, we have
${\rm Tr}\big((c)_2(c)_2^t\big)=U_{\ell_n^c}(q)$, where $\ell_n^c$ is the 
middle height of the diagram $c$, and we finally get
\eqn\pinv{\enca{\big[{\cal P}_n(q)^{-1}\big]_{c,a}~=~ {{\rm Tr}\big( (a)_1 (c)_2^t \big)
\over  {\rm Tr}\big((c)_2(c)_2^t\big)}~=~{1 \over U_{\ell_n^c}(q)}
{\rm Tr}\big( (a)_1 (c)_2^t \big) }}

\subsec{Properties of ${\cal P}_n(q)^{-1}$}

The formula \pinv\
can be used to derive many properties of the matrix ${\cal P}_n(q)^{-1}$.
Let us take $a={\cal W}_n^{(n)}$ 
(i.e., $(a)_1=f_n^{(n)}=1$) in \pinv. This yields
\eqn\parone{\big[
{\cal P}_n(q)^{-1}\big]_{c,{\cal W}_n^{(n)}}~=~{{\rm Tr}\big((c)_2^t\big)\over
U_{\ell_n^c}(q)}}
writing $c^t=lr$ as a juxtaposition of a left and right half-walk, and
using \mainpro, we compute
\eqn\trone{ {\rm Tr} (lr)_2 ~=~  {\rm Tr}\big( (lr)_2 (rr)_2 \big)~=~
{\rm Tr}\big((rr)_2 (lr)_2\big)~=~\delta_{l,r}\, {\rm Tr}(rr)_2}
Hence the trace of $(c^t)_2$ vanishes, unless $c^t$ is a symmetric
diagram, i.e. with $l=r$, in which case the trace takes
the value \track\-\evalr\
\eqn\valtra{ {\rm Tr} (c^t)_2 ~=~ {\rm Tr} (rr)_2 ~=~
{\rm Tr} (\rho_{\ell_n^c}\rho_{\ell_n^c})_2~=~ U_{\ell_n^c}(q) }
Putting \parone\ and \valtra\ together, we simply find that
\eqn\pval{\enca{\eqalign{
\big[{\cal P}_n(q)^{-1}\big]_{c,{\cal W}_n^{(n)}}~&=~ 
\left\{ \matrix{1 &{\rm if}\ c \ 
{\rm is} \ {\rm symmetric} \cr
0 &{\rm otherwise} \cr} \right. \cr
&\equiv~ \delta_{c,{\rm symmetric}} \cr }} }
With the definition \devec\ of the vector $\vec{v}$, this translates into
\eqn\transp{ {\cal P}_n(q)^{-1}\, \vec{v}~=~ \vec{s} }
where the vector $\vec{s}$ has the entries
\eqn\vecss{ \vec{s}_a~=~  \delta_{a,{\rm symmetric}} }
Comparing with \smpop, this permits to rewrite the semi-meander polynomial as
\eqn\semrew{\eqalign{ {\bar m}_n(q)~&=~ \vec{s} \cdot \Gamma_n(q)\big[{\cal P}_n(q) 
\big]^{-1} \vec{u} \cr
&=~\sum_{a,b \in W_n \atop
a\ {\rm symmetric}} \big[{\cal P}_n(q)^{-1} 
\big]_{a,b} \, U_{\ell_n^a}(q) \cr} }
whereas the meander polynomial reads
\eqn\memrew{\eqalign{ m_n(q)~&=~ {\cal P}_n(q)^{-1}\vec{u}\cdot 
\Gamma_n(q){\cal P}_n(q)^{-1}\vec{u}\cr
&=~\sum_{a\in W_n} \bigg(\sum_{b \in W_n}\big[{\cal P}_n(q)^{-1}\big]_{a,b}\bigg)^2
U_{\ell_n^a}(q) \cr} }

Another interesting particular case of formula \pinv\ is obtained
by taking $c={\cal W}_{\epsilon_n}^{(n)}$, where $\epsilon_n$ is the smallest 
possible middle height in $W_n$, namely 
$\epsilon_n=\big( 1 - (-1)^n \big)/2=\delta_{n,{\rm odd}}$.
The heights of $a$ read $\ell_{2i}=0$ and $\ell_{2i-1}=1$, for all $i$.
This diagram is the smallest of all the diagrams in $W_n$, in the sense that it is 
included in all of them. It corresponds to the first entry of the bases 1 and 2,
hence to the vector
\eqn\vecw{ \vec{w}~=~ (1,0,0,\cdots,0)}
The corresponding basis 1 and 2 elements read respectively
$f_{\epsilon_n}^{(n)}$ and $\varphi_{\epsilon_n}^{(n)}$. By the definitions 
\deffi\ and \redbas\ taken at $h=\epsilon_n$ (in which case $E_{\epsilon_n}=E_0$
or $E_1$, hence $E_{\epsilon_n}=1$), we find the following relation between them
\eqn\relaot{ \varphi_{\epsilon_n}^{(n)}~=~ (\mu_1)^{[n/2]} f_{\epsilon_n}^{(n)} }
or equivalently
\eqn\relaeqi{ \big( {\cal W}_{\epsilon_n}^{(n)} \big)_2~=~(\mu_1)^{[n/2]}
\big( {\cal W}_{\epsilon_n}^{(n)} \big)_1 } 
where we have identified $\epsilon_n=n-2p$, hence $p=[n/2]$.
For the choice $c={\cal W}_{\epsilon_n}^{(n)}$, \pinv\ reads
\eqn\chotwo{\eqalign{
\big[{\cal P}_n(q)^{-1}\big]_{{\cal W}_{\epsilon_n}^{(n)},a}~&=~
{{\rm Tr}\big( (a)_1 ({\cal W}_{\epsilon_n}^{(n)})_2 \big)  
\over U_{\ell_n^c}(q)}\cr
&=~(\mu_1)^{[n/2]}\, {{\rm Tr}\big( (a)_1({\cal W}_{\epsilon_n}^{(n)})_1 \big)
\over  U_{\epsilon_n}(q)}\cr
&=~ (\mu_1)^{[(n+1)/2]} \, 
\big[ {\cal G}_n(q) \big]_{a,{\cal W}_{\epsilon_n}^{(n)}}\cr
&=~ (\mu_1)^{[(n+1)/2]-c(a,{\cal W}_{\epsilon_n}^{(n)})} \cr }}
In the second line, we have used the relation \relaeqi, whereas in the third line,
we have used the fact that $U_{\epsilon_n}(q)=q^{\epsilon_n}=(\mu_1)^{-\epsilon_n}$
and that $\epsilon_n+[n/2]=[(n+1)/2]$. The last expression uses the 
definition of the Gram matrix \gram: the quantity
$c(a,{\cal W}_{\epsilon_n}^{(n)})$ is, in the arch configuration
picture, the number of connected components of the meander obtained by superimposing
the upper configuration $a$ and the lower configuration 
$b\equiv{\cal W}_{\epsilon_n}^{(n)}$, made of a sequence of $n$ consecutive
single arches, linking the bridges $(2i-1)$ and $(2i)$, for $i=1,2,...,n$.
The (meander) polynomial corresponding to the closings of 
${\cal W}_{\epsilon_n}^{(n)}$ was computed in
\DGG\ and reads\foot{In ref.\DGG, it has been shown that the number of closings
of ${\cal W}_{\epsilon_n}^{(n)}$ with $k$ connected components 
is identical to that of arch 
configurations of order $n$ with $k$ interior arches (i.e., arches
linking two neighboring bridges $i$ and $(i+1)$). In turn, this is nothing but 
the number of walk diagrams in $W_n$ with exactly $k$ maxima (the notion
of interior arch in an arch configuration is equivalent to that of a maximum in
the corresponding walk diagram). This number is ${k \choose n} {k-1 \choose n}/n$.}
\eqn\polclo{ i_n(q)~=~\vec{w}\cdot {\cal G}_n(q) \vec{u}
~=~\sum_{a \in W_n} q^{c(a,{\cal W}_{\epsilon_n}^{(n)})}~=~
\sum_{k=1}^n {1 \over n} {k \choose n} {k-1 \choose n} q^k }
with the vectors $\vec{u}$ and $\vec{w}$ defined respectively in \devec\ and
\vecw.
Note that the polynomial $i_n(q)$ is reciprocal, i.e. $q^n i_n(1/q)=i_n(q)$.
Hence, from \chotwo, we get a sum rule for the first line of the matrix 
${\cal P}_n(q)^{-1}$
\eqn\relawsa{\eqalign{ \sum_{a \in W_n}
\big[{\cal P}_n(q)^{-1}\big]_{{\cal W}_{\epsilon_n}^{(n)},a}~&=~
(\mu_1)^{[(n+1)/2]}\, i_n({1 \over \mu_1}) \cr
&=~ i_n(\mu_1) / (\mu_1)^{[n/2]} \cr} }
by using the reciprocality of $i_n(q)$.

\subsec{Recursion relation for the matrix ${\cal Q}_n(q)^{-1}$}

The matrix ${\cal Q}_n(q)$ is constructed in a similar way as
${\cal P}_n(q)$, as the matrix of a redefinition of basis 1,
except that all the normalization factors are dropped, namely
the prefactor $(\mu_1)^p$ in the definition \deffi\ of $\varphi_h^{(n)}$
is dropped, as well as the prefactor $\sqrt{\mu_{\ell_i+2} /\mu_{\ell_i+1}}$ 
in the multiplication rule \quanti. 
This results in a diagonal of $1$'s for ${\cal Q}_n(q)$. 
${\cal Q}_n(q)$ is the matrix of change of basis 1 to the unnormalized basis 2
(denoted by basis $2'$), with elements
$(a)_{2'}=(a)_2/{\cal N}_{a,a}$.

Let us now derive recursion relations for constructing the inverse matrix 
${\cal Q}_n(q)^{-1}$.
This matrix applies the unnormalized basis 2' into the basis 1, 
according to the identity
\eqn\qinv{ (b)_{1}~=~ \sum_{a \in W_n} 
\big[{\cal Q}_n(q)^{-1}\big]_{a,b} (a)_{2'} }
Recall that the basis 1 elements are constructed by box additions (Fig.\addbox)
on the basic elements $f_h^{(n)}$, each box addition corresponding to the
multiplication by some $e_i$.

\fig{The three possibilities for the multiplication
$e_i (a)_{2'}$, represented as a box addition at the vertical of the point $i$
on a diagram $a\in W_n$. The latter may be above (i) a slope of $a$,
(ii) a maximum of $a$ or (iii) a minimum of $a$.}{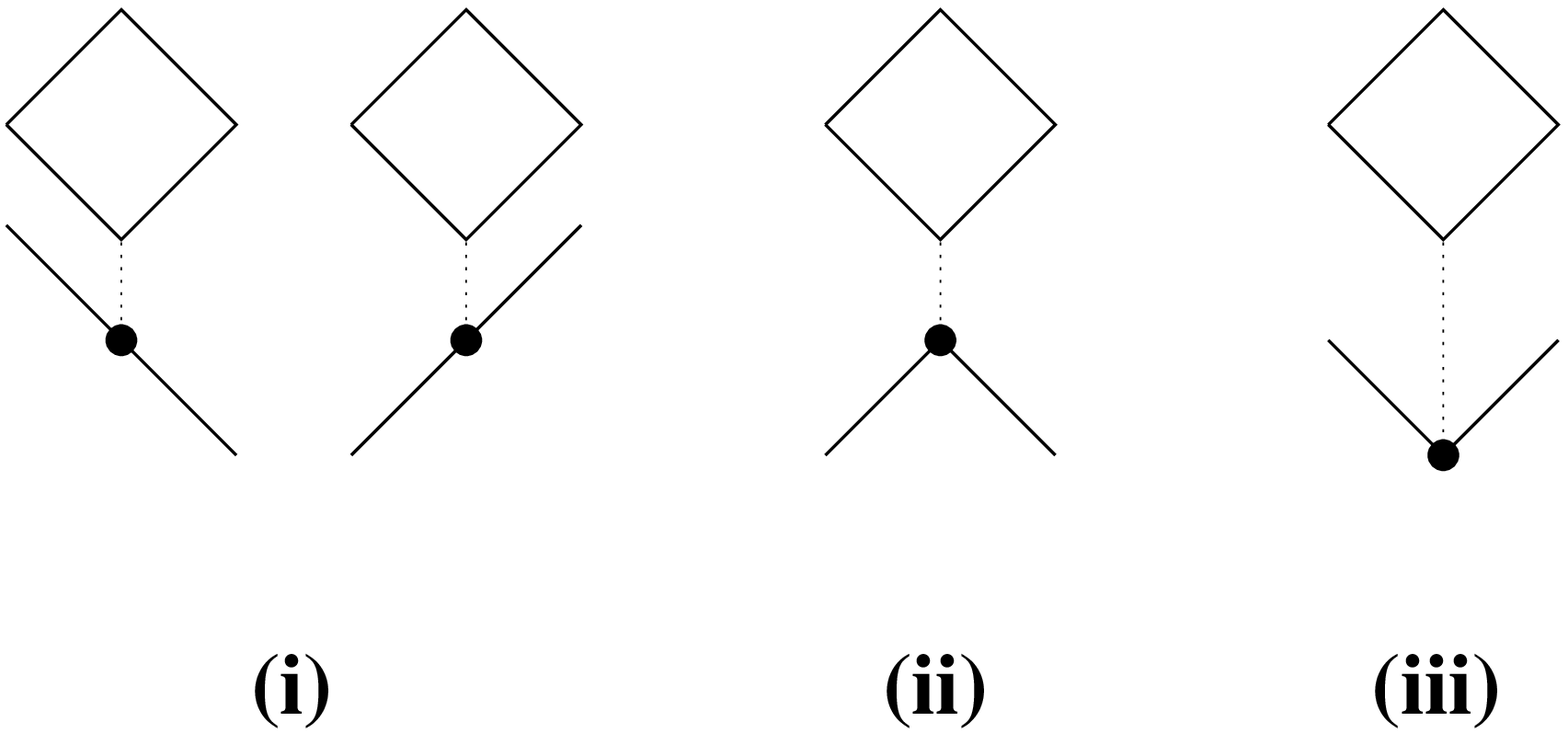}{10.cm}
\figlabel\tricase

Let us study the consequences of a left box addition on $b$, at a minimum 
$i<n$ of $b$.
Let us denote by $b+\diamond$ the resulting diagram.
Multiplying accordingly \qinv\ to the left by $e_i$, 
we find a recursion relation for the 
matrix elements of ${\cal Q}_n(q)^{-1}$. Indeed
\eqn\calpreli{\eqalign{ (b+\diamond)_1~&=~ \sum_{a \in W_n}
\big[{\cal Q}_n(q)^{-1}\big]_{a,b+\diamond} (a)_{2'} \cr
&=~e_i (b)_1~=~ \sum_{a \in W_n}
\big[{\cal Q}_n(q)^{-1}\big]_{a,b} e_i (a)_{2'} \cr} }
gives a relation between $\big[{\cal Q}_n(q)^{-1}\big]_{a,b+\diamond}$
and elements of the form $\big[{\cal Q}_n(q)^{-1}\big]_{a',b}$
by identifying the coefficients of the basis 2' elements.
Three situations may occur for $e_i (a)_{2'}$, as depicted in Fig.\tricase.

\item{(i)} The box addition is performed on a slope of $a$ 
($\ell_{i+1}^a+\ell_{i-1}^a=2 \ell_i^a$). 
Due to the vanishing property \vanslop,
we find that the resulting element vanishes, namely
\eqn\multbone{\enca{e_i \, (a)_{2'}~=~0}}

\item{(ii)} The box addition is performed on a maximum of $a$
($\ell_{i+1}^a=\ell_{i-1}^a=\ell_i^a-1$). 
For $(a)_{2'}$, this maximum is itself the result
of an (unnormalized) box addition with the rules of basis 2, hence a factor 
$(e_i -\mu_k)$, where $k=\ell_i^a-1$, according to \quanti.  
The multiplication by $e_i$ results in
\eqn\multiplimax{ e_i(e_i-\mu_k)~=~ ({1 \over \mu_1} -\mu_k) e_i~=~
{1 \over \mu_{k+1}}\times (e_i-\mu_k) + {\mu_k \over \mu_{k+1}} \times 1  }
where we have used the recursion relation \recumu\ for the $\mu$'s.
The first term in \multiplimax\  restores the box of $(a)_{2'}$, while
in the second term the box is removed, yielding $(a-\diamond)_{2'}$,
where $a-\diamond$ denotes the walk diagram $a$ with the box below the maximum
removed.
Hence
\eqn\multbtwo{\enca{ e_i (a)_{2'}~=~ {1 \over \mu_{k+1}} \big((a)_{2'}+ 
\mu_k (a-\diamond)_{2'}\big)  }}
with $k=\ell_i^a-1$.

\item{(iii)} The box addition is performed on a minimum of $a$
($\ell_{i+1}^a=\ell_{i-1}^a=\ell_i^a+1$). We are left with 
the multiplication of $(a)_{2'}$
by 
\eqn\intermin{ e_i~=~(e_i- \mu_k)+ \mu_k\times 1 }
where $k=\ell_i^a+1$. Hence
\eqn\multbthree{\enca{ e_i (a)_{2'}~=~ (a+\diamond)_{2'}+ \mu_k (a)_{2'} }}

\noindent{}Substituting \multbone\-\multbtwo\-\multbthree\ in \calpreli, we 
get
\eqn\resqmin{\eqalign{\sum_{a \in W_n} 
\big[ {\cal Q}_n(q)^{-1}\big]_{a,b+\diamond}(a)_{2'}
&=~\sum_{a \in W_n}\big[ {\cal Q}_n(q)^{-1}\big]_{a,b} \cr
&\times\bigg\{ {1 \over \mu_{\ell_i^a}}\delta_{a,{\rm max}(i)} \, 
\big( (a)_{2'}+\mu_{\ell_i^a-1} (a-\diamond)_{2'} \big) \cr
&+\delta_{a,{\rm min}(i)} \,  
\big( (a+\diamond)_{2'}+\mu_{\ell_i^a+1} (a)_{2'} \big) \bigg\} \cr } }
where we use the notation
\eqn\deddeltmm{ \eqalign{
\delta_{a,{\rm max}(i)}~=~ \left\{ \matrix{1 &{\rm if} 
\ \ell_{i+1}^a=\ell_{i-1}^a=\ell_i^a-1 \cr 
0 &{\rm otherwise} \cr} \right. \cr
\delta_{a,{\rm min}(i)}~=~ \left\{ \matrix{1 &{\rm if} 
\ \ell_{i+1}^a=\ell_{i-1}^a=\ell_i^a+1 \cr   
0 &{\rm otherwise} \cr} \right. \cr } }      
The identification of coefficients of $(a)_{2'}$ yields the
relation
\eqn\qq{\enca{\eqalign{
\big[ {\cal Q}_n(q)^{-1}\big]_{a,b+\diamond}~&=~
\delta_{a,{\rm max}(i)}
\big({1 \over \mu_{\ell_i^a}}\big[ {\cal Q}_n(q)^{-1}\big]_{a,b}+    
\big[ {\cal Q}_n(q)^{-1}\big]_{a-\diamond,b} \big) \cr
&+\delta_{a,{\rm min}(i)}\big( 
\mu_{\ell_i^a+1}\big[ {\cal Q}_n(q)^{-1}\big]_{a,b}+    
{\mu_{\ell_i^a+1} \over \mu_{\ell_i^a+2} } 
\big[ {\cal Q}_n(q)^{-1}\big]_{a+\diamond,b} \big)  \cr} } }
where we have used 
\eqn\maxdelus{\eqalign{ 
\delta_{a,{\rm max}(i)}~&=~\delta_{a-\diamond,{\rm min}(i)} \cr
\delta_{a,{\rm min}(i)}~&=~\delta_{a+\diamond,{\rm max}(i)} \cr
\ell_i^{a \pm \diamond}~&=~ \ell_i^a \pm 2 \cr}}
Together with the initial condition 
\eqn\iniq{
\big[{\cal Q}_n(q)^{-1}\big]_{a,{\cal W}_{\epsilon_n}^{(n)}}~=~
\delta_{a,{\cal W}_{\epsilon_n}^{(n)}} }
eq.\qq\ is an actual recursion
relation, yielding all the entries of ${\cal Q}^{-1}$, column by 
column starting from the left.

A first remark is in order: the entries of
${\cal Q}_n(q)^{-1}$ satisfy the property
\eqn\prof{\big[{\cal Q}_n(q)^{-1}\big]_{a,b}~\neq~0 \ 
\Rightarrow \  a \subset b}
easily proved by recursion using \qq.
This last condition has been previously derived for 
the entries of ${\cal P}_n(q)$ (cf. \predia), 
but holds as well for the inverse
matrix. 
Note that \qq\ also implies that
\eqn\diagoq{\big[ {\cal Q}_n(q)^{-1}\big]_{a,a}~=~1 } 
in agreement with the normalization of ${\cal Q}$.

\subsec{The matrix ${\cal Q}_n(q)^{-1}$}

The recursion relation \qq\ will be solved in two steps. The idea
is to treat separately the question of finding when 
$\big[ {\cal Q}_n(q)^{-1}\big]_{a,b}$ vanishes or not, and
that of determining its precise value when it does not vanish. 
This suggests to separate the matrix element 
$\big[ {\cal Q}_n(q)^{-1}\big]_{a,b}$ into a product
\eqn\redefq{\big[ {\cal Q}_n(q)^{-1} \big]_{a,b}~=~w_{a,b}\,  f_{a,b} }
where $f_{a,b}$ is subject to the recursion relation
\eqn\frec{\enca{\eqalign{ f_{a,b+\diamond}~&=~ \delta_{a,{\rm max}(i)}\big(
f_{a,b}+ f_{a-\diamond,b}\big) \cr 
&+\delta_{a,{\rm min}(i)}\big( 
f_{a,b}+ f_{a+\diamond,b} \big)\cr } } }
and 
\eqn\initf{f_{a,{\cal W}_{\epsilon_n}^{(n)}}~=~
\delta_{a,{\cal W}_{\epsilon_n}^{(n)}} }

\fig{An example of walks $a \subset b$, where $a$ is $b$-symmetric.
$b$ is represented in the arch configuration picture, and $a$
in the walk diagram picture. The dotted lines continuing the arches
of $b$ indicate the links of $a$ which have to be symmetrical: the 
two links connected to the same arch must be mirror image of 
each other.}{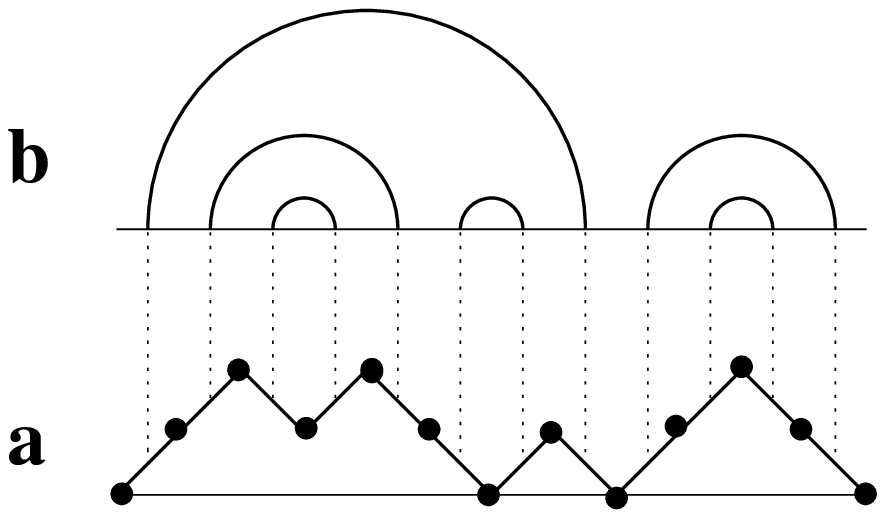}{8.cm}
\figlabel\absym

\noindent{\bf Solving for $\bf f$.} From \frec\-\initf, it
is clear that the $f$'s are nonnegative integers. In fact, the
$f$'s may only take the values $0$ or $1$, and act
as selection rules on the couples of diagrams $a \subset b$. To describe
the solution of \frec\-\initf, we need one more definition.
We will need a mixed representation of a couple $a \subset b$ of walk 
diagrams in $W_n$, namely $a\in W_n$ is represented as a walk diagram,
but $b \in A_n \equiv W_n$ is represented as an arch 
configuration of order $n$. The diagram $b$ is therefore represented
by the permutation $\sigma_b$ of the bridges, with $\sigma_b^2=1$,
describing the arches (namely $\sigma_b(i)=j$ iff the bridges $i$ and $j$
are linked by an arch).
The diagram $a \subset b$ is said to be {\it b-symmetric} iff it satisfies
\eqn\bsyma{\enca{ \ell_{\sigma_b(i)}^a-\ell_{\sigma_b(i)-1}^a~=~
-(\ell_i^a - \ell_{i-1}^a ) } }
In other words, we may represent on the same figure the arch configuration
$b$ and the walk diagram $a$, as illustrated in Fig.\absym. Each bridge $i$
of $b$ sits at the vertical of the link $(i-1,i)$ of $a$. Then $a$ is 
$b$-symmetric iff the links of $a$ are pairwise symmetrical under
the pairs of bridges linked by an arch on $b$.
In particular, if $a$ is $b$-symmetric,
then, below an interior arch of $b$ (i.e., an arch linking two
consecutive bridges $i$, $(i+1)$), $a$ must have a maximum or a minimum
(the only two left-right symmetrical link configurations around $i$).
Note also that a diagram $a$ is symmetric iff it is 
${\cal W}_n^{(n)}$-symmetric, and that the diagram 
${\cal W}_{\epsilon_n}^{(n)}$ is $b$-symmetric for all $b\in W_n$.

With this definition, the solution of the recursion relation \frec\-\initf\
reads
\eqn\solrecf{\enca{ f_{a,b}~=~ \left\{ \matrix{ 1 &{\rm if} \ a \ {\rm is}
\ b-{\rm symmetric}\cr
0 &{\rm otherwise} \cr} \right. } }
Hence, in \redefq,  $f$ selects the couples of diagrams $a \subset b$
such that $a$ is 
$b$-symmetric\foot{Note, with the above definition,
that $f_{a,b}\neq 0 \Rightarrow a \subset b$.
Indeed, if $f_{a,b}\neq 0$, $a$ cannot cross $b$, otherwise one would have
$\ell_i^a=\ell_i^b$ and $\ell_{i+1}^a=\ell_i^a+1$, $\ell_{i+1}^b=\ell_i^a-1$,
for some $i$. Take the smallest such $i$, this means that an arch of $b$
ends at the bridge $i$. Let $i'<i$ be the bridge where it starts, then by
$b$-symmetry, we must have $\ell_{i'+1}^a=\ell_{i'+1}^b$ and 
$\ell_{i'}^a=\ell_{i'+1}^a+1$, $\ell_{i'}^b=\ell_{i'+1}^a-1$, 
which contradicts the fact that $i$ is the first crossing
between $a$ and $b$.}.

With $f_{a,b}$ as in \solrecf, let us now check \frec\-\initf.
The relation \initf\ amounts to the 
fact that $a$ is $a$-symmetric. Indeed, an
arch of $a$ always starts (say, at the bridge $i$)
above an ascending link of $a$ 
($\ell_{i}^a=\ell_{i-1}^a+1$) and ends (say, at the bridge $j=\sigma_a(i)$)
over a descending link of $a$ ($\ell_{j}^a-\ell_{j-1}^a=-1$); 
these two links are therefore symmetrical.

To check \frec, let us consider a diagram $a\subset b+\diamond$, which 
is $b+\diamond$-symmetric. 
Noting that $b+\diamond$ has an interior arch
linking the bridges $i$ and $(i+1)$ (this is equivalent to a maximum 
above $i$ on the corresponding walk diagram), 
by virtue of the abovementioned property, 
the $b+\diamond$-symmetric diagram $a$ must have either a maximum or a minimum 
above $i$. These two possibilities correspond to the two lines of
\frec. To complete the check of \frec, we must prove that in either case
one and only one of the two diagrams $a$ and $a\pm \diamond$ is 
$b$-symmetric (then \frec\ simply reads $1=1$).

\fig{The bridge move $b\to b+\diamond$ on the corresponding arch
configurations. $b$ has a minimum at $i=i_2$, hence an arch
ends at the bridge $i=i_2$, and another starts at the bridge 
$i_3=(i+1)$. In $b+\diamond$, this minimum has been changed
into a maximum, hence the bridges $(i_1,i_4)$ and $(i_2,i_3)$ are connected.
All the other parts $A$, $B$, $C$ and $A'$ of $b$ are unchanged.}{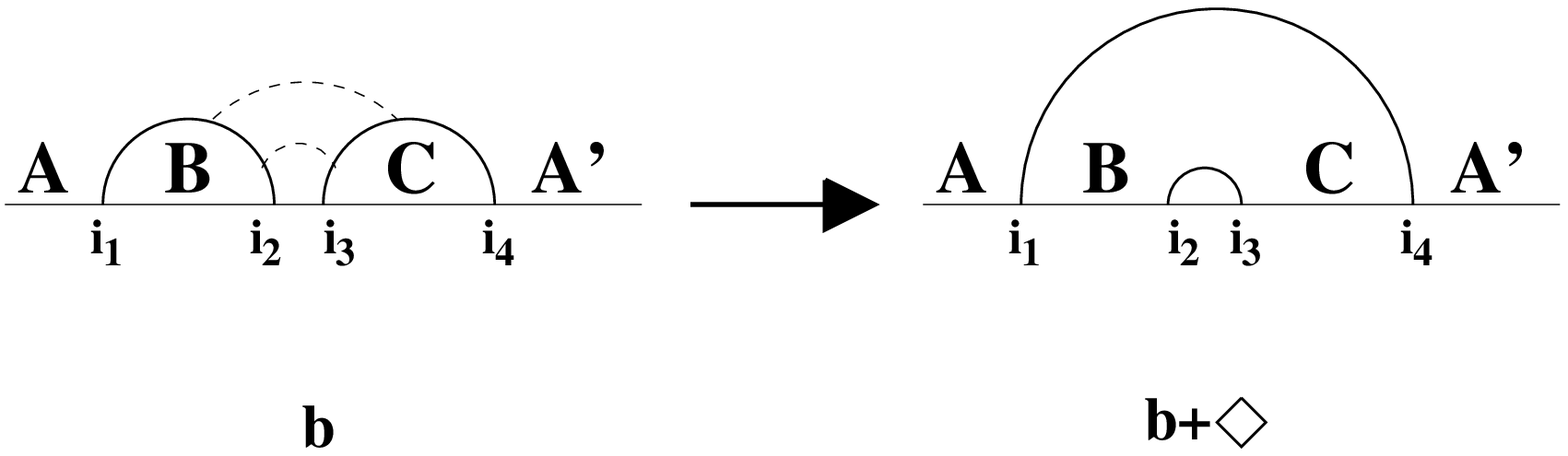}{10.cm}
\figlabel\bridge

More precisely, the box addition on $b\to b+\diamond$ is interpreted
in the arch configuration picture as the {\it bridge move} 
illustrated in Fig.\bridge. Before the box addition, $b$ has a minimum
at the vertical of $i$. 
This means that an arch (starting, say, at the bridge $i_1<i$) 
ends at the bridge $i_2=i$, and that another 
starts from the bridge $i_3=(i+1)$ ( and ends, say, at the bridge 
$i_4>(i+1)$). 
The bridge move of Fig.\bridge\ replaces these two arches by an arch
connecting the bridges $i_1$ and $i_4$, and an {\it interior} arch
connecting $i_2$ and $i_3$. The creation of an interior arch 
corresponds to that of a maximum (the top of the box) on $b$.
Let us denote by $A$, $B$, $C$, $A'$ (like in Fig.\bridge), 
the regions of $b$ lying respectively to the left of $i_1$,
between $i_1$ and $i_2$, between $i_3$ and $i_4$ and to the right of $i_4$.
Note that the regions $A$ and $A'$ may be connected to each other
by arches passing
above the $(i_1,i_2)$ and $(i_3,i_4)$ arches, but $B$ and $C$ are only 
connected to themselves.

\fig{Example of a walk $a$, which is $b+\diamond$-symmetric.
The two possibilities (i) $\sigma_1=\sigma_2=1$ and (ii)
$\sigma_1=-\sigma_2=-1$ are represented.
In both cases, one and only one of the two diagrams $a$ and 
$a-\sigma_2 \diamond$ is $b$-symmetric.}{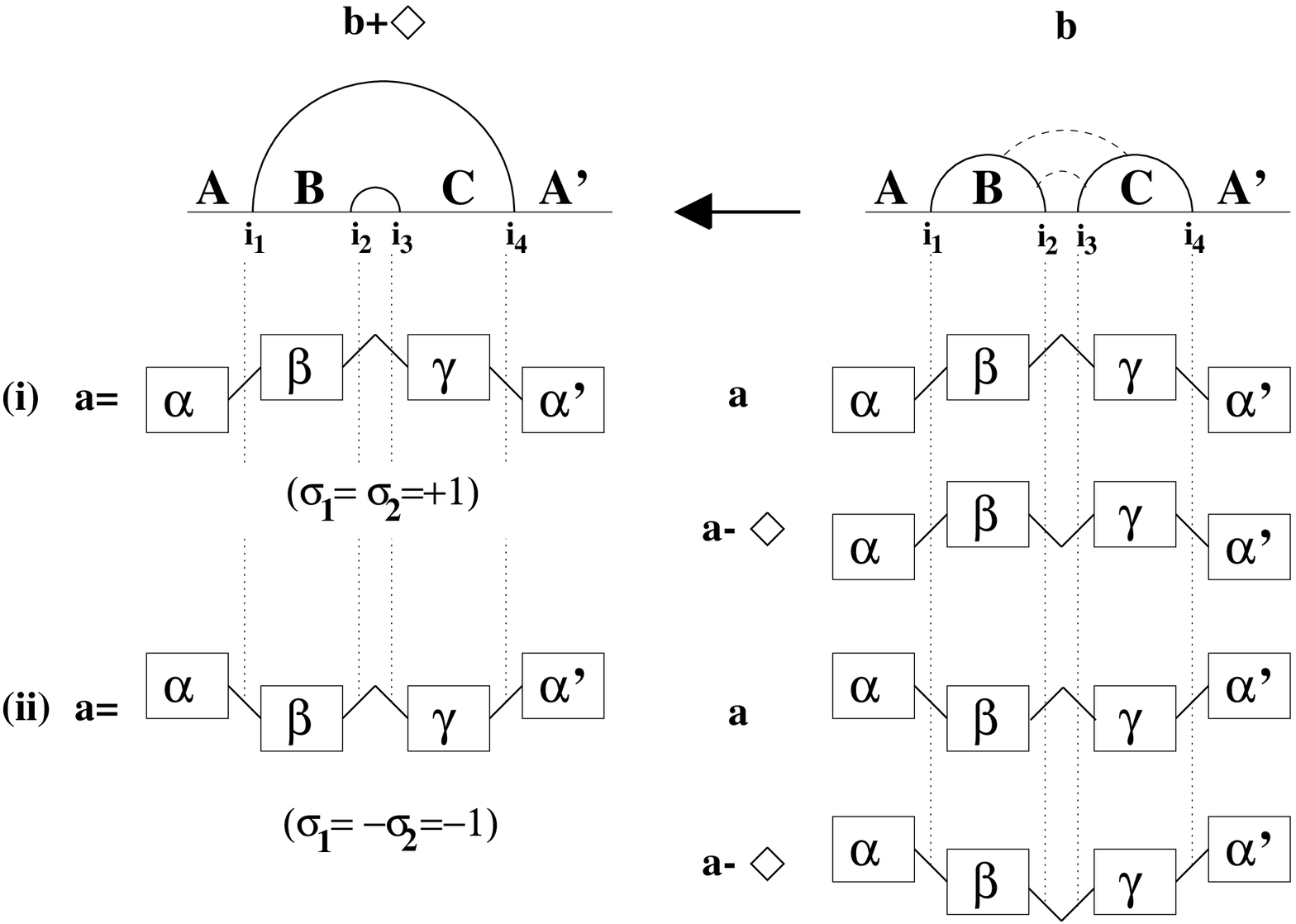}{12.cm}
\figlabel\bridgetwo

Let us consider a walk diagram $a$ which is $b+\diamond$-symmetric 
(cf. Fig.\bridgetwo).
The portions $\alpha$, $\beta$, $\gamma$, $\alpha'$ of the walk $a$
lying respectively below $A$, $B$, $C$, $A'$ satisfy the 
following properties: $\beta$ is $B$-symmetric, $\gamma$ is $C$-symmetric,
and $\alpha\alpha'$ is $AA'$-symmetric\foot{Here we extend 
slightly the notion of respective symmetry to walks $c \subset d$,
with initial and final heights not necessarily equal to $0$, 
by still imposing the condition \bsyma.}.  
All these portions of $a$ 
remain untouched in $a \pm \diamond$. Only the two
links $(i_2-1,i_2)$ and $(i_3-1,i_3)$
of $a$ will be affected.
The $b+\diamond$-symmetry of $a$ implies that 
\eqn\symba{\eqalign{ 
(\ell_{i_1}^a -\ell_{i_1-1}^a)~&=~ (\ell_{i_4-1}^a-\ell_{i_4}^a)
\equiv \sigma_1~=~ \pm 1 \cr
(\ell_{i_2}^a -\ell_{i_2-1}^a)~&=~ (\ell_{i_3-1}^a-\ell_{i_3}^a) 
\equiv \sigma_2~=~ \pm 1 \cr} }
as the bridges $(i_1,i_4)$ and $(i_2,i_3)$ are connected in $b+\diamond$.
Two situations may now occur, according to the relative values of 
$\sigma_1$ and $\sigma_2$.

\item{(i)} $\sigma_1=\sigma_2$: $a$ is not $b$-symmetric, because the links 
$(i_1-1,i_1)$ and $(i_2-1,i_2)$ of $a$ are not symmetrical
(the same holds for the links $(i_3-1,i_3)$ and $(i_4-1,i_4)$).
On the contrary, $a-\sigma_2 \diamond$ is $b$-symmetric, because both links
$(i_2-1,i_2)$ and $(i_3-1,i_3)$ are flipped by the box 
addition$/$subtraction. This is illustrated on Fig.\bridgetwo-(i).

\item{(ii)} $\sigma_1=-\sigma_2$: $a$ is $b$-symmetric, but $a-\sigma_2 \diamond$
is not, as the situation of the previous case is reversed. This 
is illustrated in Fig.\bridgetwo-(ii).

\noindent{}Hence, we have shown that, when
$a$ is $b+\diamond$-symmetric,
one and only one of the two diagrams
$a$ and $a-\sigma_2 \diamond$ appearing on the rhs of \frec\
is $b$-symmetric.
This completes the check of the recursion relation \frec\ (which reduces 
in both cases $\sigma_2=\pm 1$ to $1=1$).
Eq. \solrecf\ is the unique solution to \frec\-\initf.

\fig{A particular folding of the walk diagram $b \in W_n$, leading
to an $a\in W_n$, such that $a$ is $b$-symmetric. 
The solid horizontal lines represent the unfolded folding lines, 
while the horizontal dashed lines represent the lines along which $b$ is
effectively folded (lines number 3,5,6). 
The total number of folding lines is $n$, the order of
the diagrams ($n=6$ here).}{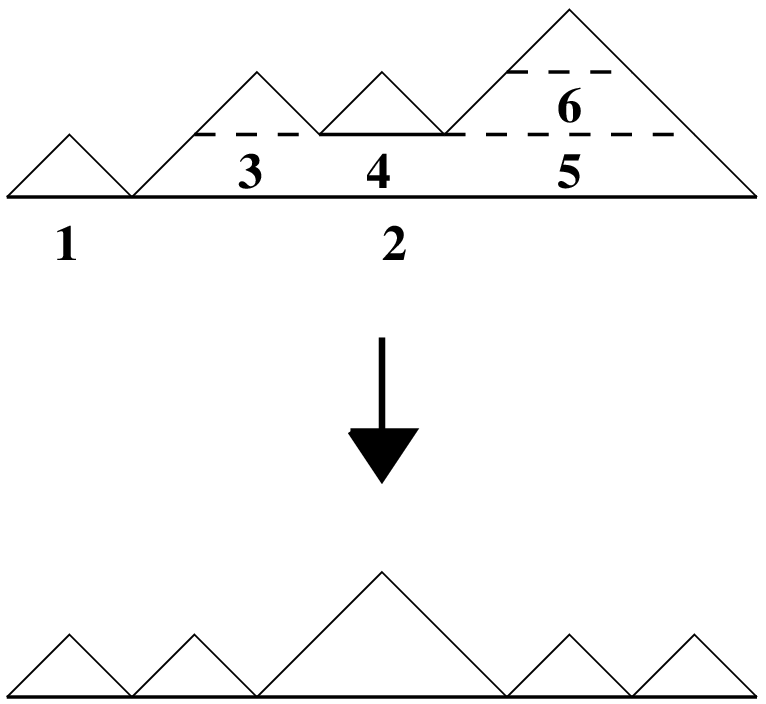}{6.5cm}
\figlabel\foldingue

In addition to their defining recursion relation, the $f$'s satisfy a number
of interesting properties, which will prove crucial in the study of 
meander and semi-meander polynomials. 
Among the many interpretations of the condition $f_{a,b}=1$, the set of 
$a$'s such that $f_{a,b}=1$ for a given $b\in W_n$, may be obtained 
as shown in Fig.\foldingue. 
First represent $b$ as a walk diagram of $2n$ steps. 
Then draw horizontal lines joining the couples of points 
(of the form $(i,\ell_i^b)-(j,\ell_j^b\equiv \ell_i^b)$, $i$, $j\geq 1$) 
corresponding to the beginning and end of all arches of $b$ (the arch starts
at the bridge $(i+1)$ and ends at the bridge $j$). It is easy to see that 
there are exactly $n$ such lines.
The set of admissible $a$'s 
is simply obtained by {\it folding} the path $b$ arbitrarily along these lines
(see Fig.\foldingue).
Indeed, the folding operation preserves the $b$-symmetry of $a$, by simply reversing
all the quantities $(\ell_{i+1}^a-\ell_i^a)$ along the folding line.
If no additional constraint was imposed on the $a$'s, we would get 
$2^n$ possible foldings for each diagram $b$. However, $a$ is further constrained
to have nonnegative heights, which reduces this number, but we expect it to still
behave as $2^n$ for most $b$'s, in the large $n$ limit.  

\fig{For fixed $a$, the $b$'s such that $f_{a,b}=1$ are the arch
configurations connecting bridges with the same value of
$\sigma_i(a)=(-1)^{i-1} t_i(a)$, where $t_i(a)=\ell_{i}^a-\ell_{i-1}^a$,
for $i=1,2,..,2n$. Here $n=6$, and we have represented 
one of the admissible $b$'s.}{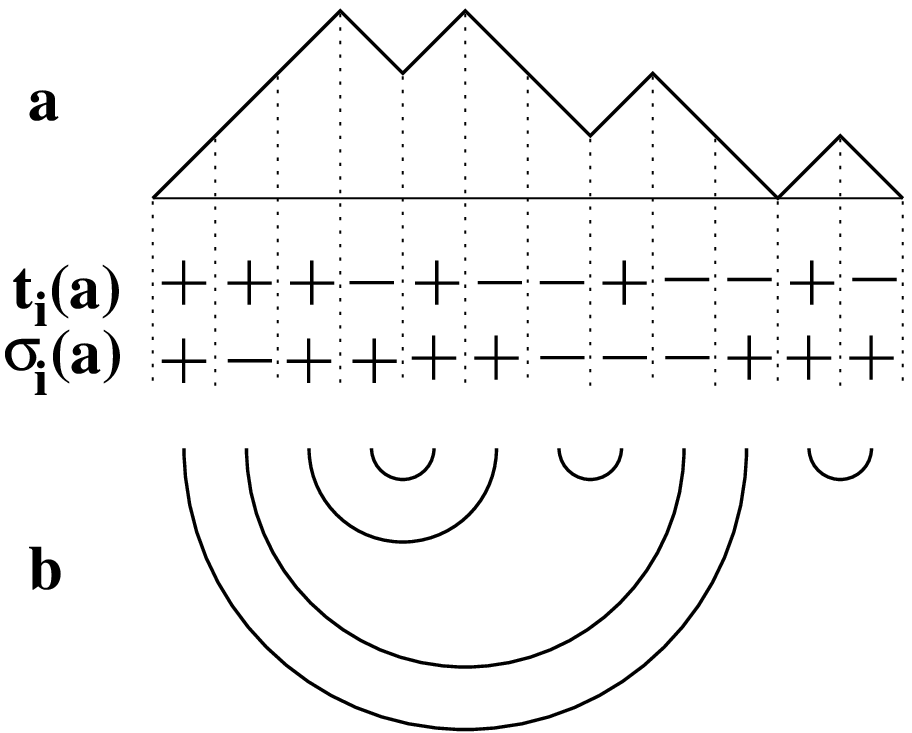}{7.cm}
\figlabel\unfoldingue

Conversely, here is an algorithm to generate, for fixed $a\in W_n$, all the
walks $b \in W_n$ such that $f_{a,b}=1$. The path $b=a$ is always admissible.
Let us represent it by the sequence of signs $t_i(a)=\ell_{i}^a-\ell_{i-1}^a$,
$i=1,2,...,2n$, and consider the modified sequence
\eqn\modiseq{ \sigma_i(a)~=~(-1)^{i-1} \, t_i(a)~=~ 
(-1)^{i-1} \,(\ell_{i}^a-\ell_{i-1}^a)}
Interpreting these indices $i$ as bridge numbers 
(from $1$ to $2n$), the set of $b$'s such that $f_{a,b}=1$ is simply the
set of arch configurations linking these $2n$ bridges, such that each arch
connects two bridges with the {\it same} value of the sign $\sigma_i(a)$. 
An example is displayed in Fig.\unfoldingue. 
The number of admissible $b$'s for fixed $a$
seems to depend strongly on $a$.

Let us finally mention the following sum rule, 
proved in detail in Appendix C
\eqn\sumfab{\enca{ \sum_{a,b \in W_n} f_{a,b}~=~ 
3 {2^{n-1} (2n)! \over n! (n+2)!}~=~3\, {2^{n-1} \over n+2}\, c_n } }
expressing the total number of couples $(a,b)\in W_n\times W_n$, 
where $a$ is $b$-symmetric. 
By Stirling's formula, we see that
\eqn\asyfabtot{\sum_{a,b \in W_n} f_{a,b}~\sim {3 \over 2 \sqrt{\pi}}
\, {8^n \over n^{5/2}} }
The leading behavior $8^n$ agrees with the expectation that the number of admissible
$a$'s for fixed $b$ behave like $2^n$ for most $b$'s 
(whose number is of the order of $4^n$).

\noindent{\bf Solving for $\bf w$.} To complete the solution of \qq, 
we have to compute the weight 
$w_{a,b}=\big[{\cal Q}^{-1}\big]_{a,b}$ when $a$ is $b$-symmetric.
The form of $w_{a,b}$ is entirely dictated by the coefficients of 
the recursion relation \qq.
The result reads 
\eqn\valw{\enca{\eqalign{ w_{a,b}~&=~\prod_{i=1}^{2n-1} 
\big( w(\ell_{i-1}^a,\ell_i^a,\ell_{i+1}^a)
\big)^{{1 \over 4}(\ell_i^b -\ell_i^a)} \cr
w(k,\ell,m)~&=~{\mu_{\ell+1} \over \mu_{\ell}} 
\big( \mu_{\ell} \mu_{\ell+1} \big)^{{1 \over 2}(k+m)-\ell} \cr} } }
In order to check that this is compatible with \qq, we note that, 
with the form \redefq, and when $a$ is $b+\diamond$-symmetric,
one and only one of the four terms in the r.h.s. 
of \qq\ is non-zero. Assuming for instance that $a$ has a maximum
at $i$, it is sufficient to check that
\eqn\checwa{\eqalign{
{w_{a,b+\diamond} \over w_{a,b}}~&=~ {1 \over \mu_{\ell_i^a}} \cr
{w_{a,b+\diamond} \over w_{a-\diamond,b}}~&=~ 1 \cr } }
irrespectively of which term survives. 
Eqs.\checwa\ follow directly from 
\valw, and are exactly what is needed to absorb the
coefficients in \qq.
Similarly, if $a$ has a minimum at $i$,
one easily checks that the sufficient conditions
\eqn\checwat{\eqalign{
{w_{a,b+\diamond} \over w_{a,b}}~&=~ \mu_{\ell_i^a+1} \cr
{w_{a,b+\diamond} \over w_{a+\diamond,b}}~&=~ 
{\mu_{\ell_i^a+1} \over \mu_{\ell_i^a+2}} \cr } }
are fulfilled. 
Note also that $w_{a,a}=1$ as required.

\fig{An example of computation of $w_{a,b}$, for $a \subset b$.
$b$ is obtained from $a$ by six box additions. The box
weights are computed using the rules (i)-(ii)-(iii). Here
we have $w_{a,b}=(\mu_1 \mu_2)^{3/2}$.}{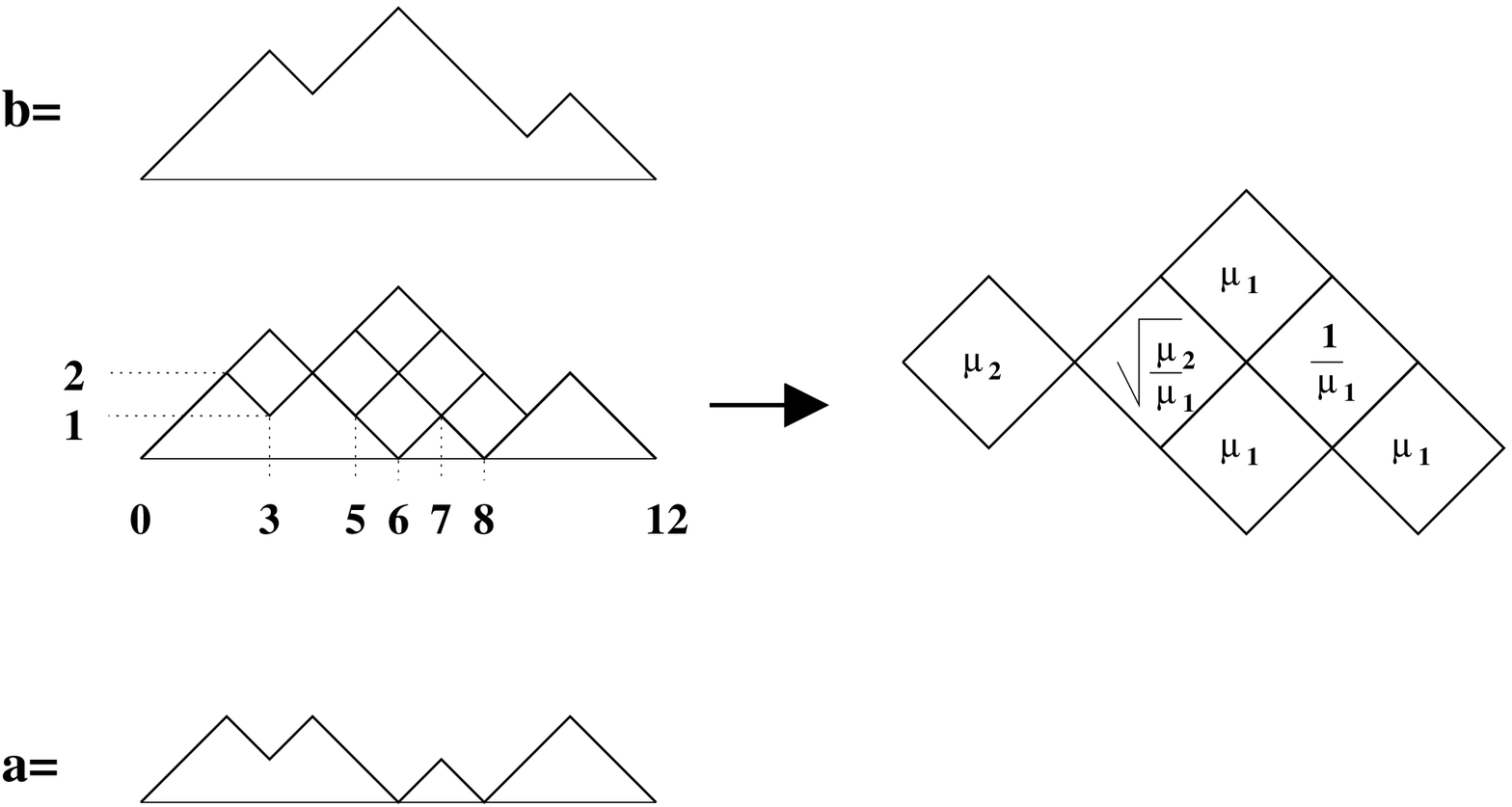}{10.cm}
\figlabel\boxad

Practically, for $a \subset b$ irrespectively of whether $a$ is $b$-symmetric,
the weight \valw\ can be computed as a product of box factors over
all the boxes which must be added to $a$ to build $b$.

\item{(i)} If the box is added at the vertical of a maximum of $a$,
the weight $w$ is multiplied by $1/\mu_{\ell_i^a}$.

\item{(ii)} If the box is added at the vertical of a minimum of $a$,
the weight $w$ is multiplied by $\mu_{\ell_i^a+1}$.

\item{(iii)} If the box is added at the vertical of a slope of $a$,
the weight $w$ is multiplied by $\sqrt{\mu_{\ell_i^a+1}/\mu_{\ell_i^a}}$.

This is a direct consequence of the expression of $w(k,\ell,m)$ in \valw,
where the power ${1 \over 2}(k+m)-\ell$ distinguishes between maxima (value $-1$),
minima (value $1$) and slopes (value $0$).
An explicit example is given in Fig.\boxad.

\subsec{The normalization matrix ${\cal N}_n(q)$}

Let us reexpress the (diagonal) matrix elements of ${\cal N}_n(q)$
\nor\ in the language of weights $w_{a,b}$ \valw. The
result reads simply 
\eqn\resnw{ \enca{\big[ {\cal N}_n(q) \big]_{a,a}~=~ w_{a,{\cal W}_n^{(n)}} }}
where ${\cal W}_n^{(n)}$ is the largest walk diagram of $W_n$, i.e.
containing all the others, with heights 
\eqn\maxdiag{ \ell_i^{\rm max}~=~{\rm min}(i,2n-i)}

This is easily proved as follows. Let us first consider a
{\it symmetric} diagram $a$. As mentioned before, this diagram is also  
${\cal W}_n^{(n)}$-symmetric, hence $f_{a,{\cal W}_n^{(n)}}=1$.
By \pval, we find that
\eqn\rpval{\eqalign{ 
1~&=~\big[ {\cal P}_n(q)^{-1}\big]_{a,{\cal W}_n^{(n)}}\cr
&=~\big[{\cal N}_n(q)^{-1}\big]_{a,a} \big[
{\cal Q}_n(q)^{-1}\big]_{a,{\cal W}_n^{(n)}}\cr  } }
hence 
\eqn\calcnw{\big[{\cal N}_n(q)\big]_{a,a}~=~
\big[{\cal Q}_n(q)^{-1}\big]_{a,{\cal W}_n^{(n)}}~
=~w_{a,{\cal W}_n^{(n)}}   }
as $f_{a,{\cal W}_n^{(n)}}=1$.
This proves \resnw\ for any symmetric diagram $a$.
However, the expression \nor\ for arbitrary $a$ is clearly factorized
into products pertaining to the left and right halves of $a=lr$, namely
\eqn\factonlr{\big[{\cal N}_n(q)\big]_{a,a}~=~
\prod_{{\rm left}\ {\rm strips}\atop {\rm on}\ l} \sqrt{\mu_{\ell}} 
\prod_{{\rm right}\ {\rm strips}\atop {\rm on} \ r} 
\sqrt{\mu_{\ell}}~=~n(l)\, n(r)}
Analogously, the quantity $w_{a,{\cal W}_n^{(n)}}$ factorizes into
two products pertaining respectively to the left and right 
halves of $a=lr$, namely
\eqn\factow{ w_{a,{\cal W}_n^{(n)}}~=~ w(l)\, w(r)}
where
\eqn\wlr{
w(l)~=~\big( x(\ell_{n-1}^l,\ell_n^l)\big)^{n-\ell_n^l \over 4}
\prod_{i=1}^{n-1} 
\big(w(\ell_{i-1}^l,\ell_i^l,\ell_{i+1}^l)
\big)^{{1 \over 4}(\ell_i^{\rm max}-\ell_i^l)}}
and 
\eqn\defx{\eqalign{ x(k,\ell)~&=~ \big( \mu_{{\rm max}(k,\ell)} 
\big)^{k-\ell}\cr 
&=~\big( \mu_{(k+\ell+1)/2} \big)^{k-\ell}\cr } }
In the above, we have used the formula
${\rm max}(k,\ell)=(|k-\ell|+k+\ell)/2=(k+\ell+1)/2$, as $k-\ell=\pm 1$.
The weights $x$ separate the middle box factors in \valw\
into left and right halves. More precisely, each box factor
\eqn\halfw{\enca{ w(k,\ell,m)~=~ x(k,\ell) \, x(m,\ell) } }
is factorized into a left half $(k,\ell)$ (described 
from left to right on $a$) and a right half $(m,\ell)$ (described
from right to left on $a$). Eq.\halfw\ results from the identity
\eqn\idwmu{ {\mu_{\ell+1} \over \mu_{\ell}} ( \mu_{\ell} \mu_{\ell+1}
)^{{1 \over 2}(k+m)-\ell} ~=~ \mu_{{\rm max}(k,\ell)}^{k-\ell}\, 
\mu_{{\rm max}(m,\ell)}^{m-\ell} }
easily proved by inspection.
Now the result for symmetric diagrams $a=lr$, $r=l$, reads
\eqn\symresw{ \big[{\cal N}_n(q)\big]_{a,a}~=~n(l)^2~=~w(l)^2 }
hence $n(l)=w(l)$. For arbitrary diagrams
$a=lr$, we have
\eqn\reswtot{\big[{\cal N}_n(q)\big]_{a,a}~=~n(l)n(r)~=~w(l)w(r)~=~ 
w_{a,{\cal W}_n^{(n)}}}
which completes the proof of \resnw.

As a by-product of the previous analysis, the local 
factorization property \halfw\ and the obvious relation
$x(k,\ell) x(\ell,k)=1$, enable us to rewrite the general
expression \valw\ as
\eqn\wval{\enca{\eqalign{ w_{a,b}~&=~\prod_{i=1}^{2n-1} \big[x(\ell_{i-1}^a,\ell_i^a)
x(\ell_{i+1}^a,\ell_i^a )\big]^{{1 \over 4}(\ell_i^b-\ell_i^a)}\cr
&=~\prod_{i=0}^{2n-1} \big[ x(\ell_i^a,\ell_{i+1}^a) 
\big]^{{1 \over 4}\big((\ell_{i+1}^b-\ell_i^b)-
(\ell_{i+1}^a -\ell_i^a)\big)} \cr 
&=~\prod_{i=0}^{2n-1} \big( \mu_{(1+\ell_i^a+\ell_{i+1}^a)/2} 
\big)^{{1\over 4}\big(1- (\ell_{i+1}^a-\ell_i^a)
(\ell_{i+1}^b-\ell_i^b)\big)}    \cr} } }
This last expression reduces directly to \nor\ in the case 
$b={\cal W}_n^{(n)}$, providing us with an alternative proof of
\resnw. Indeed, 
$\ell_{i+1}^b-\ell_i^b=1$ if $i<n$, and $-1$ if $i\geq n$,
and the factor 
${1 \over 2}\big(1-(\ell_{i+1}^a-\ell_i^a)(\ell_{i+1}^b-\ell_i^b)\big)$
takes the value $1$ on $a$'s (left and right) tops of strips, and $0$
everywhere else, while $(1+\ell_i^a+\ell_{i+1}^a)/2$ is the corresponding
length of strip.

\subsec{The path formulation of (semi)-meanders}

We now have all the elements to write alternative expressions for the
meander and semi-meander polynomials, as weighted sums over paths.  
The entries of
the matrix ${\cal P}_n(q)^{-1}$ read, using \valw\
\eqn\expinv{\eqalign{\big[ {\cal P}_n(q)^{-1} \big]_{a,b}~&=~ 
\big[{\cal N}_n(q)^{-1}\big]_{a,a} \, \big[ {\cal Q}_n(q)^{-1} \big]_{a,b}\cr
&=~ f_{a,b} \, {w_{a,b} \over w_{a,{\cal W}_n^{(n)}}} \cr
&=~ f_{a,b} \, \prod_{i=1}^{2n-1} \big[ w(\ell_{i-1}^a,\ell_i^a,\ell_{i+1}^a)
\big]^{{1 \over 4}(\ell_i^b - \ell_i^{\rm max})} \cr
&=~ f_{a,b} \, e^{{1 \over 4}\sum_{i=1}^{2n-1} 
(\ell_i^b - \ell_i^{\rm max}) \alpha(\ell_{i-1}^a,\ell_i^a,\ell_{i+1}^a; 
q)} \cr}}
where $\ell_i^{\rm max}$ is defined in \maxdiag, and 
\eqn\alphadef{ \alpha(k,\ell,m;q)~=~
\log \left({\mu_{\ell+1} \over \mu_{\ell}}\right) +{k+m -2\ell \over 2}
\log \big( \mu_{\ell} \mu_{\ell+1} \big) }

Using the alternative expression of $w_{a,b}$ \wval, we may also write
\eqn\otherpinv{\enca{\big[ {\cal P}_n(q)^{-1} \big]_{a,b}= 
f_{a,b}\, e^{{1 \over 4}\sum_{i=0}^{2n-1} 
\big[(\ell_{i+1}^{\rm  max}-\ell_i^{\rm max})- (\ell_{i+1}^b -\ell_i^b)\big]
(\ell_{i+1}^a -\ell_i^a)\,
\log \, \mu_{(\ell_i^a+\ell_{i+1}^a+1)/2}}  } }
The passage from \expinv\ to \otherpinv\ may be viewed as a discrete integration
by parts in the sum over $i$.

Substituting the expression \otherpinv\ in \semrew\ and \memrew, we get the following
expressions for the semi-meander and meander polynomials
\eqn\smeffect{\enca{\eqalign{
{\bar m}_n(q)~&=~\sum_{a,b \in W_n} f_{a,b}\, f_{a,{\cal W}_n^{(n)}}
\, U_{\ell_n^a}(q) \cr
&\times \, e^{ {1 \over 4}\sum_{i=0}^{2n-1} 
\big[(\ell_{i+1}^{\rm  max}-\ell_i^{\rm max})- (\ell_{i+1}^b -\ell_i^b)\big]
(\ell_{i+1}^a -\ell_i^a)\, 
\log \mu_{(\ell_i^a+\ell_{i+1}^a+1)/2}} \cr} }}
\eqn\meffect{\enca{\eqalign{ m_n&(q)=\sum_{a,b,b'\in W_n} f_{a,b}\, f_{a,b'} \, 
U_{\ell_n^a}(q)\cr
&\times \, e^{ {1 \over 4}\sum_{i=0}^{2n-1} 
\big[2(\ell_{i+1}^{\rm  max}-\ell_i^{\rm max})- (\ell_{i+1}^b -\ell_i^b)
-(\ell_{i+1}^{b'} -\ell_i^{b'})\big]
(\ell_{i+1}^a -\ell_i^a)
\log \mu_{(\ell_i^a+\ell_{i+1}^a+1)/2}} \cr} } }
Note that the semi-meander expression \smeffect\ may be viewed as \meffect\
in which $b'$ is fixed to be ${\cal W}_n^{(n)}\equiv r_n$, the walk diagram corresponding to the rainbow arch configuration of order $n$, which restricts the
sum to symmetric walk diagrams $a$.

The expressions \smeffect\-\meffect\ should permit a detailed asymptotic
study of the semi-meander and meander polynomials for large $n$.

\subsec{Connected components in meanders}

For any $b\in A_n\equiv W_n$, let $\vec{v_b}$ be the vector with entries
$(\vec{v_b})_a=\delta_{a,b}$. The matrix elements of ${\cal G}_n(q)$ can
be expressed as
\eqn\expreg{ \big[ {\cal G}_n(q) \big]_{b,b'}~=~ \vec{v_{b'}} \cdot 
{\cal G}_n(q)\vec{v_b}~=~ \big({\cal P}_n(q)^{-1}\vec{v_{b'}}\big)
\cdot \Gamma_n(q){\cal P}_n(q)^{-1}\vec{v_{b}}~=~q^{c(b,b')} }
where $c(b,b')$ \tramea\ is the number of connected components of the meander
obtained by superimposing the arch configurations $b$ and $b'$.
Hence we can write a refined version of \meffect\ for fixed $b$ and $b'\in A_n$
\eqn\refeff{\enca{\eqalign{ q^{c(b,b')}~&=~  
\sum_{a \in W_n}f_{a,b}\, f_{a,b'} \, 
U_{\ell_n^a}(q)\cr
&\times \, e^{ {1 \over 4}\sum_{i=0}^{2n-1} 
\big[2(\ell_{i+1}^{\rm  max}-\ell_i^{\rm max})- (\ell_{i+1}^b -\ell_i^b)
-(\ell_{i+1}^{b'} -\ell_i^{b'})\big]
(\ell_{i+1}^a -\ell_i^a)
\log \mu_{(\ell_i^a+\ell_{i+1}^a+1)/2}} \cr} } }
Note that the highly non-local quantity $c(b,b')$ is expressed as a sum
of {\it local} weights. However, the non-locality reemerges in a weaker form
through the selection factors $f$, which induce mutually 
non-local constraints on the walks summed over.

This formula gives an interesting expression for $c(b,b')$ in the limit
of large $q$. Indeed, we have, for $q \to \infty$
\eqn\limitsq{ U_{\ell}(q)~\sim~ q^{\ell} \qquad 
\mu_{\ell} ~\sim ~ {1 \over q} }
hence \refeff\ becomes
\eqn\neweff{ \eqalign{ q^{c(b,b')}~&\sim~ 
\sum_{a \in W_n}f_{a,b}\, f_{a,b'} \,
q^{ {1 \over 4}\sum_{i=0}^{2n-1} \big((\ell_{i+1}^b -\ell_i^b)+ 
(\ell_{i+1}^{b'} -\ell_i^{b'})\big) (\ell_{i+1}^a -\ell_i^a)} \cr} }
where we the contributions of the $\ell_i^{\rm max}$'s and 
that of the Chebishev polynomial have cancelled each other, thanks
to the identity
\eqn\cantche{ \ell_n^a -{1 \over 2}\sum_{i=0}^{2n-1}
(\ell_{i+1}^{\rm max}-\ell_i^{\rm max})(\ell_{i+1}^a-\ell_i^a)~=~0}
For large $q$'s, the sum in the rhs of \neweff\ is dominated by some 
$a \in W_n$
for which the exponent of $q$  is maximal. Such a maximum is unique, 
as the coefficient of $q^{c(b,b')}$ is $1$. This yields
the following formula for the number of connected components $c(b,b')$
\eqn\formaxc{\enca{ c(b,b')={1 \over 4}{{\ } \atop {\displaystyle {\rm max} \atop 
{{\scriptstyle a \in W_n,} \atop
{\scriptstyle b\ {\rm and} \ b'-{\rm symmetric}}  } }} 
\left\{\sum_{i=0}^{2n-1}\big[(\ell_{i+1}^b -\ell_i^b)+ 
(\ell_{i+1}^{b'} -\ell_i^{b'})\big] (\ell_{i+1}^a -\ell_i^a)\right\} }}
A particular case corresponding to semi-meanders consists in
taking $b'={\cal W}_n^{(n)}\equiv r_n$ the rainbow configuration of order $n$.
Using \cantche, we find
\eqn\formaxsm{\enca{ 
c(b)=c(b,{\cal W}_n^{(n)})={1 \over 4}{{\ } \atop {\displaystyle 
{\rm max} \atop {{\scriptstyle a \in W_n, \ {\rm symmetric}} \atop
{\scriptstyle {\rm and}\ b-{\rm symmetric}} } }} \left\{ 2 \ell_n^a+
\sum_{i=0}^{2n-1}(\ell_{i+1}^b -\ell_i^b)(\ell_{i+1}^a -\ell_i^a)\right\} } }

Another interesting consequence of the expression \refeff\ is obtained if we take
$b=b'$, in which case $c(b,b)=n$. It takes the form of a sum rule for $f_{a,b}$,
namely, for any $b\in W_n$ 
\eqn\rulsumf{ q^n~=~\sum_{a \in W_n} f_{a,b}\,U_{\ell_n^a}(q)\, 
e^{{1\over 2}\sum_{i=0}^{2n-1}
\big[(\ell_{i+1}^{\rm  max}-\ell_i^{\rm max})- (\ell_{i+1}^b -\ell_i^b)\big]
(\ell_{i+1}^a -\ell_i^a)
\log \mu_{(\ell_i^a+\ell_{i+1}^a+1)/2} } }
In particular, for $b={\cal W}_n^{(n)}$, hence $\ell_i^b=\ell_i^{\rm max}$ for
all $i$, we find, with $f_{a,{\cal W}_n^{(n)}}=\delta_{a,{\rm symmetric}}$:
\eqn\sumruff{\eqalign{ 
q^n~&=~ \sum_{a \in W_n \atop a \ {\rm symmetric}} U_{\ell_n^a}(q) \cr
&=~ \sum_{p=0}^{[n/2]} b_{n,n-2p}\, U_{n-2p}(q) \cr} }
which is easily proved by recursion on $n$ (the coefficient $b_{n,n-2p}$,
computed in \halwa, is indeed 
the number of symmetric diagrams with middle height $h=n-2p$).

\subsec{Asymptotics for $q \geq 2$}

In this section, we use the expressions \smeffect\-\meffect\
to derive asymptotic formulas for the semi-meander and meander polynomials
for large $n$. Such formulas can only be inferred when all the
terms in the sums \smeffect\-\meffect\ over walk diagrams are positive.
This is the case for all $q \geq 2$, for which $U_m(q)>0$ and $\mu_m>0$
for all $m$.

\noindent{\bf q=2.}
As a preliminary exercise, let us start by taking the limit $q \to 2$ 
of the sum rule \rulsumf.  
Due to the definition \eqtc, we have
\eqn\limtwo{ U_\ell(2)~=~ (\ell+1) \qquad \mu_\ell(2)~=~ {\ell \over \ell+1} }
therefore, when $q \to 2$,  \rulsumf\ becomes
\eqn\limtwosrf{ 2^n~=~ \sum_{a \in W_n} f_{a,b} \, (\ell_n^a+1)
e^{{1\over 2}\sum_{i=0}^{2n-1}
\big[(\ell_{i+1}^{\rm  max}-\ell_i^{\rm max})- (\ell_{i+1}^b -\ell_i^b)\big]
(\ell_{i+1}^a -\ell_i^a)
\log {\ell_i^a+\ell_{i+1}^a+1 \over \ell_i^a+\ell_{i+1}^a+3} } }
Note that, summing \limtwosrf\ over $b \in W_n$ we get the result
\eqn\assy{\sum_{a,b \in W_n} f_{a,b} \, (\ell_n^a+1)
e^{{1\over 2}\sum_{i=0}^{2n-1}
\big[(\ell_{i+1}^{\rm  max}-\ell_i^{\rm max})- (\ell_{i+1}^b -\ell_i^b)\big]
(\ell_{i+1}^a -\ell_i^a)
\log {\ell_i^a+\ell_{i+1}^a+1 \over \ell_i^a+\ell_{i+1}^a+3} }~=~ 2^n c_n}
which behaves, for large $n$, like
\eqn\toassy{ {8^n \over n^{3/2}} ~\sim~
n \sum_{a,b \in W_n}  f_{a,b}  } 
by making use of the asymptotics \asyfabtot.  Comparing \assy\ and \toassy,
we are led to the following scaling hypothesis for the values
of $\ell_i^b$ and $\ell_i^a$ dominating the sum \assy:
\eqn\scahyp{ \ell_i^a~\sim~ n^\nu \ell^a(x) \qquad \ell_i^b~\sim~ n^\nu \ell^b(x)}
where $x=i/n$ and $\nu\in [0,1]$ is an exponent characterizing the average 
height of the walk diagrams $a$, $b$. 
For this hypothesis to be compatible with \toassy, we must necessarily have
$\nu=1$, in which case the exponential in \assy\ tends to a 
constant\foot{To see why, note that for large $n$ and $\ell$'s the sum in 
the exponential may be approximated by 
$$\eqalign{{1 \over 2}\sum_{i=0}^{2n-1}
\big[(\ell_{i+1}^{\rm  max}-\ell_i^{\rm max})&- (\ell_{i+1}^b -\ell_i^b)\big]
{\ell_{i+1}^a -\ell_i^a \over \ell_i^a+1} \cr
&\sim~ -\sum_{i=0}^{2n-1}\bigg[\delta_{i,n}-\big({\ell_{i+1}^b+
\ell_{i-1}^b \over 2}-\ell_i^b \big) \bigg]\, \log (\ell_i^a+1) \cr}$$
where we have performed a discrete integration by parts.
Hence the exponential of this sum is equivalent to
$$ (\ell_n^a+1) \times {\prod_{i\ {\rm min.}\ {\rm of}\ b} (\ell_i^a+1) \over 
\prod_{i\ {\rm max.}\ {\rm of}\ b} (\ell_i^a+1)} ~\sim~{\rm const.} $$
The products extend respectively over the $i$'s which are minima and maxima of the walk $b$
and as there is always one more maximum than minima, the above ratio is exactly balanced,
hence is of order $1$ for large $\ell_i^a$'s.}
(the sum over $i$ is of order $n$, but the logarithm is of order $1/n$), 
and the factor $(\ell_i^a+1)$ tends to const.$\times n$, which yields
\toassy.
This is an example of use of a scaling hypothesis on the $\ell$'s 
dominating the sum \assy, leading to large $n$ asymptotics.

Analogously, if we make the same scaling hypothesis \scahyp, with $\nu=1$, 
on the 
$\ell$'s dominating the sums \smeffect\-\meffect, for $q=2$, we find
the asymptotic relations, valid for large $n$
\eqn\valasym{\enca{\eqalign{ {\bar m}_n(2)~&\sim~ n \sum_{a,b \in W_n \atop 
a\ {\rm symmetric}} f_{a,b} \cr
m_n(2)~&\sim~ n \sum_{a,b,b' \in W_n} f_{a,b}\, f_{a,b'} \cr} } }
This expresses the asymptotics of the meander and semi-meander polynomials 
at $q=2$ in terms of $f_{a,b}$ only.
In going from \toassy\ to \valasym, we have assumed that configurations of
the {\it same} order of magnitude dominate both sums.  In fact, we have
made a scaling hypothesis on the matrix elements of ${\cal P}_n^{-1}(q=2)$
and $\Gamma_n(q=2)$, namely that the configurations
with 
\eqn\scap{ \big[{\cal P}_n^{-1}(2)\big]_{a,b}~\sim~ f_{a,b} \qquad 
\big[{\Gamma}_n(2)\big]_{a,a}~=~(\ell_n^a+1)~\sim~n^\nu }
dominate the three sums
\eqn\bothsu{\eqalign{ {\rm Tr}\big({\cal G}_n(2)\big)&~\sim~ n^\nu\, 
\sum_{a,b \in W_n} f_{a,b} \cr
\vec{v} \cdot {\cal G}_n(2) \vec{u}&~\sim~ n^\nu \sum_{a,b \in W_n
\atop a\ {\rm symmetric}} f_{a,b} \cr
\vec{u} \cdot {\cal G}_n(2) \vec{u}&~\sim~ n^\nu \sum_{a,b,b' \in W_n} 
f_{a,b}\, f_{a,b'} \cr
}}
with the same value of $\nu=1$.  
Let us stress, however, that the scaling hypothesis \scap\ leads to a
wrong result for the meander determinant, $D_n(2)$, for large $n$.
Indeed, from \scap, we would conclude that 
\eqn\wrongdet{D_n(2)~\sim~\prod_{a \in W_n} f_{a,a}^2  n^{\nu}
~\sim~ n^{\nu c_n} }
whereas, from the exact result \mainres\ for $D_n(2)$, we extract the
large $n$ asymptotics
\eqn\rightdet{ \log \, D_n(2)~=~\sum_{j=1}^n a_{n,j} \, \log (j+1)
~\sim~ \sqrt{\pi n}\, c_n}
by the standard saddle point technique 
(note that we find exactly twice the previous result \lodet\ for the large $n$
asymptotics of $\log \det D_n'(0)$). 
The correct asymptotics \rightdet\ contradict \wrongdet. 
This simply means that the configurations of 
$a \in W_n$ dominating the meander determinant are very different from those 
dominating the trace of the Gram matrix or the (semi-)meander polynomial.

\noindent{\bf q$>$2.}
We start again from the sum rule \rulsumf, with $q=e^\theta+e^{-\theta}$,
$\theta>0$. We again make the hypothesis that, when summed over $b\in W_n$,
the sum \rulsumf\ is dominated by large $\ell$'s for large $n$. 
Noting that
\eqn\larnmu{ U_m(e^\theta +e^{-\theta})~
\sim~ {e^{m \theta} \over 1-e^{-2\theta}} \qquad
\mu_m ~\sim ~ e^{-\theta} }
for large $m$, 
this gives the asymptotic formula
\eqn\asfortic{\eqalign{
c_n \, (e^\theta + e^{-\theta})^n ~&\sim~
\sum_{a,b \in W_n} f_{a,b} {e^{\theta\ell_n^a} \over 1 -e^{-2\theta}} 
e^{-{\theta \over 2} \sum_{i=0}^{2n-1}
\big[(\ell_{i+1}^{\rm  max}-\ell_i^{\rm max})- (\ell_{i+1}^b -\ell_i^b)\big]
(\ell_{i+1}^a -\ell_i^a)} \cr
&=~ {1 \over 1 -e^{-2\theta}}\sum_{a,b \in W_n} f_{a,b} 
e^{{\theta \over 2} \sum_{i=0}^{2n-1}
(\ell_{i+1}^b -\ell_i^b)(\ell_{i+1}^a -\ell_i^a)} \cr
&\sim~ {4^n \over n^{3/2}}\, (e^\theta + e^{-\theta})^n \cr}}
where we have used \cantche.  This gives an asymptotic sum rule involving
the $f_{a,b}$'s and $q$.

Assuming that the same scaling hypothesis holds for the sums \smeffect\-\meffect,
we find the following asymptotic formulas
\eqn\assasimp{\enca{ \eqalign{
{\bar m}_n(e^\theta+e^{-\theta})~&\sim~\sum_{a,b\in W_n \atop
a\ {\rm symmetric}}f_{a,b}
e^{{\theta \over 2}\big[ \ell_n^a +{1\over 2}
\sum_{i=0}^{2n-1} (\ell_{i+1}^b -\ell_i^b)(\ell_{i+1}^a -\ell_i^a) \big]} \cr
m_n(e^\theta+e^{-\theta})~&\sim~\sum_{a,b,b'\in W_n}f_{a,b}f_{a,b'}
e^{{\theta \over 4}\sum_{i=0}^{2n-1} 
\big[(\ell_{i+1}^b -\ell_i^b)+(\ell_{i+1}^{b'} -\ell_i^{b'})\big]
(\ell_{i+1}^a -\ell_i^a)} \cr } } }
where we have dropped the prefactor $1/(1-e^{-2\theta})$, subleading for $\theta>0$.
Indeed, the limits $\theta \to 0$ and $n \to \infty$ do not commute,
hence \assasimp\ is only valid for $\theta>0$. On the other hand, in
the limit $\theta \to \infty$, we recover the large $q$
asymptotics 
\eqn\asymmto{\eqalign{
{\bar m}_n(q)&~\sim ~q^n~\sim~e^{n \theta} \cr
m_n(q)~&\sim~ c_n \, q^n~\sim~{(4 e^\theta)^n \over n^{3/2}} \cr } }
by using the two formulas \formaxsm\-\formaxc. 

As before, we can test the scaling hypothesis used above against the large
$n$ asymptotics of the meander determinant for $q>2$.  This hypothesis amounts
to writing
\eqn\hyptwo{\eqalign{ \big[{\cal P}_n^{-1}(e^\theta+e^{-\theta})\big]_{a,b}~&\sim~ 
f_{a,b}\, e^{{\theta \over 4}\big[-2 \ell_n^a+\sum_{i=0}^{2n-1}
(\ell_{i+1}^b -\ell_i^b)(\ell_{i+1}^a -\ell_i^a)\big]} \cr 
\big[{\Gamma}_n(e^\theta+e^{-\theta})\big]_{a,a}~&\sim~e^{\theta \ell_n^a} \cr} }
The corresponding large $n$ estimate of the meander determinant reads
\eqn\larnest{ D_n(e^\theta+e^{-\theta})~\sim~\prod_{a \in W_n} f_{a,a}^2
e^{n \theta} ~\sim~ e^{n c_n \theta} }
whereas the exact formula \mainres\ leads to the asymptotics
\eqn\truasy{ \log \, D_n(e^\theta+e^{-\theta})~=~\theta \sum_{j=1}^n a_{n,j}
\, \log { \sinh (j+1)\theta \over \sinh \theta} ~\sim~
n c_n \theta }
by the standard saddle point method.
The agreement between the two estimates \larnest\-\truasy\ is a 
confirmation {\it a posteriori} that the scaling hypothesis \hyptwo\ holds for
a very large class of properties of the gram matrix ${\cal G}_n(q)$, for
$q>2$ and large $n$.

Finally, in view of the assumed $q=2$ value $\nu(2)=1$, and the exact $q\to \infty$
value $\nu(\infty)=1$ (the semi-meander polynomial \asymmto\ is indeed dominated
by the single diagram $b={\cal W}_n^{(n)}$, with winding 
$\ell_n^b=n\sim n^{\nu(\infty)}$),
it is reasonable to infer that $\nu(q)$ is identically equal to $1$ for all $q \geq 2$.

\subsec{Meander and semi-meander polynomials as SOS partition functions}

The asymptotic formulas \assasimp\ are to be compared with the following 
exact formulas
\eqn\exacmm{\enca{\eqalign{
{\bar m}_n(e^\theta+e^{-\theta})~&=~\sum_{a\in P_n, b\in W_n \atop
a\ {\rm symmetric}}f_{a,b}
e^{{\theta \over 2}\big[ \ell_n^a +{1\over 2}
\sum_{i=0}^{2n-1} (\ell_{i+1}^b -\ell_i^b)(\ell_{i+1}^a -\ell_i^a) \big]} \cr
m_n(e^\theta+e^{-\theta})~&=~\sum_{a\in P_n\atop b,b'\in W_n}f_{a,b}f_{a,b'}
e^{{\theta \over 4}\sum_{i=0}^{2n-1} 
\big[(\ell_{i+1}^b -\ell_i^b)+(\ell_{i+1}^{b'} -\ell_i^{b'})\big]
(\ell_{i+1}^a -\ell_i^a)} \cr } } }
where $a$ runs now over the set $P_n$
of {\it all} closed paths of $(2n)$ steps
(with $\ell_0^a=\ell_{2n}^a=0$) {\it not subject to the constraint}
$\ell_i^a\geq 0$.
The relations \exacmm\ may indeed be obtained as consequences of the 
following alternative formula for $q^{c(b,b')}$, $b$, $b' \ \in W_n$ (to be compared
with \refeff)
\eqn\alterego{\enca{ 
(e^\theta+e^{-\theta})^{c(b,b')}~=~\sum_{a \in P_n} f_{a,b}f_{a,b'}
e^{{\theta \over 4}\sum_{i=0}^{2n-1} 
\big[(\ell_{i+1}^b -\ell_i^b)+(\ell_{i+1}^{b'} -\ell_i^{b'})\big]
(\ell_{i+1}^a -\ell_i^a)} } }
Let us now prove \alterego.
On the one hand, as $a$ is both $b$ and $b'$-symmetric,
the values of $t_i(a)=(\ell_{i+1}^a -\ell_i^a)$ are fixed, up to an overall sign,
along each connected component of the meander $(b,b')$, and alternate on successive
bridges along the connected component.

\fig{The four possible local environments of the $(i+1)$-th bridge together with the
corresponding value $s_i(b,b')=\pm 1,0$.}{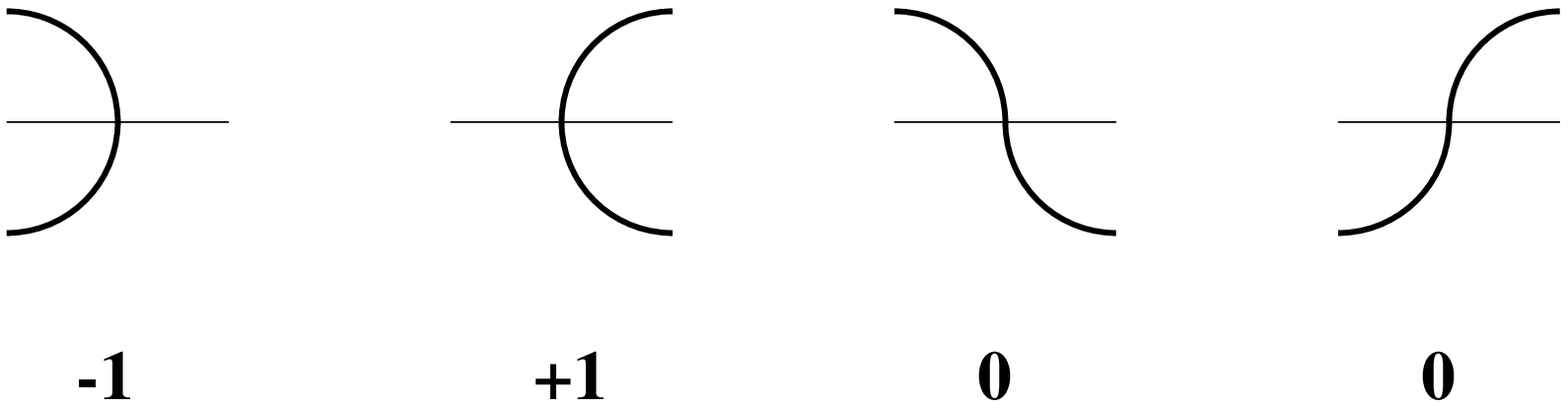}{10.cm}
\figlabel\locenv

On the other hand, the quantity 
$s_i(b,b')=\big[(\ell_{i+1}^b -\ell_i^b)+(\ell_{i+1}^{b'} -\ell_i^{b'})\big]/2$ 
may only take the three
values $-1$, $0$ and $+1$, corresponding to the four possibilities of local
environment of the $(i+1)$-th bridge of the meander $(b,b')$, depicted in Fig.\locenv.
Along any connected component of $(b,b')$, the variable $s_i(b,b')$ alternates as long
as it remains nonzero, and discarding all the zeros 
leaves us with an alternating sign. 

\fig{An oriented connected component $K$ with $10$ bridges. Starting from
bridge $1$, the sequence of visited bridges is 1, 8, 9, 10, 3, 4, 7, 6, 
5, 2.}{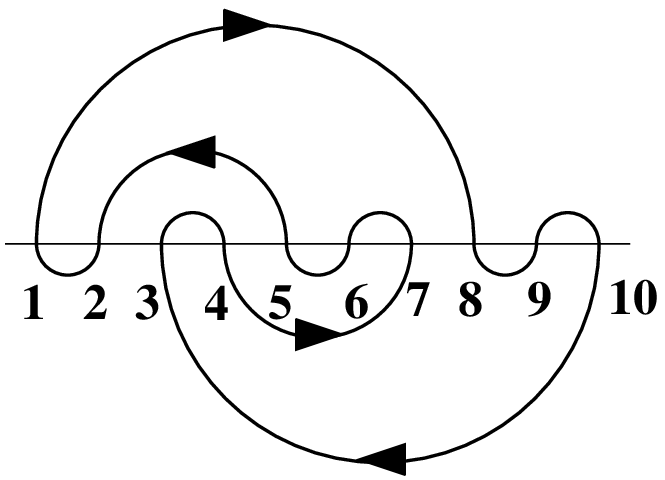}{6.cm}
\figlabel\connectex

For illustration, with the connected component
depicted in Fig.\connectex, this gives the sequence, starting from the bridge $1$
$$\enca{\matrix{ {\rm bridge}\ i & 1 & 8 & 9 & 10 & 3 & 4 & 7 & 6 & 5 & 2 \cr
t_i(a)    & + & - & + & - & + & - & + & - & + & - \cr
s_i(b,b') & + & 0 & 0 & - & + & 0 & - & 0 & 0 & 0 \cr
t_i(a) s_i(b,b') & + & 0 & 0 & + & + & 0 & - & 0 & 0 & 0 \cr
{\rm turn}       & R & - & - & R & R & - & L & - & - & - \cr} }$$
where we also indicated the type of turn (right=R, left=L) taken on the corresponding
bridge. 
The global sign $t_i(a) s_i(b,b')$ is thus constant between two zeros
and is reversed through each zero. Since a zero indicates a transition from turning
left to right and vice versa along the meander, the quantity
\eqn\sommeg{ {1 \over 2} \sum_{i\ {\rm along}\ K}(\ell_{i+1}^a -\ell_i^a) 
\big[(\ell_{i+1}^b -\ell_i^b)+(\ell_{i+1}^{b'} -\ell_i^{b'})\big] }
summed along any connected component $K$ of the meander $(b,b')$, is simply equal, up to a
sign, to the total number of right turns minus that of left turns $(n_R-n_L)$,
taken on the bridges along $K$. As on any closed loop we have $(n_R-n_L)=\pm 2$, we
compute
\eqn\calk{\eqalign{ 
f(K)~&=~
\sum_{t_i(a)=\pm 1\atop i\ {\rm along}\ K} f_{a,b} f_{a,b'} e^{{\theta \over 2}
\sum_{i \ {\rm along}\ K} t_i(a) s_i(a) } \cr
&=~ \sum_{\epsilon=\pm 1} e^{\theta \epsilon (n_R-n_L)/2} \cr
&=~ e^\theta + e^{-\theta} \cr} }
where the sum over $\epsilon=\pm 1$ corresponds to the only overall sign ambiguity
left on the $t_i(a)$ after taking into account the $b$ and $b'$-symmetry of $a$ on $K$.
The final result \alterego\ is simply the product over all the connected components $K$ of
$(b,b')$ of the weight $f(K)$ above, which completes the proof of the result.

More generally, the above analysis can be carried over to $q=z+1/z$, for any
complex number $z$, resulting in
\eqn\egoalter{\enca{ (z+1/z)^{c(b,b')}~=~ \sum_{a \in P_n}
f_{a,b}f_{a,b'} z^{{1 \over 4}\sum_{i=0}^{2n-1} 
\big[(\ell_{i+1}^b -\ell_i^b)+(\ell_{i+1}^{b'} -\ell_i^{b'})\big]
(\ell_{i+1}^a -\ell_i^a)}  } }
This yields the following general expressions for semi-meander and meander polynomials 
at $q=z+1/z$ for arbitrary complex $z$
\eqn\examme{\enca{\eqalign{
{\bar m}_n(z+1/z)~&=~\sum_{a\in P_n, b\in W_n \atop
a\ {\rm symmetric}}f_{a,b}\,
z^{{1 \over 2}\big[ \ell_n^a +{1\over 2}
\sum_{i=0}^{2n-1} (\ell_{i+1}^b -\ell_i^b)(\ell_{i+1}^a -\ell_i^a) \big]} \cr
m_n(z+1/z)~&=~\sum_{a\in P_n\atop b,b'\in W_n}f_{a,b}f_{a,b'}\,
z^{{1 \over 4}\sum_{i=0}^{2n-1} 
\big[(\ell_{i+1}^b -\ell_i^b)+(\ell_{i+1}^{b'} -\ell_i^{b'})\big]
(\ell_{i+1}^a -\ell_i^a)} \cr } } }

\fig{An example of SOS configuration attached to a meander. 
We display the value of the height $\ell$. Note that it is entirely dictated
by the choices of orientation of the connected components of the meander, and the
fact that $\ell=0$ at infinity.}{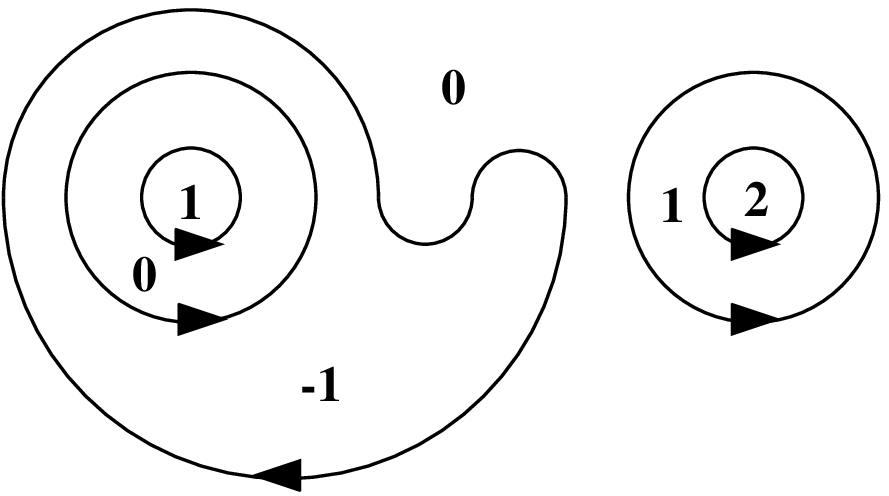}{7.cm}
\figlabel\exsos

This analysis suggests to interpret the quantity $q^{c(b,b')}$ as the Boltzmann weight
of a particular configuration, formed by the meander $(b,b')$,
of a suitably defined {\it SOS model}. 
Indeed, the $b$ and $b'$-symmetry of $a \in P_n$ implies that 
the variable $\ell_i^a$ takes identical values on all segments of river which can be
connected to each other without crossing any arch of $b$ or $b'$.
Therefore, the variable $\ell_i^a$ may be considered as a height
variable in the plane, constant on each connected component delimited by one ore 
several roads, and undergoing a jump discontinuity of $\pm 1$ across each road
(see Fig.\exsos\ for an example), and continuous across the river. In particular,
$\ell=0$ at infinity, due to the boundary condition $\ell_0=\ell_{2n}=0$.
Such an height configuration induces a unique orientation of the various
connected components
of $(b,b')$, by 
taking the convention that $\ell \to \ell+1$ (resp. $\ell \to \ell-1$)
across a road pointing to the right (resp. left). Conversely, a choice
of orientation of the connected components of $(b,b')$ specifies uniquely the height configuration, by further demanding
that $\ell=0$ at infinity.
The Boltzmann weight
\eqn\boweit{z^{{1 \over 4}\sum_{i=0}^{2n-1} 
\big[(\ell_{i+1}^b -\ell_i^b)+(\ell_{i+1}^{b'} -\ell_i^{b'})\big]
(\ell_{i+1}^a -\ell_i^a)} }
corresponds to attaching
to each bridge of $(b,b')$ one of the following Boltzmann weights
\eqn\bowesos{ \matrix{ \figbox{1.2cm}{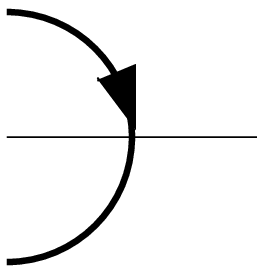} & \figbox{1.2cm}{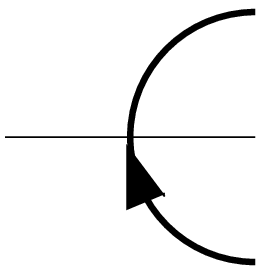} & 
\figbox{1.2cm}{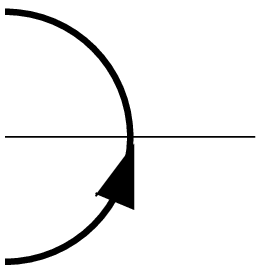} & \figbox{1.2cm}{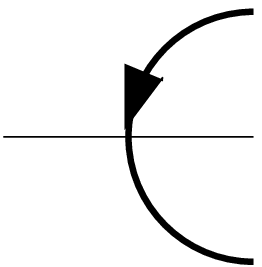} \cr
z^{1 \over 2} & z^{1 \over 2} &  z^{-{1 \over 2}} & z^{-{1 \over 2}} \cr
\figbox{2.6cm}{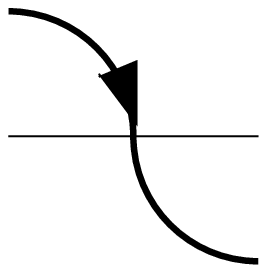}
& \figbox{2.6cm}{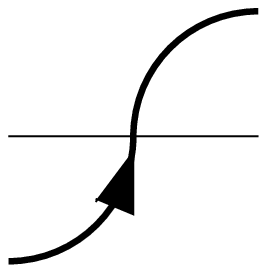} & \figbox{2.6cm}{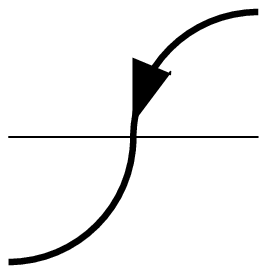} & 
\figbox{2.6cm}{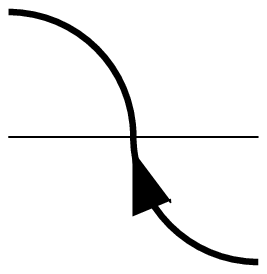} \cr 
1 & 1 & 1 & 1 \cr} }
according to the local environment of the bridge, and taking the product over all 
the bridge weights. 
Again, summing over the two orientations of each connected component
$K$ of $(b,b')$ results in a total weight per connected component
\eqn\soswei{ \sum_{\epsilon=\pm 1} z^{\epsilon (n_R -n_L)/2} ~=~ z+1/z~=~q}
where $n_R$ (resp. $n_L$) is the number of right (resp. left) turns of the road 
on the bridges of $K$, and $\epsilon=\pm 1$ accounts for the global orientation of $K$.
In the language of SOS models, the expression \refeff\ corresponds to a {\it Restricted
SOS} version, in which the height variable is further restricted to be non-negative
(in particular the configuration of Fig.\exsos\ is ruled out).

As a first element of comparison with the results of the previous section, 
if we write \examme\ at $z=1$, hence $q=2$, we see that
\eqn\compapre{\enca{\eqalign{ {\bar m}_n(2)~&=~\sum_{a\in P_n, b\in W_n \atop
a\ {\rm symmetric}}f_{a,b} \cr
m_n(2)~&=~\sum_{a\in P_n\atop b,b'\in W_n}f_{a,b} \, f_{a,b'} \cr}} }
to be compared with the asymptotic estimates \valasym: this gives a relation
between sums over $P_n$ and over $W_n$, involving the same combinations of $f$. 
Note that the same type of relation links the cardinals of the two sets
over which $a$ is summed, namely
\eqn\carcomp{ {\rm card}(P_n)~=~ {n \choose 2n} ~=~ (n+1) c_n~=~ (n+1)\,
{\rm card}(W_n) }
and also, using \sumfab\
\eqn\compaff{ \sum_{a \in P_n, b \in W_n} f_{a,b}~=~2^n\, c_n~=~
{2 \over 3} (n+2) \sum_{a,b \in W_n} f_{a,b} }
The reader could wonder in what the restricted expressions \smeffect\-\meffect\
of the previous
section are really different from the simple SOS expressions \exacmm\ obtained above.
Actually, the considerations of the previous section 
on the heights $\ell$ dominating the 
expressions \smeffect\-\meffect\
for the meander and semi-meander polynomials,
eventually leading to an exponent $\nu=1$ for $q=2$, could not be carried over
here, because of the lack of an explicit prefactor proportional to $(\ell+1)$.
Hence, in some sense, the formulas \smeffect\-\meffect\ (at least for $q=2$)
give us access to more
precise details on the path formulation.

More generally, it is interesting to compare the $q>2$ formulas \exacmm\ and
\assasimp. We see that these are identical, except for the range of summation over $a$
($W_n$ in \assasimp\ and $P_n$ in \exacmm).  
We conclude that the restriction condition that $\ell_i^a \geq 0$ in \assasimp\
is not important in the large $n$  limit, for $q>2$.

\newsec{Generalization: the semi-meander determinant}

In this section, we consider a possible generalization of the meander determinant
to semi-meanders in the following way. 

\fig{Any semi-meander may be viewed as the superimposition of an upper and a lower
open arch configurations. Here the initial semi-meander has winding $3$.
The two open arch configurations on the right have $h=3$ open arches. To
recover the initial semi-meander, these open arches must be connected two by two,
fom the right to the left
(the arches number 5,4,1 of the upper configuration are respectively connected to
the arches number 5,4,3 of the lower configuration).}{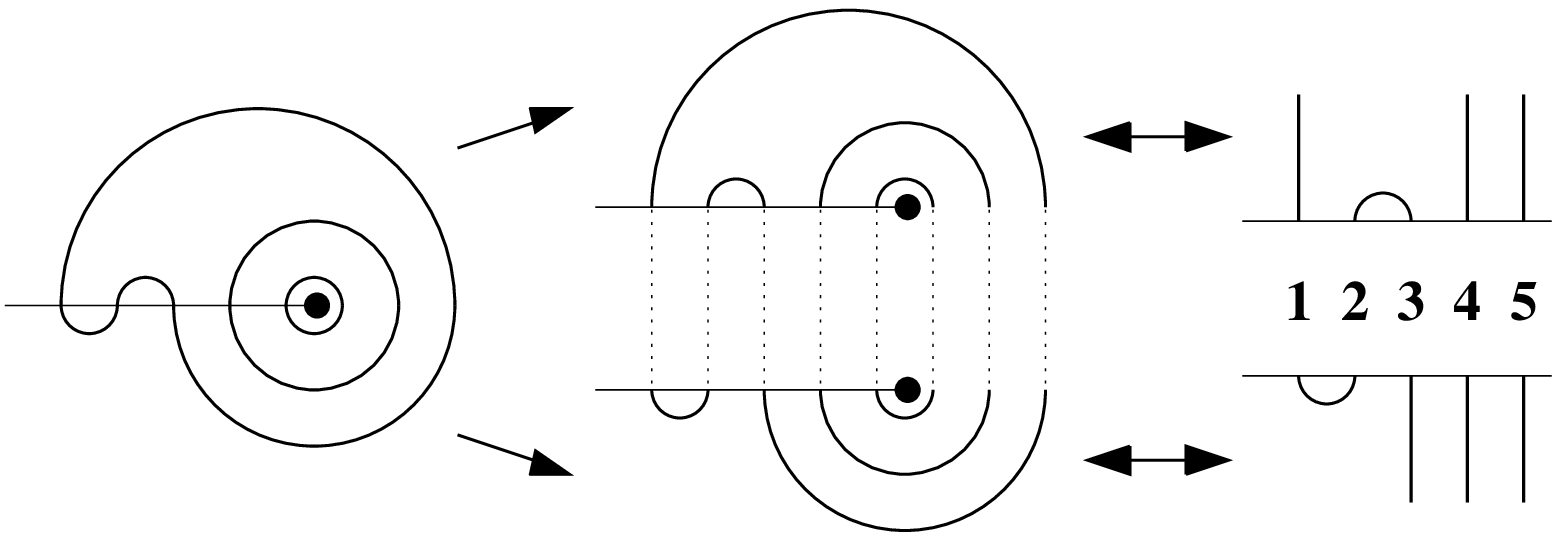}{10.cm}
\figlabel\smopen

Going back to the original river/road formulation
of semi-meanders, we see on Fig.\smopen\
that any given semi-meander, with winding number $h$, 
is obtained as the superimposition of two (upper and
lower) {\it open arch configurations} of order
$n$, with $h$ open arches. By this, we mean that $h$ semi-infinite 
vertical roads originate from $h$ of the $n$ bridges, otherwise connected by 
pairs through $(n-h)/2$ nonintersecting arches (the winding
$h$ has always the same parity as the order $n$ in the semi-meanders). 
The semi-meander is re-built in a unique way by connecting the upper
and lower open arches from the right to the left. In particular, only open arch configurations with the same number of open arches may be superimposed to yield
a semi-meander.  Let $A_n^{(h)}$ denote the set of open arch configurations of
order $n$ with $h$ open arches.  It is a simple exercise to show that
\eqn\cardaa{ {\rm card}(A_n^{(h)})~=~b_{n,h}~=~{n \choose {n-h \over 2}}-
{n \choose {n-h \over 2}-1} } 
Indeed, the open arch configurations of order $n$ with $h$ open arches are
in one-to-one correspondence with the half-walk diagrams of $n$ steps, with final
height $h$, namely with $\ell_0=0$, $\ell_i\geq 0$ and $\ell_n=h$. 
Let $W_n^{(h)} \equiv A_n^{(h)}$ denote the set of half-walks of order $n$ with 
final height $h$.  
The number of such half-walks has been derived in eq.\halwa\ above.
We now define the semi-meander determinant of order $n$ and winding $h$, as the
determinant $D_n^{(h)}(q)$ of the matrix ${\cal G}_n^{(h)}(q)$ with entries
\eqn\matgrasm{ \big[ {\cal G}_n^{(h)}(q) \big]_{l,l'}~=~ q^{c(l,l')} \qquad
l,\ l'\ \in W_n^{(h)}\equiv A_n^{(h)} }
where $c(l,l')$ denotes the number of connected components of the semi-meander
obtained by superimposing the open arch configurations $l$ and $l'$ and
connecting their $h$ open arches.
For illustration, we list below the matrices corresponding to $n=4$, $h=0,2,4$
\eqn\exmatsm{ {\cal G}_4^{(0)}(q)~=~\pmatrix{q^2 &q \cr
q &q^2 \cr} \qquad {\cal G}_4^{(2)}(q)~=~\pmatrix{q^3 &q^2 &q\cr
q^2 &q^3 &q^2 \cr
q &q^2 &q^3 \cr}\qquad {\cal G}_4^{(4)}(q)~=~ q^4 }
with the following ordering of open arch configurations
\eqn\orderarch{ h=0~:\ \ \figbox{2.cm}{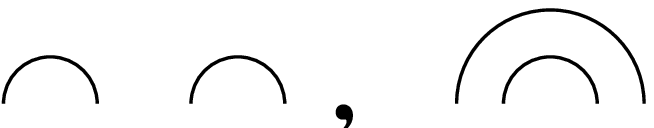}\quad 
h=2~:\ \ \figbox{3.cm}{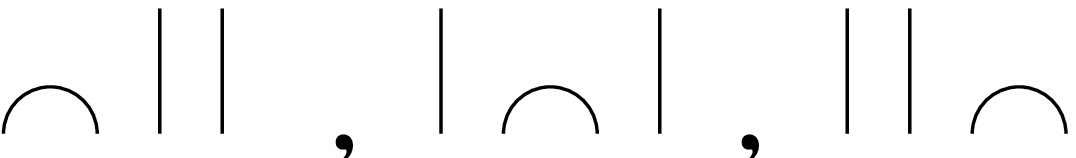}\quad h=4~:\ \ \figbox{.8cm}{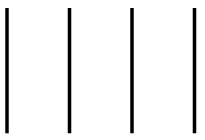} }
Note also that ${\cal G}_{2n}^{(0)}(q)={\cal G}_{2n-1}^{(1)}(q)={\cal G}_{n}(q)$, 
hence the formula \mainres\ applies to the winding zero and one cases.  More
generally, we conjecture that 
\eqn\conjdetsm{\enca{ D_n^{(h)}(q)~=~\det \,{\cal G}_n^{(h)}(q)~=~ 
\prod_{j=1}^{{n-h \over 2}+1} U_j(q)^{\alpha_{n,j}^{(h)} } } }
where the numbers $\alpha_{n,j}^{(h)}$ read, in terms of the $a_{n,j}$
of \mainres\
\eqn\numalph{\enca{\eqalign{ \alpha_{2n,j}^{(2h)}~&=~a_{n,j+h}+ 2h\, a_{n,j+h-1} \cr
\alpha_{2n-1,j}^{(2h+1)}~&=~ a_{n,j+h}+2h\, (a_{n-1,j+h}+a_{n-1,j+h-1})
\cr}}}
We checked the validity of this conjecture up to $n=9$. 
For instance, for $n=8,9$, we have
\eqn\neight{\eqalign{ 
&\enca{n~=~8 \qquad \matrix{ D_8^{(0)} &=~U_1^8 &U_2^{13} &U_3^6 &U_4 \cr
D_8^{(2)} &=~U_1^{29} &U_2^{32} &U_3^{13} &U_4^2 \cr
D_8^{(4)} &=~U_1^{58} &U_2^{25} &U_3^4 & \cr
D_8^{(6)} &=~U_1^{37} &U_2^{6} &  & \cr
D_8^{(8)} &=~U_1^8 & & & \cr}} \cr
&\enca{n~=~9 \qquad 
\matrix{ D_9^{(1)} &=~U_1^{15} &U_2^{40} &U_3^{26} &U_4^8 &U_5 \cr
D_9^{(3)} &=~U_1^{82} &U_2^{64} &U_3^{22} &U_4^3 \cr
D_9^{(5)} &=~U_1^{102} &U_2^{36} &U_3^5 & \cr
D_9^{(7)} &=~U_1^{50} &U_2^{7} &  & \cr
D_9^{(9)} &=~U_1^9 & & & \cr} }
\cr}}
in agreement with \conjdetsm\-\numalph. We have performed various checks on 
the numbers $\alpha_{n,j}^{(h)}$ \numalph. In particular, the
term of highest degree of $D_n^{(h)}(q)$, as a polynomial of $q$,
is given by the product of the diagonal terms in 
${\cal G}_n^{(h)}(q)$, namely
\eqn\prodiasm{ q^{{\rm deg}(D_n^{(h)})}~=~\prod_{l \in W_n^{(h)} } q^{n+h \over 2} }
hence
\eqn\degsmp{ {\rm deg}(D_n^{(h)})~=~{n+h \over 2} b_{n,h} }
This can actually be derived from \numalph.

We expect that \conjdetsm\-\numalph\ can be proved by diagonalizing 
the matrix ${\cal G}_n^{(h)}(q)$.
This matrix has again a simple interpretation as the Gram matrix of a certain
subspace of $TL_n(q)$, generated by some particular basis 1 elements.   
Inspired by the one-to-one correspondence between walk diagrams of order $n$ and 
the elements of the basis 1, we attach to any half-walk $l$ of $n$ steps and final 
height $h$ in $W_n^{(h)}$ the basis 1 element $(a)_1$
corresponding to the walk diagram 
$a=lr\in W_n$, where we have completed the half-walk $l$ with a particular choice
of right half-walk $r$ of final height $h$, namely with $\ell_i^r=[1 + (-1)^i]/2$,
$i=0,1,...,n-h$, and $\ell_i^r = i+h-n$ for $i=n-h+1,n-h+2,...,n$.  
This corresponds to
only retaining basis 1 elements which are obtained by acting on $f_h^{(n)}$
(defined in \redbas)
through {\it left} multiplications by $e_i$. 
In this new basis, the scalar product between two elements reads
\eqn\scanew{ (lr,l'r)~=~ {\rm Tr}\big( (lr)_1 (l'r)_1^t \big)~=~
q^{c(lr,l'r)} ~=~ q^{n-h \over 2} \, q^{c(l,l')}  }
which coincides with \matgrasm\ up to an overall prefactor of $q^{(n-h)/2}$ due
to the addition of $(n-h)/2$ trivial loops to the semi-meander $ll'$.  
A proof of \conjdetsm\-\numalph\ should 
follow the lines of that of \mainres, by writing a change of basis which 
diagonalizes the Gram matrix \matgrasm.  Note also that like in the meander case, 
the formula \conjdetsm\-\numalph\ gives the multiplicities of the zeros of
$D_n^{(h)}(q)$.  

Finally, the product 
over all the possible windings of
the semi-meander determinants takes the simple form
\eqn\prowin{\enca{{\bar D}_{n}(q)~=~ \prod_{h=0 \atop n-h=0\ {\rm mod} \ 2}^n
D_n^{(h)}(q)~=~ \prod_{j=1}^n U_j(q)^{\beta_{n,j}} } }
where
\eqn\betsm{\enca{\eqalign{ \beta_{2n,j}~&=~ 3 {2n \choose n-j} -{2n \choose n-j-1} \cr
\beta_{2n-1,j}~&=~3 {2n \choose n-j} -2 {2n-1 \choose n-j}-{2n \choose n-j-1} \cr}}}
as a direct consequence of \numalph, with $\beta_{n,j}=\sum_h \alpha_{n,j}^{(h)}$.
Eq.\prowin\ may be viewed as the semi-meander counterpart of \mainres.  

The semi-meander gram matrix \matgrasm\ also gives access to refined properties of the
semi-meanders. Indeed, we may compute
\eqn\meanwin{\eqalign{ 
{\bar m}_n^{(h)}(q^2)~&=~ {\rm Tr}\left( {\cal G}_n^{(h)}(q)^2 \right)\cr
&=~\sum_{k=1}^n {\bar M}_n^{(k)}(h)\, q^{2k} \cr}}
where ${\bar M}_n^{(k)}(h)$ denotes the total number of semi-meanders
of order $n$ with winding $h$ and $k$ connected components. An asymptotic
study of these numbers should be made possible by the explicit 
diagonalization of ${\cal G}_n^{(h)}(q)$.

\newsec{Conclusion}

In this paper, we have extensively studied the representation of
the meander and semi-meander enumeration problems within the framework
of the Temperley-Lieb algebra $TL_n(q)$. 
This representation is induced by the existence of a map between the 
reduced elements of $TL_n(q)$ and the arch configurations of order $n$ used
to build meanders and semi-meanders.  Moreover, we have seen that the 
standard trace over $TL_n(q)$ provides a tool for counting the number
of connected components of meandric objects.  
The first result of this paper is a direct computation of the meander determinant
\mainres,
interpreted as the Gram determinant of the basis of reduced elements of
$TL_n(q)$, and the exact study of its zeros \rewdet\
and associated multiplicities \gendn\-\genmudn.

Beyond the meander determinant, we have been able to rewrite the change of
basis diagonalizing the Gram matrix in terms of local height variables 
defining a restricted SOS model (see \refeff). 
We also derived an unrestricted
SOS model interpretation (see \egoalter) of the Gram matrix elements.  These
lead to various expressions for the meander and semi-meander polynomials,
as weighted sums over discrete paths (walk diagrams).  It is tempting to
try to approximate these sums by continuous path integrals, in the limit
of large number of bridges.  In the case $q \geq 2$, where all
the SOS Boltzmann weights are positive, this path integral might even 
be dominated by a simple subset of configurations, obtained
for instance through a saddle point approximation.

A generalization of this approach to the semi-meanders with fixed winding
(number of times the roads wind around the source of the river) 
should be possible, in view of the conjectured form \conjdetsm\ for the
corresponding (fixed winding) semi-meander determinants. 
A proof of \conjdetsm\ should be at hand, by a 
simple adaptation of the proof of \mainres\ presented here. 
This will be addressed elsewhere.

\noindent{\bf Acknowledgements}

We thank A. Zvonkin for bringing Ref.\KOSMO\ to our knowledge, R. Balian
for helpful discussions, S. Legendre for interesting historical remarks
and J.-B. Zuber for a careful reading of the manuscript.

\appendix{A}{Proof of the formula \gendn\ for the multiplicities of the
zeros of the meander determinant}

In order to prove \gendn, we note that
\eqn\delmod{ \delta_{j+1,0\ {\rm mod}\ (k+1)}~=~{1 \over k+1} \sum_{m=0}^k
(\omega_{k+1})^{m(j+1)} }
where $\omega_{k+1}=e^{2i\pi/(k+1)}$, and rewrite
\eqn\rewdnca{\eqalign{
d_n(z_{k,l})~&=~{1 \over k+1} \sum_{m=0}^k \sum_{j=1}^n (\omega_{k+1})^{m(j+1)}
a_{n,j} \cr
&=~ {1 \over k+1} \sum_{m=0}^k \sum_{j=1}^n {2n \choose n-j} \bigg[ 
(\omega_{k+1})^{m(j+1)}-2 (\omega_{k+1})^{mj}+(\omega_{k+1})^{m(j-1)}\bigg]\cr
&-{2n \choose n-1} \cr
&=~-{1 \over k+1} \sum_{m=0}^k (2 \sin {\pi m \over k+1})^2
\sum_{j=1}^n {2n \choose n-j}(\omega_{k+1})^{mj} -{2n \choose n-1}\cr
&=~-{1 \over 2(k+1)} \sum_{m=0}^k (2 \sin {\pi m \over k+1})^2
\bigg[ \big(\sqrt{\omega_{k+1}}+{1 \over \sqrt{\omega_{k+1}}}\big)^{2n}-
{2n \choose n} \bigg] \cr
&-{2n \choose n-1}\cr
&=~c_n - {1 \over 2(k+1)} \sum_{m=1}^k (2 \sin {\pi m \over k+1})^2
(2 \cos {\pi m \over k+1})^{2n} \cr}}
which is equivalent to \gendn. In the second line of \rewdnca, we have performed
two discrete integrations by parts, which have produced the boundary term
${2n \choose n-1}$. In the fourth line of \rewdnca, we have used the reality of
$d_n(z_{k,l})$ to express the sum over $j$ as
\eqn\interstep{ \sum_{j=1}^{n} {2n \choose n-j} {\omega^j+ \omega^{-j}\over 2}=
{1 \over 2}\left[
\big( \sqrt{\omega} +{1 \over \sqrt{\omega}} \big)^{2n} -{2n \choose n}\right] }
In the last line of \rewdnca, we have used the sum rule
\eqn\smrcos{ {1 \over 2(k+1)} \sum_{m=0}^k \left( 2 \sin {\pi m \over k+1} \right)^2
~=~ 1 }
and recombined ${2n \choose n} -{2n \choose n-1}=c_n$.

\appendix{B}{The Gram matrix at $q=\sqrt{2}$}

Let us illustrate the conjecture \conjex\ in the case $k=3$, $l=1$, namely
$q=z_{3,1}=\sqrt{2}$. 
For $n=3,4$ we have the following identities relating the last line
of ${\cal G}_n(\sqrt{2})$ to those corresponding to diagrams of maximal
height $2$
\eqn\exahtwo{\eqalign{
\figbox{1.5cm}{onew.eps}~&=~ \sqrt{2} \left( \figbox{1.5cm}{e1w.eps} +
\figbox{1.5cm}{e2w.eps} \right) - \left( 
\figbox{1.5cm}{e21w.eps} +\figbox{1.5cm}{e12w.eps} \right) \cr 
\figbox{1.5cm}{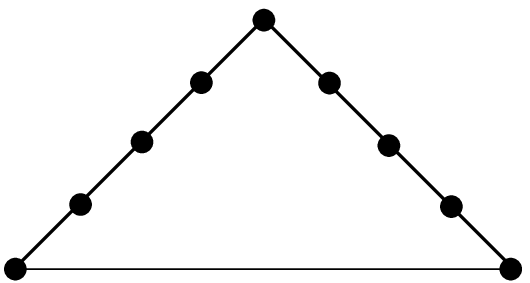}~&=~\sqrt{2} \left( \figbox{1.5cm}{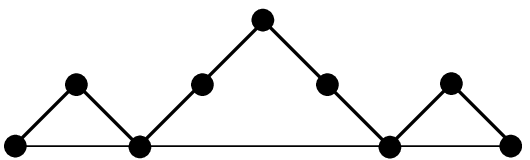} +
\figbox{1.5cm}{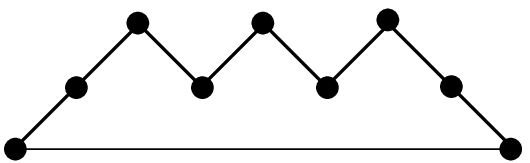} \right) - \left( 
\figbox{1.5cm}{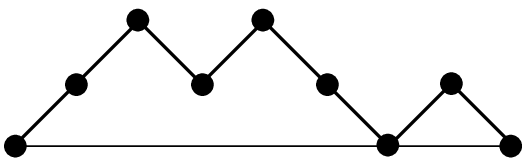} +\figbox{1.5cm}{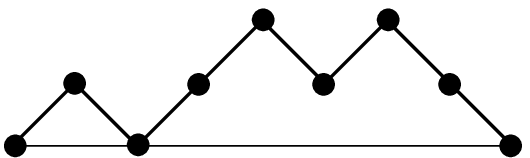} \right) \cr }}
where each line vector is represented by its labeling diagram.
In turn, the labeling diagram represents a basis 1 element for $TL_n(q=\sqrt{2})$.
The equations \exahtwo\ translate into the fact that the element
\eqn\eltl{E_3(e_1,e_2)~=~
1  -\sqrt{2} (e_1+e_2) +(e_2 e_1+e_1 e_2) }
is {\it orthogonal} (with respect to the scalar product \defsca) 
to all the elements of respectively $TL_3(\sqrt{2})$ and  $TL_4(\sqrt{2})$.
This is a direct consequence of the following identities
\eqn\idenftwo{\eqalign{ 
e_1\, E_3(e_1,e_2)~&=~e_2\, E_3(e_1,e_2) ~=~0 \cr
{\rm Tr} \big(1 \, E_3(e_1,e_2) \big)~&=\eta U_3(\sqrt{2})~=~0 \cr
{\rm Tr} ( e_3 E_3(e_1,e_2) )~&=~\sqrt{2}\,U_3(\sqrt{2})~=~0 \cr } }
where the first and second lines are valid in both $TL_3(\sqrt{2})$
($\eta=1$) and $TL_4(\sqrt{2})$ ($\eta=\sqrt{2}$),
and the third line holds only in $TL_4(\sqrt{2})$.

More generally, the element \eltl\ is orthogonal to all the elements of
$TL_n(\sqrt{2})$ for any $n \geq 5$ as the $e_i$ commute with
$E_3(e_1,e_2)$ for $i \geq 4$.
For $n \geq 5$ however, all the linear combinations we get involve
diagrams with some heights $\geq 3$.
For instance, for $n=5$, the first combination reads 
\eqn\fivecomb{\figbox{1.5cm}{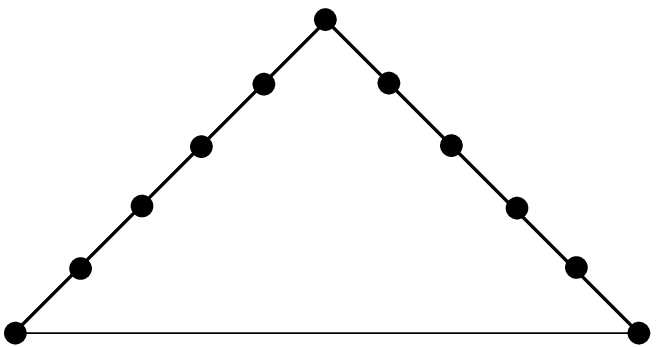}~=~\sqrt{2} 
\left( \figbox{1.5cm}{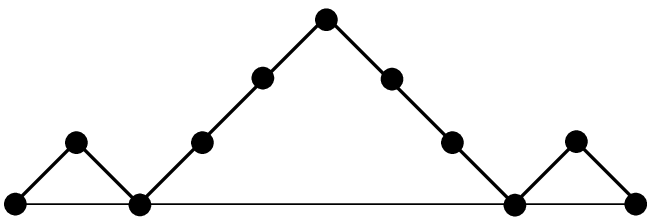} +
\figbox{1.5cm}{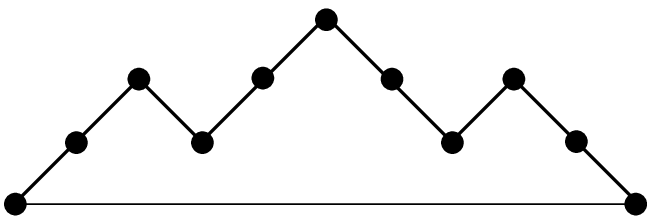} \right) - \left( 
\figbox{1.5cm}{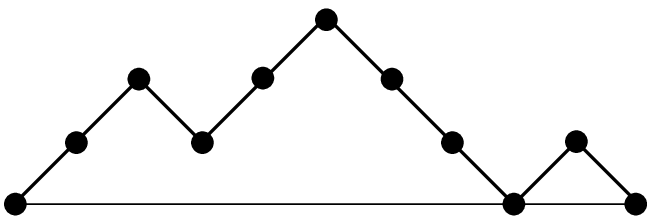} +\figbox{1.5cm}{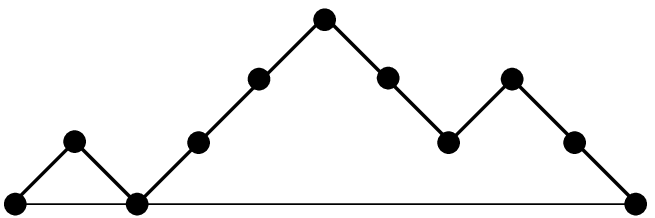} \right)}
\fig{The enhancement transformation of a walk diagram. The
walk diagram $a=AB \in W_n$ is enhanced at the point marked by a dot,
by simply inserting a maximum at this point. Here $A=l$ and
$B=r^t$, as the marked point lies in the middle of the diagram.
The enhanced diagram belongs to $W_{n+1}$.}{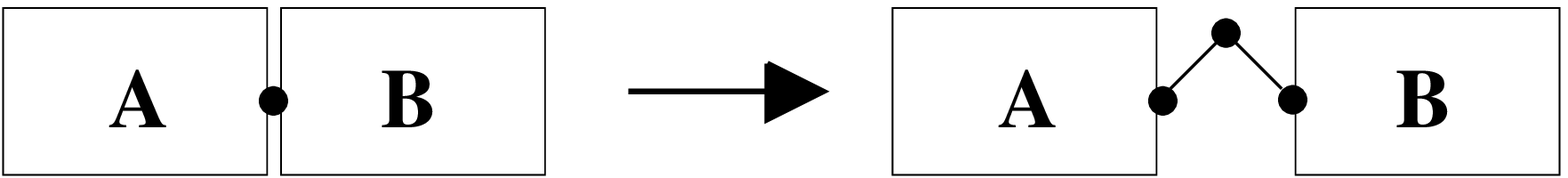}{9.cm}
\figlabel\enhance
Note that going from $TL_4$ to $TL_5$ (as well as going from $TL_3$ to $TL_4$)
amounts simply to enhancing the middle 
part of the diagrams, as depicted in Fig.\enhance, which results 
in a middle height $\ell_4=2 \to \ell_5'=3$ for the
four diagrams on the r.h.s. of \fivecomb.  
To reexpress the combination
\fivecomb\ in terms of diagrams of $W_{5,2}$, we note that the four
diagrams appearing in the r.h.s. of \fivecomb\ contain a middle 
sequence of heights of
the form $(\ell_3=1,\ell_4=2,\ell_5=3,\ell_6=2,\ell_7=1)$, as the result of two
successive enhancements. 
Using the first line of \exahtwo, we may rewrite this central part 
as a linear combination of four diagrams with central height $\leq 2$, 
which results in the four combinations
\eqn\fourco{\eqalign{ \figbox{1.5cm}{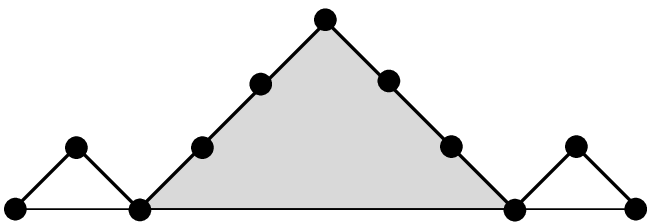}~&=~\sqrt{2} 
\left( \figbox{1.5cm}{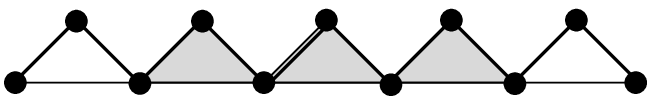} +
\figbox{1.5cm}{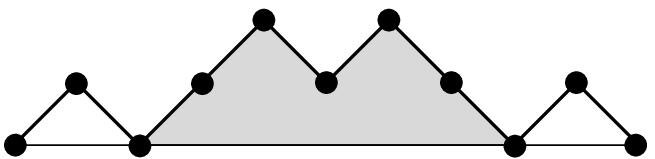} \right) - \left( 
\figbox{1.5cm}{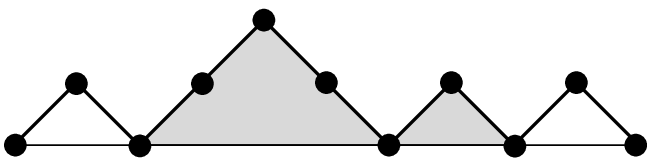} +\figbox{1.5cm}{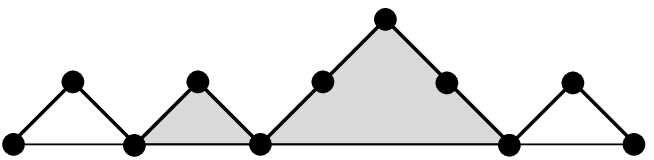} \right) \cr
 \figbox{1.5cm}{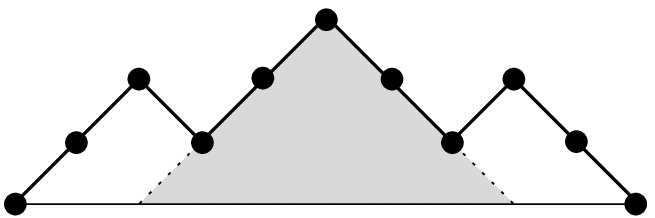}~&=~\sqrt{2} 
\left( \figbox{1.5cm}{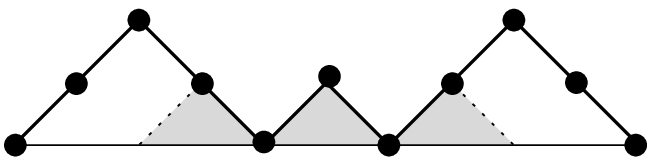} +
\figbox{1.5cm}{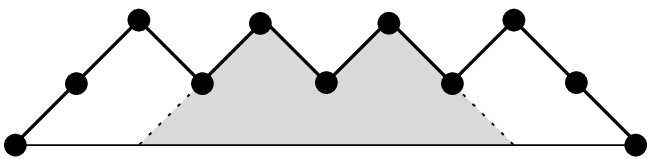} \right) - \left( 
\figbox{1.5cm}{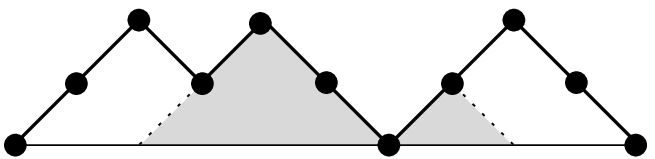} +\figbox{1.5cm}{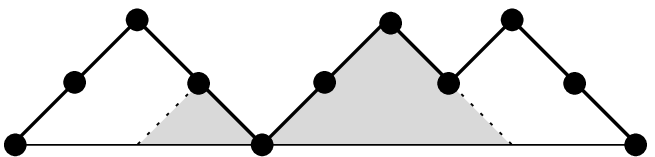} \right) \cr
 \figbox{1.5cm}{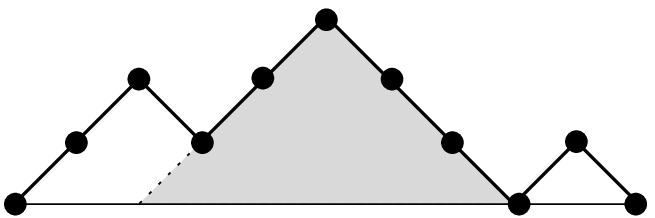}~&=~\sqrt{2} 
\left( \figbox{1.5cm}{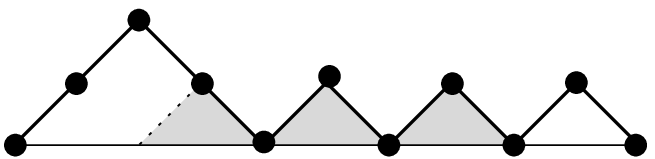} +
\figbox{1.5cm}{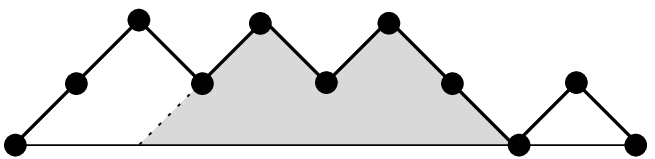} \right) - \left( 
\figbox{1.5cm}{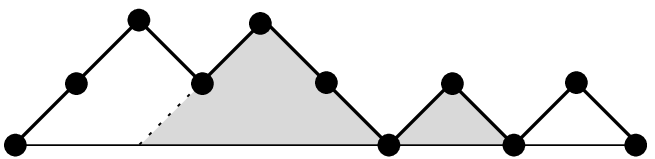} +\figbox{1.5cm}{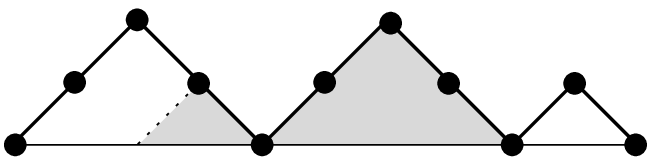} \right) \cr
 \figbox{1.5cm}{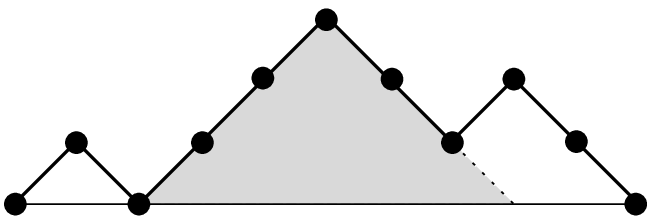}~&=~\sqrt{2} 
\left( \figbox{1.5cm}{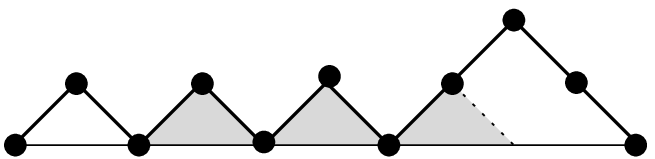} +
\figbox{1.5cm}{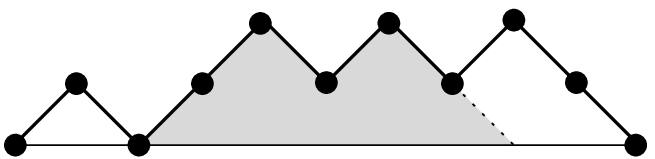} \right) - \left( 
\figbox{1.5cm}{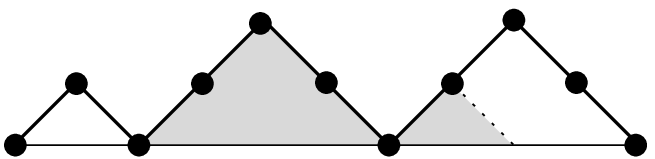} +\figbox{1.5cm}{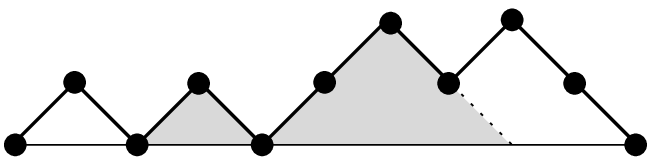} \right) \cr } }
which, upon substitution into \fivecomb,
yield the desired expression of the last line of 
${\cal G}_5(\sqrt{2})$ as a linear combination of the $2^{5-1}=16$ lines
corresponding to the elements of $W_{5,2}$. Note that all these diagrams 
have middle height $1$.  
For general $n$, we have the following recursive algorithm 
to generate the desired linear combination expressing the
last line of ${\cal G}_n(\sqrt{2})$ in terms of the lines
$a \in W_{n,2}$, denoted by $K_n=\sum_{a \in W_{n,2}} \lambda_a^n (a)$. 
The combinations $K_3$, $K_4$ and $K_5$ have been constructed above.
Suppose we have constructed $K_n$. Two situations may occur for $K_{n+1}$.

\item{(i)} If $n=2p-1$, the combination $K_{2p}$ is simply obtained by 
enhancing (see Fig.\enhance) the middle of all the diagrams of $W_{2p-1,2}$
appearing in $K_{2p-1}$, and keeping the
coefficients of the combination fixed. But as the middle
heights always satisfy $\ell_n=n \ {\rm mod}\ 2$, for all $n$, 
the diagrams of $W_{2p-1,2}$ have all middle height $\ell_{2p-1}=1$. 
Therefore, the combination $K_{2p}$ only contains elements of 
$W_{2p,2}$, with middle height $\ell_{2p}=2$.

\item{(ii)} If $n=2p$, the combination $K_{2p+1}$ is obtained in two steps.
First enhance the middle of all the diagrams in $K_{2p}$ to get another
linear combination $L_{2p+1}$. According to the
previous discussion, the enhanced diagrams in $L_{2p+1}$ have all middle height
equal to $3$. But they actually arise from the diagrams appearing in 
$K_{2p-1}$, after {\it two successive} enhancements. 
This means that they all contain a middle sequence of heights of the form
$(\ell_{n-1}=1,\ell_{n}=2,\ell_{n+1}=3,\ell_{n+2}=2,\ell_{n+3}=1)$.
The second step consists in using the first line of \exahtwo\ to reexpress
this middle piece as a linear combination of diagrams with middle height 
$1\leq 2$. This yields $K_{2p+1}$ after substitution in $L_{2p+1}$.

By carefully following the above algorithm, we find the following compact 
expression for the linear combination $K_{2p+1}$.
\eqn\compK{\enca{ ({\cal W}_{2p+1}^{(2p+1)})~=~K_{2p+1}~=~ 
\sum_{j=0}^{p} (-1)^j (\sqrt{2})^{p-j}
\sum_{a \in I_j} (a) }}
where the sets $I_j \subset W_{2p+1,2}$ are constructed recursively as follows.
$I_0$ is the set of symmetric diagrams of $W_{2p+1,2}$. 
$I_k$ is the set of diagrams of $W_{2p+1,2}$ which may be obtained
from diagrams in $I_{k-1}$ by {\it one} box addition, and which 
are not already elements of some $I_{k-l}$, $l \geq 1$.
One can easily show that card$(I_j)=2^p {p \choose j}$. 
The reader will easily check 
\compK\ for $n=3,4,5$, with the previous expressions 
\exahtwo\-\fivecomb\-\fourco. The expression for $K_{2p+2}$ is easily
obtained by enhancing $K_{2p+1}$ (case (i) above).

This leads to the relation \conjex\ linking the semi-meander polynomial
of degree $(2p+1)$
at $q=\sqrt{2}$ to the polynomials \popoli\ 
corresponding to the closures of all 
$a \in W_{2p+1,2}$, at the same value of $q$
\eqn\polexconj{\enca{ {\bar m}_{2p+1}(\sqrt{2}) ~=~ \sum_{j=0}^p
(-1)^j (\sqrt{2})^{p-j} \sum_{a \in I_j} {\bar m}(a,\sqrt{2}) }}
This proves the conjectured relation \conjex\ in the case $k=3,l=1$.
Note also that changing $\sqrt{2} \to -\sqrt{2}$ in \polexconj\
gives an analogous relation in the case $k=3,l=2$.

More generally, the element $\varphi_n^{(n)}=E_n(e_1,...,e_{n-1})$ \deffi\ 
is orthogonal to all the elements
of $TL_n(q=2 \cos \pi/(n+1))$, as a consequence of the identities
\eqn\idenn{\eqalign{ e_i\, \varphi_n^{(n)}~&=~0 \quad {\rm for}\ i=1,2,...,n-1 \cr
{\rm Tr} \varphi_n^{(n)} ~&=~ U_n(q=2 \cos {\pi\over n+1})~=~0 \cr} }
This permits to
express the last line of ${\cal G}_n(q=2 \cos \pi/(n+1))$ (corresponding
to the diagram ${\cal W}_n^{(n)}$ or equivalently to
the element $({\cal W}_n^{(n)})_1=1$) 
as a linear combination of the $(c_n-1)$ other lines,
corresponding to diagrams with heights $\leq (n-1)$, and middle
height $(n-2)$. 
This implies in particular that $r_n\big(2 \cos \pi/(n+1)\big)\leq c_n-1$,
and agrees with the conjectured relation \mulrank, which reads here
\eqn\agre{ d_n(2 \cos {\pi \over n+1})~=~ 1 
\qquad r_n(2 \cos {\pi \over n+1})~=~ c_n-1}
For $m>n$, $E_n(e_1,...,e_{n-1})$ remains orthogonal to all the elements of
$TL_m(2 \cos \pi/(n+1))$. This results in an expression of the last line of
${\cal G}_m(2 \cos \pi/(n+1))$ as a linear combination of the $(c_n-1)$
repeated ($m-n$ times)
enhancements of the elements of $W_{n,n-1}$, which belong to $W_{m,m-1}$.
For $m=n+1$, the elements of the enhanced linear combination still lie in 
$W_{n,n-1}$ as only the middle heights have been affected, and changed from 
$(n-2)$ to $(n-1)$. Hence all the linear combinations corresponding to
$m=kn+1$ are the trivial enhancements of the linear combination at 
$m=kn$. In all the other cases, many reductions must be applied to 
the diagrams to eventually get a linear combination of elements of 
$W_{m,n-1}$ only. We will not discuss the details of this mechanism here.

\appendix{C}{Proof of the sum rule \sumfab\ }

We wish to establish the following result
\eqn\resfab{\enca{ 
\sum_{a,b \in W_n} f_{a,b}~=~ 2^n c_n -{1 \over 8} 2^{n+1} c_{n+1} }}
valid for $n \geq 1$ (we set the number on the lhs of \resfab\ to be $1$ when $n=0$). 
By a simple rearrangement of factorials, this
is readily seen to be equivalent to \sumfab.
Our strategy will be the following. 
First we write a system of recursion relations linking the numbers \resfab\ 
to other numbers, to be defined below. We proceed and show that this set completely
determines all the numbers, provided we take some suitable boundary conditions.
Finally, we solve the system explicitly, and extract back the exact value \resfab.

Like in Sect. 6.9, we denote by $P_n$ the set of unrestricted walks $a$, such that
$\ell_0^a=\ell_{2n}^a=0$, without the positivity constraint on the $\ell_i^a$'s.
Let $P_n^{(-k)}$ denote the set of walks $a\in P_n$, whose 
(possibly negative) heights are 
bounded from below by $-k$, $k$ a given nonnegative integer.
\eqn\bounbe{P_n^{(-k)}~=~\{ a \in P_n, {\rm s.t.}\, \ell_0^a=\ell_{2n}^a=0 \,
{\rm and}\, \ell_i^a \geq -k,\, \forall\, i \} }
In particular, $P_n^{(0)}=W_n$.
Note also that if $k \geq n$, the above restriction amounts to no restriction
at all, hence $P_n^{(-k)}=P_n$.
We define
$\eta_n^{(k)}$ to be the total number of couples $(a,b)$, 
$a\in P_n^{(-k)}$ and $b \in W_n$, such that $a$ is $b$-symmetric
\eqn\defeta{ \eta_n^{(k)}~=~ \sum_{a\in P_n^{(-k)}, \, b\in W_n} f_{a,b} }
and $E_k$ the generating function
\eqn\geneta{ E_k(x)~=~ \sum_{n=0}^\infty \eta_n^{(k)}\, x^n }
Again, whenever $k \geq n$, we simply have 
\eqn\bounda{\eta_n^{(k)}~=~\sum_{a \in P_n, \, b \in W_n} 
f_{a,b}~=~2^n \, c_n}
as shown in \compaff.

The desired result \resfab\ amounts to writing that
\eqn\desir{ E_0(x)~=~ C(2x)-{1 \over 8x}\big( C(2x)-1-2x \big) }
where $C(x)$ denotes the generating function \catge\ of the Catalan numbers
(the subtractions in the second term are {\it ad hoc} to yield the 
initial value $\eta_0^{(0)}=1$).

\fig{The recursion for $\eta_{n+1}^{(k)}$. The diagram $b\in W_{n+1}$ is represented
as an arch configuration, and we have represented its leftmost arch, 
separating its interior piece $b_1\in W_j$ from its exterior piece $b_2\in W_{n-j}$.
The $a$'s $\in P_{n+1}^{(-k)}$ which are $b$-symmetric are of either form depicted. 
In the first case, $\ell_1^a=\ell_{2j+1}^a=1$. 
The piece $a_1$ of $a$ between these two points is $b_1$-symmetric, 
and has its restriction condition lowered by $1$: $a_1\in P_j^{(-k-1)}$
(the dashed line represents the $\ell=0$ line in the $a_i$'s). There
are $\eta_j^{(k+1)}$ such couples $(a_1,b_1)$. 
In the second case, $\ell_1^a=\ell_{2j+1}^a=-1$. $a_1$ is $b_1$-symmetric, 
but now its restriction condition is raised by $1$: $a_1\in P_j^{(-k+1)}$.
There are $\eta_j^{(k-1)}$ such couples $(a_1,b_1)$.
The piece $a_2$ is $b_2$-symmetric and has its restriction condition 
unchanged in both cases: $a_2\in P_{n-j}^{(-k)}$. The are $\eta_{n-j}^{(k)}$
couples $(a_2,b_2)$.}{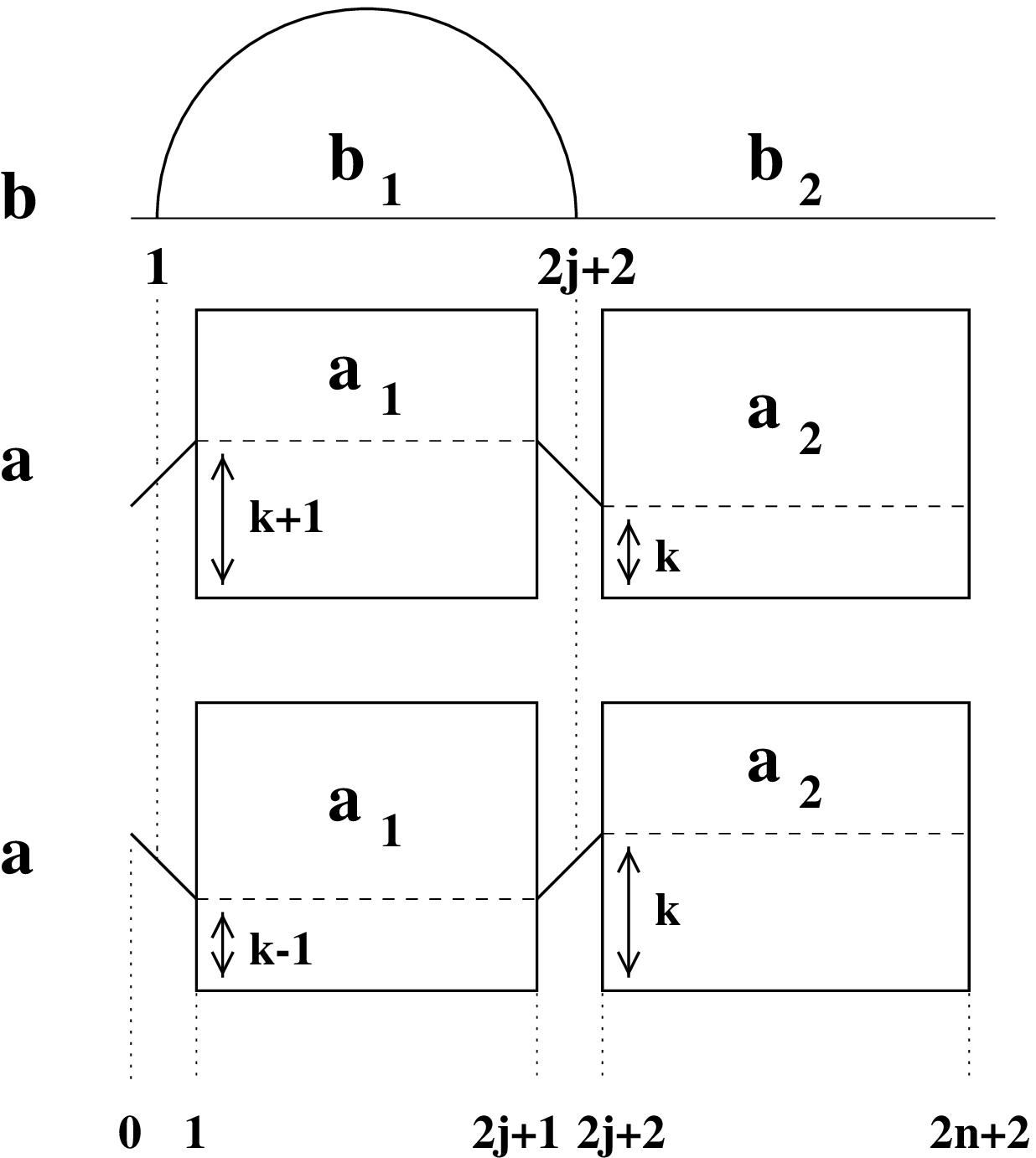}{7.cm}
\figlabel\recure

Let us now derive a system of recursion relations for the numbers
$\eta_n^{(k)}$.  Let us count the pairs of walk diagrams 
$(a\in W_{n+1},b\in W_{n+1})$ such that $a$ is $b$-symmetric. Representing $b$
in the arch configuration picture as in Fig.\recure, let us concentrate 
on its leftmost arch, 
connecting the first bridge (1) to, say, the bridge $(2j+2)$ 
(the bridge number must clearly be even).  
This arch isolates its interior, corresponding to the bridges 
$2$, $3$, ..., $(2j+1)$ from its exterior, corresponding to the bridges 
$(2j+3)$, ..., $(2n+2)$: these two sets of bridges cannot be connected to each other.
Let us now count the $a$'s which are $b$-symmetric, and consider
an $a\in W_{n+1}$, such that $f_{a,b}=1$. The part $a_1$ of $a$ 
corresponding to the interior $b_1$ of the leftmost arch of $b$ 
is symmetric w.r.t. this piece of $b$. 
The same holds for the part $a_2$ of $a$ corresponding to the exterior $b_2$
of this arch, which may be simply seen as a walk with $2(n-j)$ steps, i.e.
an element of $W_{n-j}$. In addition, we also have $\ell_1^a-\ell_0^a=1=\ell_{2j+1}^a-\ell_{2j+2}^a$
by symmetry w.r.t. the leftmost arch of $b$, which implies that 
$\ell_1^a=\ell_{2j+1}^a=1$, while $\ell_i^a \geq 0$ for $i=1,2,...,2j+1$. 
Therefore, by a trivial translation of the heights and bridge 
numbers $\ell_i'=\ell_{i+1}-1$,
the part of $a$ corresponding to the interior of the arch may be seen as a walk of
$(2j)$ steps with ${\ell_0^a}'={\ell_{2j}^a}'=0$, but with the constraint that 
${\ell_i^a}'\geq -1$ for $i=0,1,...,2j$, hence as an element of $P_j^{(-1)}$.
Conversely, we may build any $a$ which is $b$-symmetric by the juxtaposition
of a walk in $P_j^{(-1)}$ and one in $W_{n-j}$, with the respective conditions
that they are $b$-symmetric w.r.t. the corresponding portions of $b$, and elevating
the interior portion by shifting the $\ell^a$'s of $P_j^{(-1)}$ by $+1$, and adding
$\ell_0^a=\ell_{2j+2}^a=0$.  This is summarized in the following recursion relation
\eqn\recufirst{ \eta_{n+1}^{(0)}~=~ \sum_{j=0}^n \eta_{j}^{(1)}\,
\eta_{n-j}^{(0)} }
More generally, the same reasoning applies to $\eta_{n+1}^{(k)}$, with the
result (see Fig.\recure)
\eqn\recubelle{  \eta_{n+1}^{(k)}~=~ \sum_{j=0}^n \big(\eta_{j}^{(k+1)}+
\eta_j^{(k-1)} \big) \,
\eta_{n-j}^{(k)} }
where two situations may now occur for the part of $a$ corresponding 
to the interior of the arch: either $\ell_1^a=\ell_{2j+1}^a=1$, in which case the restriction condition on $a$ is lowered by $1$ (term $\eta_j^{(k+1)}$), 
or $\ell_1^a=\ell_{2j+1}^a=-1$, which may occur as soon as $k\geq 1$, in
which case the restriction condition is raised by $1$ (term $\eta_j^{(k-1)}$).
The exterior part of $a$ is unaffected and keeps the restriction condition 
at level $-k$ (term $\eta_{n-j}^{(k)}$).
We may take \recubelle\ as generic recursion relation, also valid for $k=0$,
provided we define $\eta_n^{(-1)}\equiv 0$ for all $n \geq 0$.
In addition to this boundary condition, we set $\eta_0^{(k)}=1$ for all $k$
(there is exactly one walk diagram of $0$ steps, with $\ell_0=0$,
whatever the restriction $k$).

The recursion relations \recubelle\ together with the boundary conditions
\eqn\boucon{ \eta_n^{(-1)}~=~ 0 \qquad  \eta_0^{(k)}~=~1 }
determine all the numbers $\eta_n^{(k)}$ completely. Indeed, \recubelle\
expresses $\eta_{n+1}$ in terms of $\eta_j$, $j\leq n$, hence by repeated
applications, we may express all the numbers $\eta_n^{(k)}$ in terms of the
collection of numbers $\eta_0^{(k)}$. This establishes the uniqueness of the
solution to \recubelle\-\boucon, provided it exists. To show the existence, 
we next exhibit the solution explicitly.
It is best expressed in terms of the generating functions $E_k(x)$ \geneta,
in terms of the variable
\eqn\varexp{ y={C(2x)-1 \over 2}=\sum_{n=1}^\infty 2^{n-1} \, c_n \, x^n }
easily invertible as 
\eqn\inverty{ x~=~{y \over (2y+1)^2} }
by use of \catge.
The general solution reads
\eqn\solrec{\enca{\eqalign{ 
E_{2k}(x)~&=~2y+1- {y+1 \over U_k(1/y)U_{k+1}(1/y) } \cr
E_{2k+1}(x)~&=~2y+1 - {(2y+1)(y+1) \over y \big(U_k(1/y)+U_{k+1}(1/y)\big)
\big(U_{k+1}(1/y)+U_{k+2}(1/y)\big) } \cr}}}
where $U_k(z)$ denote the Chebishev polynomials \eqtc.
Note in particular that for $k=0$, we recover $E_0(x)~=~1+2y -y(y+1)=1+y-y^2$,
which yields the desired result \desir, and therefore proves \resfab.
The first few generating functions read
\eqn\firstfew{\eqalign{
E_0(x)~&=~1+y-y^2 \cr
E_1(x)~&=~{(1-y)(2y+1)^2 \over 1+y-y^2}\cr
E_2(x)~&=~{ 1+y-2 y^2 - y^3 \over 1-y} \cr
E_3(x)~&=~{(1-2y^2)(2y+1)^2 \over (1+y-y^2)(1+y-2 y^2 - y^3)} \cr}}
Note also that the expressions \solrec\ make it clear that the $E_k(x)$
converge uniformly towards $(2y+1)=C(2x)$ when $k \to \infty$, for small
enough $x$ (indeed, when expanded at small $y$, \solrec\ reads
$E_k(x)=2y+1 +O(y^{k+1})\to 2y+1$ when $k \to \infty$). 
This is not surprising, as letting $k$ tend to infinity amounts to 
progressively removing the constraints on the counted paths, whose numbers
tend to  $2^n c_n$ (they are actually exactly equal to this for all $n \leq k$),
and $2y+1=C(2x)$ is precisely the generating function for unconstrained paths.

To prove \solrec, let us rephrase the recursion relations \recubelle\ in terms
of generating functions. We have
\eqn\generecu{ E_k(x)-1~=~ x E_k(x) \big( E_{k+1}(x)+E_{k-1}(x) \big) }
where we have used the boundary condition $\eta_0^{(k)}=1 \Rightarrow E_k(0)=1$.
The remainder of \boucon\ implies that
\eqn\bougenef{ E_{-1}(x)~=~0 }
It is now a straightforward but tedious exercise to check that \generecu\
is satisfied by \solrec.

For odd $k=2p+1$, we have
\eqn\calca{ \eqalign{1- x( E_{2p+2}(x)+&E_{2p}(x)) \cr
&=~ 1-{2y \over 2y+1}+{y(y+1) \over (2y+1)^2}{U_{p+2}(1/y)+U_p(1/y) \over
U_p(1/y) U_{p+1}(1/y) U_{p+2}(1/y) }\cr
&=~ {1 \over 2y+1}+{y+1 \over (2y+1)^2 U_p(1/y) U_{p+2}(1/y) } \cr
&=~ {(2y+1)U_p(1/y) U_{p+2}(1/y)+ y+1 \over  
(2y+1)^2 U_p(1/y) U_{p+2}(1/y) } \cr
} }
where, in the second line, we have used the recursion relation \tche. 
On the other hand, we compute
\eqn\compucon{ 
{1 \over E_{2p+1}(x)}~=~{y\big(U_p+U_{p+1}\big)\big(U_{p+1}+U_{p+2}\big)\over
(2y+1)\bigg( y\big(U_p+U_{p+1}\big)\big(U_{p+1}+U_{p+2}\big)-y-1\bigg)} }
Using the multiplication rule
\eqn\mulrule{ U_k(t) \, U_m(t)~=~\sum_{j=|m-k|\atop j=m+k\, {\rm mod}\, 2}^{m+k}
U_j(t) }
easily proved by recursion, and implying in particular that 
$U_{p+1}^2=U_p U_{p+2}+1$, we reexpress
\eqn\reeexp{ \eqalign{
\big(U_p(t)+U_{p+1}(t)\big)\big(U_{p+1}(t)&+U_{p+2}(t)\big)\cr
&=~ U_{p+1}( U_p+U_{p+2}) + U_{p+1}^2 +U_p U_{p+2}\cr
&=~ (t+1) U_{p+1}^2 + U_p U_{p+2}\cr
&=~ (t+1) ( U_p U_{p+2} +1)+ U_p U_{p+2}\cr 
&=~ (t+2)U_p(t) U_{p+2}(t) +t+1 \cr}}
by various applications of \mulrule.
Substituting this into \compucon, with $t=1/y$, this gives exactly \calca,
thus proving \generecu\ for $k=2p+1$.

For even $k=2p$, we have 
\eqn\calcul{\eqalign{1- x\big( E_{2p+1}(x)+&E_{2p-1}(x)\big) \cr
&=~{1 \over 2y+1}+ {y+1 \over 2y+1}{U_{p-1}+U_p+U_{p+1}+U_{p+2} 
\over (U_{p-1}+U_p)(U_p+U_{p+1})(U_{p+1}+U_{p+2})} \cr
&=~{1 \over 2y+1}+ {y+1 \over y(2y+1)(U_{p-1}+U_p)(U_{p+1}+U_{p+2})} \cr
&=~{(U_{p-1}+U_p)(U_{p+1}+U_{p+2})+(y+1)/y \over 
(2y+1)(U_{p-1}+U_p)(U_{p+1}+U_{p+2})}   \cr}}
We then compute
\eqn\compupac{\eqalign{ (U_{p-1}(t)+&U_p(t))(U_{p+1}(t)+U_{p+2}(t))+t+1\cr
&=~
\big( U_{p-1} U_{p+1}+U_p U_{p+2}  \big)+ U_{p-1} U_{p+2}+ U_p U_{p+1}+t+1 \cr
&=~ \big( t U_p U_{p+1} -1\big)+\big( U_p U_{p+1} -U_1 \big)+U_p U_{p+1}+t+1\cr
&=~ (t+2) U_p(t) U_{p+1}(t) \cr}}
Finally, we write
\eqn\fincalcu{ {1 \over E_{2p}(x)}~=~{U_p(1/y)U_{p+1}(1/y) \over 
(2y+1) U_p(1/y) U_{p+1}(1/y) -y-1}  }
which, upon the substitution of \compupac, with $t=1/y$, is equal to
\calcul. This completes the proof of \generecu\ for $k=2p$.

\listrefs


\bye